%% file: DIDINT.tex
\documentclass[12pt]{article}
\usepackage{setspace} 
\usepackage[utf8]{inputenc}
\usepackage[english]{babel}
\usepackage[dvipsnames]{xcolor}
\usepackage[]{appendix}
\usepackage{booktabs}
\usepackage{tabularx}
\usepackage{caption}
\usepackage{array}

\usepackage{tikz}
\usetikzlibrary{arrows.meta, positioning, shapes.geometric}

\tikzstyle{startstop} = [rectangle, rounded corners, minimum width=3.5cm, minimum height=1cm, text centered, draw=black, fill=blue!10]
\tikzstyle{decision} = [diamond, aspect=2, text centered, draw=black, fill=green!20, inner sep=1pt]
\tikzstyle{process} = [rectangle, minimum width=3.5cm, minimum height=1cm, text centered, draw=black, fill=blue!20]
\tikzstyle{arrow} = [thick,->,>=stealth]

\usepackage{subcaption}
\usepackage{graphicx}
\usetikzlibrary{arrows.meta}

\makeatletter
  \newcommand*\l@authors{\@dottedtocline{1}{0pt}{0pt}}
\makeatother

\usepackage[a4paper,top=2cm,bottom=2cm,left=2cm,right=2cm,marginparwidth=2cm]{geometry}

\usepackage[parfill]{parskip}    		

\usepackage{graphicx}
\usepackage[colorlinks=true, allcolors=blue]{hyperref}
\usepackage{pdflscape}
\usepackage{rotating}
\usepackage{multirow}
\usepackage{caption}
\usepackage{subcaption} 
\usepackage[flushleft]{threeparttable}
\usepackage{multicol}
\usepackage{wrapfig}
\renewcommand{\arraystretch}{1.5}
\usepackage{amsmath}
\usepackage{amssymb}

\usepackage{amsthm}

\newtheorem{theorem}{Theorem}

\theoremstyle{definition}

\newtheorem{assumption}{Assumption}
\theoremstyle{definition}

\theoremstyle{definition}

\theoremstyle{definition}

\usepackage{enumerate}

\usepackage{comment}

\usepackage{pifont}
\usepackage{placeins}

\usepackage{pgfplots}
\usepgfplotslibrary{groupplots}
\pgfplotsset{compat=1.18}

\usepackage{tikz}
\usetikzlibrary{arrows.meta, positioning, shapes.geometric}

\tikzstyle{startstop} = [rectangle, rounded corners, minimum width=3.5cm, minimum height=1cm, text centered, draw=black, fill=blue!10]
\tikzstyle{decision} = [diamond, aspect=2, text centered, draw=black, fill=green!20, inner sep=1pt]
\tikzstyle{process} = [rectangle, minimum width=3.5cm, minimum height=1cm, text centered, draw=black, fill=blue!20]
\tikzstyle{arrow} = [thick,->,>=stealth]

\usepackage[]{natbib}
\setcitestyle{authoryear,open={(},close={)}}
\title{Good Controls Gone Bad:\\ Difference-in-Differences with Covariates\thanks{We are grateful to the Canadian Institutes of Health Research (CIHR) for funding this project: grant number PJT-175079.  Thanks to Nichole Austin, Petyo Bonev,Thomas Russell, Erin Strumpf, and Patrick Wilson for helpful comments.  We are grateful to Eric B. Jamieson for help with the software packages. Thanks to audience members at Bank of Mexico, the Canadian Economics Association conference, and Carleton Center for Monetary and Financial Economics conference, and the 2024 Southern Economics Association conference for helpful suggestions.}}
\author{Sunny Karim\thanks{Corresponding author, Carleton University, Sunny.Karim@cmail.carleton.ca } \and Matthew D. Webb \thanks{Carleton University, matt.webb@carleton.ca}}
\date{\today}

\begin{document}
\maketitle

This paper introduces the common causality of covariates (CCC) assumption in difference-in-differences with covariates. This assumption stipulates that covariates must have common effects on untreated potential outcomes across regions and periods. We also introduce Integrated Parallel Trends, a weaker alternative to conditional parallel trends that does not require point-wise equality of trends across covariate values, and can accommodate CCC violations. Pre-trend diagnostics for integrated parallel trends may reveal plausibly parallel trends that are obscured under conventional conditional parallel trends diagnostics, thereby broadening the set of empirical settings in which credible DID identification may be possible. Furthermore, we propose a new estimator called the Integrated Difference-in-Differences (DID-INT) estimator, which can identify the ATT when integrated parallel trends holds and the CCC assumption is violated. The DID-INT estimator can also accommodate treatment-effect heterogeneity with respect to covariates and compositional changes, while avoiding forbidden comparisons in staggered adoption settings.

\section{Introduction}
\label{sec:intro}

Difference-in-differences (DID) is a widely used method for assessing the effectiveness of a policy implemented non-randomly at the regional level. In the simplest two-region, two-period setting, DID compares the difference in outcomes before and after treatment between the region that received treatment and the region that did not. This simple setup serves as the building block for estimating the average treatment effect on the treated (ATT) within more complex staggered treatment rollout frameworks, as in methods proposed by \citet*{callaway2021difference, de2023prooftwfe, borusyak2017revisiting, deb2024flexible} and \citet*{sun2021estimating}.

Both conventional and modern DID approaches rely on well-documented identifying assumptions that allow researchers to estimate the ATT. Among these, the parallel trends assumption is the most crucial \citep*{roth2022s, abadie2005semiparametric, de2020twott, callaway2021difference}. It requires that, in the absence of treatment, the average outcomes of the treated and control regions would have evolved in parallel. Since we do not observe the untreated potential outcome of the treated group after treatment, researchers examine pre-intervention trends between the treated and control regions to assess the plausibility of parallel trends after the intervention. To improve the plausibility of parallel trends, researchers impose the parallel trends assumption conditional on covariates. This is known as the conditional parallel trends assumption, which requires that, conditional on the observed covariates, the average untreated potential outcomes of the treated and control regions would have evolved in parallel over time.

Define $s = 1,2,\ldots,S$ as the region index and $t=1,2,...,T$ as the period index. Additionally, let $S_i$ denote the region to which observation $i$ belongs. Conventional DID estimation strategies with $S$ regions and $T$ periods involve running the following two-way fixed effects (TWFE) regression with covariates \citep*{bertrand2004much}:
\begin{equation}
    \label{equation: TWFE}
        Y_{i,s,t} = \alpha_s + \delta_t + \beta_{D} D_{s,t} + \sum_k \gamma^k X^k_{i,s,t} + \epsilon_{i,s,t}.
\end{equation}  
Here, $Y_{i,s,t}$ is the outcome variable of interest for individual $i$ in region $s$ in period $t$. $\alpha_s$ represents region fixed effects that account for unobserved heterogeneity, and $\delta_t$ denotes time fixed effects, which capture common shocks across periods. $D_{s,t}$ is a treatment indicator, which takes a value of one when region $s$ is exposed to treatment in period $t$. $X^k_{i,s,t}$ denotes the $k$th covariate, which can either be time-invariant or time-varying. 

Current methods of including covariates implicitly assume that the effect of the covariates on the untreated potential outcome remains constant between regions and periods, but do not address this explicitly. These methods are discussed in Section \ref{sec:olddid}. In this paper, we explicitly introduce this assumption, which we call the \textbf{common causality of covariates (CCC)}. Specifically, we distinguish between three types of CCC assumptions: region-invariant CCC, time-invariant CCC, and two-way CCC. To illustrate the empirical relevance of these assumptions, we consider three examples corresponding to different forms of CCC violations. The first, based on water-use restrictions in Calgary, illustrates a violation of the region-invariant CCC assumption, where the effects of seasonal controls differ between treated and control neighborhoods. The second considers a hypothetical intervention aimed at reducing deaths from heart attacks and illustrates a violation of the time-invariant CCC assumption, since the relationship between age and heart-attack mortality changes substantially over time as individuals age. The third considers wage outcomes and education using the Canadian Labour Force Survey (LFS), and illustrates how the relationship between education and earnings can vary across both provinces and years, making violations of the two-way CCC assumption plausible. Together, these examples show that restrictions requiring covariate effects to remain constant across regions, periods, or both may be empirically restrictive and can have important consequences for DID identification.

In this paper, we first derive the two jointly necessary and sufficient conditions for conditional parallel trends under a linearity assumption on the untreated potential outcome model. The first condition requires that changes in the untreated outcome intercepts be the same across treated and control regions. The second requires that changes in the effects of the covariates between the pre- and post-intervention periods be the same across treated and control regions. Importantly, two-way CCC is sufficient, but not necessary, for the second condition. Therefore, conditional parallel trends can continue to hold when CCC is violated, provided that changes in the covariate effects satisfy the required restriction across the treated and control regions.

A key limitation of conditional parallel trends is that it is a pointwise restriction, requiring the trends in untreated potential outcomes to be parallel for every covariate value. Since there is no formal test for parallel trends, researchers usually evaluate the plausibility of conditional parallel trends using pre-treatment outcomes. This typically involves regressing outcomes on covariates and visually comparing the residualized trends of treated and control regions. However, these diagnostics often assume common covariate coefficients across regions and periods, as the possibility of conditional parallel trends holding under CCC violations had not been formally considered prior to this paper. Moreover, researchers generally do not evaluate conditional parallel trends separately at every value of the covariates, even though it is a pointwise restriction. 

Motivated by these observations, we introduce an alternative identifying assumption, which we call \textbf{Integrated Parallel Trends (IPT)}. Integrated parallel trends requires the difference in untreated potential outcome trends between the treated and control regions to equal zero after the conditional trends are integrated over the covariate distribution of the treated region. Unlike conditional parallel trends, IPT is no longer a pointwise restriction that must hold separately for every value of the covariates. Instead, untreated outcome trends may differ between the treated and control regions at individual covariate values, provided that these differences cancel after integration. Under linearity, we derive the necessary and sufficient condition under which integrated parallel trends holds. We show that integrated parallel trends can also accommodate CCC violations by allowing the effects of the covariates to vary across regions and periods. We further show that conditional parallel trends implies integrated parallel trends, while the converse need not hold. Therefore, integrated parallel trends can support identification in settings where the stronger pointwise restrictions imposed by conditional parallel trends are implausible.

We examine whether existing DID estimators can identify the ATT under integrated parallel trends. For TWFE, identification under integrated parallel trends requires a more flexible specification of the covariates that correctly accommodates the relevant CCC violations. Specifically, rather than imposing a common covariate coefficient, the flexible TWFE interacts the covariates with region indicators, period indicators, or both, depending on whether the covariate effects vary across regions, periods, or both. However, even with the correct covariate specification, identification under integrated parallel trends requires the treatment effects to be homogeneous with respect to the covariates. When treatment effects vary with the covariates, the flexible TWFE treatment coefficient generally need not identify the ATT, even when integrated parallel trends holds. Moreover, in staggered treatment-adoption settings, the flexible TWFE remains susceptible to forbidden comparisons and negative weighting under heterogeneous treatment effects \citep{goodman2021difference}. We also find several limitations of the \cite{callaway2021difference} DID (CS-DID) estimator, which was proposed to overcome forbidden comparisons and negative weighting under heterogeneous treatment effects. First, the propensity score-based implementations of CS-DID, such as inverse probability weighting and doubly robust DID, generally cannot accommodate more flexible covariate specifications, since these specifications violate the overlap condition. The outcome-regression implementation can identify the ATT under integrated parallel trends and treatment effect heterogeneity with respect to covariates. However, in repeated cross-sectional settings, this additionally requires the distribution of the covariates within each region to remain unchanged over time. Therefore, compositional changes can prevent the outcome regression from identifying the ATT under integrated parallel trends.

Given the limitations of existing DID estimators, we propose a new estimator, called the \textbf{Integrated Difference-in-differences (DID-INT)}, that can identify the ATT even when the CCC assumption is violated. The DID-INT estimator is also applicable in settings with staggered treatment adoption. Our estimator requires integrated parallel trends rather than conditional parallel trends, but remains consistent under conditional parallel trends when the two-way CCC condition is violated, without imposing additional restrictions on covariates. Moreover, identification under DID-INT can accommodate treatment effect heterogeneity with respect to covariates and compositional changes in covariates. It is also robust to forbidden comparisons and negative weighting in staggered treatment-adoption settings.

Our Monte Carlo simulations further illustrate these theoretical results. The simulations show that standard TWFE, which imposes a common covariate coefficient across regions and periods, can become inconsistent under CCC violations. Flexible TWFE specifications that allow covariate effects to vary across regions and periods can also exhibit bias, while the two-way DID-INT continues to recover the ATT. In particular, the CS-DID, imputation estimator \citep*{borusyak2024revisiting}, and FLEX estimator \citep*{deb2024flexible} can exhibit bias under the data-generating processes with CCC violations considered in our simulations. Additionally, we consider DGPs in which the covariate distribution changes over time or treatment effects depend on covariates. When the covariate distribution changes over time and covariate effects vary by state, DID-INT remains unbiased, whereas TWFE, the imputation estimator, FLEX, and CS-DID are biased. When treatment effects depend on covariates, DID-INT, TWFE, and the imputation estimator remain unbiased, whereas flexible TWFE is biased and CS-DID and FLEX exhibit small biases.

The rest of the paper is organized as follows. Section \ref{sec:olddid} reviews existing approaches to including covariates in DID. Section \ref{sec:setup} introduces the notation and setup used throughout the paper. Section \ref{sec: ccc} introduces the three CCC assumptions and discusses settings in which they may be violated. Section \ref{sec:nature} categorizes covariates according to the corresponding CCC restrictions. Section \ref{section: PT} derives the necessary and sufficient conditions for parallel trends, conditional parallel trends, and integrated parallel trends, and discusses the relationships between these assumptions. Section \ref{sec: multgroups} generalizes the analysis to multiple regions, periods, and staggered treatment adoption. Section \ref{sec: didint} introduces the Integrated Difference-in-Differences estimator and its alternative covariate specifications. Section \ref{sec:twfeproof} studies the identification properties of the standard and flexible TWFE estimators under CCC violations. Section \ref{sec:other} examines alternative DID estimators, including the Callaway and Sant'Anna estimator in Section \ref{sec:CS-DID}, the imputation estimator in Section \ref{sec:imputation}, and the FLEX estimator in Section \ref{sec:flex}. Section \ref{section: MC} presents the Monte Carlo experiments, Section \ref{sec:empex} presents an empirical example and Section \ref{sec:conclusion} concludes.
    
\section{Existing Approaches for Including Covariates}\label{sec:olddid}

While empirical researchers usually include covariates to ensure the plausibility of conditional parallel trends in DID, they might also use them to control for factors that can influence outcomes. The literature highlights the need to choose covariates carefully in DID analyses. Notably, covariates that are affected by participation in treatment, known as \textbf{bad controls}, should be excluded \citep{caetano2024PTholds}. The DID literature also recommends using either time-invariant or pre-treatment covariates when covariates vary over time \citep{caetano2024PTholds}. However, researchers might still want to include time-varying covariates, even though they are not needed to maintain parallel trends. For instance, consider a study in which we are interested in the effect of a hypothetical treatment on reducing cardiac arrests, and the treatment is implemented at the provincial level. In such a study, researchers may want to control for time-varying covariates like age and smoking status. Age, in particular, is unlikely to be affected by the treatment, and being older increases the frequency of cardiac arrests. Including pre-treatment age values in this analysis may lead to counterintuitive results, as we are unable to capture the effect of higher age on the frequency of cardiac arrests.  Additionally, many datasets are repeated cross-sections rather than true panels, and pre-treatment values are typically unavailable in these datasets.

Including time-varying covariates in DID analyses is more complex and has been examined extensively in the recent literature. \cite{caetano2024PTholds} demonstrated that the TWFE regression can produce inconsistent estimates when incorporating time-varying covariates because fixed effects not only remove nuisance factors but also alter the covariates themselves. This bias is known as the \textit{hidden linearity bias}. To clarify this bias, a simple example is provided below. Suppose a researcher is studying the effect of a hypothetical policy aimed at improving wages. The researcher is concerned that the treated and untreated regions differ because the treated region has a younger workforce, and younger workers might have experienced different wage trajectories compared to older workers, even in the absence of the policy. To account for this, the researcher includes age as a covariate in the TWFE regression to compare regions with similar age distributions. However, the issue is that after applying fixed effects, the model may condition on changes in age over time rather than on age levels themselves. Consequently, the comparison shifts from “regions with a younger population versus similar untreated regions with a younger population” to “regions with similar changes in age composition," which are not the same. This can have important consequences for standard pre-test diagnostics, where the researcher may have provided misleading evidence of conditional parallel trends across age levels, rather than changes in age, whereas the TWFE regression conditions on age-compositional changes. In staggered adoption designs with heterogeneous treatment effects, bias in TWFE is further exacerbated by negative weighting and forbidden comparisons \citep{goodman2021difference}.

For designs involving staggered adoption, the Callaway and Sant'Anna DID (CS-DID) offers a feasible alternative, which estimates the ATT without the forbidden comparisons \citep{callaway2021difference}. Estimating the ATT using this method involves two steps. First, the dataset is partitioned into several ``2x2 comparison'' blocks, each consisting of a treated region and an untreated (or not yet treated) region. The ATT for each ``2x2 comparison'' block, denoted as $ATT(g,t)$, is estimated using the DR-DID estimator developed by \cite{sant2020doubly} by default, although it can also use OR and IPW methods as well. Second, these estimates are aggregated by taking a weighted average of the $ATT(g,t) $'s from the first step to obtain an overall estimate of the ATT. Note $g$ here comprises all individuals who are first treated in the same period.  This is usually, but not always, a collection of $s$ with the same treatment timing.

A second alternative, called the static imputation estimator \citep{borusyak2017revisiting}, is done in four steps. First, a TWFE model is estimated using data for the control group. Next, this fitted model is used to predict the untreated counterfactuals for the treated observations in the post-intervention period. In step 3, the unit specific treatment effects are estimated by taking the difference between the observed outcome and the predicted counterfactual from the previous step. Finally, these unit specific treatment effects are aggregated together to get an overall estimate of the ATT. While \cite{borusyak2024revisiting} primarily focuses on event study designs, the paper also includes a section on the static version of the imputation estimator, which we examine in our paper. 

A third alternative, called FLEX, uses a linear regression with a flexible functional form.  This estimator was developed in part to allow for time-varying covariates.  The estimator includes group-by-time treatment effects, two-way fixed effects, and interaction terms with leads and lags \citep{deb2024flexible}. 

\begin{table}[h!]
\centering
\small
\setlength{\tabcolsep}{4pt}
\resizebox{\textwidth}{!}{
\begin{tabular}{lccccc}
    \toprule
    & \begin{tabular}[c]{@{}c@{}}Avoids forbidden\\ comparisons /\\ negative weights\end{tabular}
    & \begin{tabular}[c]{@{}c@{}}CCC\\ violations\end{tabular}
    & \begin{tabular}[c]{@{}c@{}}Integrated\\ parallel trends\end{tabular}
    & \begin{tabular}[c]{@{}c@{}}Treatment effects\\ vary with covariates\end{tabular}
    & \begin{tabular}[c]{@{}c@{}}Compositional\\ changes\end{tabular} \\
    \midrule

    \textbf{Standard TWFE}
    & \ding{55}
    & \ding{55}
    & \ding{55}
    & \ding{55}
    & \ding{55} \\

    \textbf{Flexible TWFE}
    & \ding{55}
    & \ding{51}$^{*}$
    & \ding{51}$^{*}$
    & \ding{55}
    & \ding{51} \\

    \textbf{CS-DID: OR}
    & \ding{51}
    & \ding{51}
    & \ding{51}$^{\dagger}$
    & \ding{51}
    & \ding{55} \\

    \textbf{CS-DID: IPW / DR-DID}
    & \ding{51}
    & \ding{55}
    & \ding{55}
    & \ding{55}
    & \ding{55} \\

    \textbf{Imputation}$^{\ddagger}$
& \ding{51}
& \ding{55}
& \ding{55}
& \ding{51}
& \ding{55} \\

\textbf{FLEX}$^{\ddagger}$
& \ding{51}
& \ding{55}
& \ding{55}
& \ding{55}
& \ding{55} \\

    \textcolor{ForestGreen}{\textbf{DID-INT}}
    & \textcolor{ForestGreen}{\ding{51}}
    & \textcolor{ForestGreen}{\ding{51}}
    & \textcolor{ForestGreen}{\ding{51}}
    & \textcolor{ForestGreen}{\ding{51}}
    & \textcolor{ForestGreen}{\ding{51}} \\

    \bottomrule
\end{tabular}
}

\vspace{0.5em}

\caption*{\footnotesize
\textit{Note}: \ding{51} = can accommodate the feature or identifying condition;
\ding{55} = cannot generally accommodate the feature or identifying condition.
$^{*}$ Requires the correct flexible covariate specification and additional restrictions for identification under integrated parallel trends.
$^{\dagger}$ Requires no compositional changes in repeated cross-sectional settings.
$^{\ddagger}$ For the Imputation and FLEX estimators, we do not provide theoretical results; the reported conclusions are based solely on the Monte Carlo experiments considered in this paper.
}

\caption{Properties of DID Estimators}
\label{table:comparison}
\end{table}

Table \ref{table:comparison} compares the identification properties of the estimators considered in this paper. Standard TWFE does not generally accommodate CCC violations or integrated parallel trends and remains susceptible to forbidden comparisons and negative weighting under heterogeneous treatment effects in staggered adoption settings \citep{goodman2021difference}. A flexible TWFE specification can accommodate CCC violations and integrated parallel trends under additional restrictions, but it does not generally identify the ATT when treatment effects vary with covariates and remains susceptible to the problems associated with staggered treatment adoption. The CS-DID estimator avoids forbidden comparisons and negative weighting, but its ability to accommodate CCC violations depends on the implementation. As we show in Section \ref{sec:CS-DID}, the outcome-regression implementation can accommodate CCC violations under integrated parallel trends when the no compositional changes assumption holds, whereas propensity-score-based implementations face additional overlap restrictions. We do not study the imputation and FLEX estimators theoretically; instead, their performance under the CCC violations considered in this paper is examined only through the Monte Carlo simulations. The imputation estimator performs well in several of our Monte Carlo designs, but exhibits bias under the CCC violations considered in the simulations, including settings with state-varying and two-way covariate effects. FLEX also exhibits bias under these CCC violations and shows a small bias when treatment effects vary with covariates.

\section{Setup} \label{sec:setup}

In this section, we introduce the notation used throughout this paper. We consider a staggered adoption setting in which some regions implement treatment at different times. Within each period, each region can be classified into two broad categories: the treated region and the control region. The control region includes either \textbf{never-treated} regions or both \textbf{never-treated} and \textbf{not-yet-treated} regions. Never-treated regions are regions that have not been treated throughout the available dataset, while not-yet-treated regions have not yet implemented treatment in period $t$. In this paper, we give researchers the flexibility to choose whether to use never-treated regions or both never-treated and not-yet-treated regions as controls. We use the treatment indicator $D_{s,t}$ defined in Section \ref{sec:intro} to classify region $s$ in period $t$, where $D_{s,t} = 1$ for treated regions and $D_{s,t} = 0$ otherwise. 
\[ D_{s,t} = \begin{cases} 
      1 & \mbox{if region $s$ is treated in period \textit{t}}.\\
      0 & \mbox{if region $s$ is never treated or not-yet treated in period \textit{t}}. \\
       \end{cases}
\]

In a staggered design, different regions can be treated in the same period $t$. Therefore, we define $\mathbf{s}^t = \{s \in S \mid D_{s,t} = 1\}$ as a \textbf{vector} of regions which are treated in period $t$. Similarly, $\mathbf{s}^c = \{s \in S \mid D_{s,t} = 0\}$ is a \textbf{vector} of regions which are either \textbf{never treated} or both \textbf{never treated and not-yet-treated} in period $t$, depending on what types of controls researchers want to use. As in \cite{callaway2021difference}, we treat the period immediately before region $s$ is treated as the pre-intervention period, which we index by the \textbf{scalar} $t^s-1$. 

\begin{figure}
            \centering
            \includegraphics[width=0.5\textwidth]{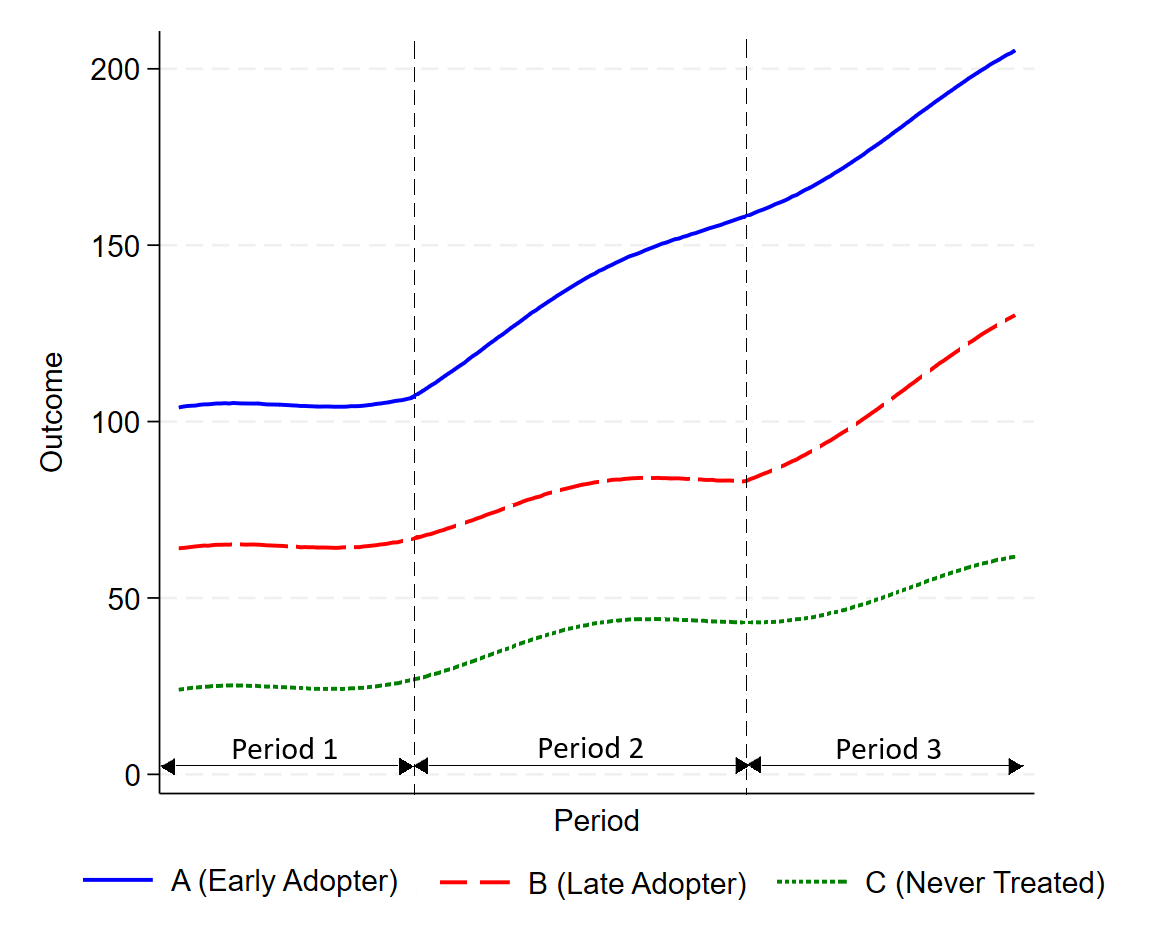}
            \caption{A three region, three period staggered design}
            \label{fig: staggereddraw}
        \end{figure}

To clearly present the notation used, consider the simple three-region and three-period case shown in Figure \ref{fig: staggereddraw}. In this example, the set of regions $S = \{A,B,C\}$ and the set of periods $T = \{1,2,3\}$. Region A receives treatment in period 2, and region B receives treatment in period 3. In period 2, Region B is categorized as a \textbf{not-yet-treated} region because it receives treatment in a later period in the available data. Region C, which remains untreated throughout the dataset, is categorized as a \textbf{never-treated} region. In period 2, $D_{A,2}$ = 1, while $D_{B,2} = D_{C,2} = 0$, and the vector of treated regions $\mathbf{s}^2 = (A)$, since only A is treated in period 2. Similarly, in period 3, $D_{A,3} = D_{B,3} = 1$, while $D_{C,3} = 0$, and $\mathbf{s}^3 = (A, B)$, since both A and B are treated in period 3.      

\section{Common Causality of Covariates}
\label{sec: ccc}

In this section, we formalize the common causality of covariates assumption, and distinguish between three types of CCC assumptions: the two-way CCC, the region-invariant CCC and the time-invariant CCC. Each assumption imposes different restrictions on the effect of the covariates across regions and periods. To clarify the CCC assumptions, we first present a model of untreated potential outcomes and then introduce the underlying assumptions made for this model. We then define the three CCC assumptions in detail and illustrate, with three examples, how these assumptions may be violated in real-world datasets.  

Let the untreated potential outcome of an individual $i$ in region $s$ at period $t$ be given by:
    \begin{equation}
                Y_{i,s,t}(0) = \sum_{\ell=1}^{S} \sum_{q=1}^{T} \theta^{\circ}_{\ell,q} Z_{i,\ell,q} + \sum_{k=1}^K \gamma_{s,t}^{\circ, k} X^k_{i,s,t}(0) + \epsilon_{i,s,t}
        \label{eq:po}
    \end{equation}

Here, there are $K$ covariates, and $X^k_{i,s,t}(0)$ represents the untreated potential value of the $k^{th}$ covariate. $Z_{i,s,t}$ denotes the set of the region-time indicator variables. The parameters $\theta^{\circ}_{s,t}$ are the coefficients associated with these region-time indicators, and $\gamma_{s,t}^{\circ, k}$ represents the true effect of the $k^{th}$ covariate on the untreated potential outcome. $\epsilon_{i,s,t}$ denotes the error term,  which satisfies the conditional mean zero assumption: $ \mathbb{E}[\epsilon_{i,s,t} \mid S_i = s, X_{i,s,t} = x] = 0$. The model for the untreated potential outcome in Equation \eqref{eq:po} imposes the following assumptions. 

\begin{assumption}[Random Sampling]
\label{as: randomsampling}
For each region-period cell $(s,t)$, the observations
\begin{equation}
\left\{Y_{i,s,t},X^k_{i,s,t}\right\}_{i=1}^{N_{s,t}}
\end{equation}
are independently and identically distributed draws from the corresponding population.
\end{assumption}

\begin{assumption}[Linearity] 
    \label{as: linearity}
        The conditional mean of the untreated potential outcome $Y_{i,s,t}(0)$ can be expressed as a linear combination of covariates $X_{i,s,t}(0)$ within each region-period cell.
\end{assumption}

\begin{assumption}[Covariates Unaffected by Treatment]
\label{as: covariateunaffected}
Treatment does not affect the covariates. That is, for all individuals, regions, and periods,
\begin{equation}
X^k_{i,s,t}(1)=X^k_{i,s,t}(0)
\qquad
\text{for all } k=1,\ldots,K.
\end{equation}
\end{assumption}

Assumption \ref{as: covariateunaffected} implies that $\mathbb{E}\biggr[X_{t}(0) \mid D_s = 1 \biggr] = \mathbb{E}\biggr[X_{t}(1) \mid D_s = 1 \biggr]$, and rules out \textbf{bad controls}, or controls which are affected by treatment. For simplicity, let us assume that there is a single covariate $X_{i,s,t}(0)$, and omit the $(0)$ since the covariate is unaffected by treatment due to Assumption \ref{as: covariateunaffected}. Therefore, we re-write equation \eqref{eq:po} as:
    \begin{equation}
                Y_{i,s,t}(0) = \sum_{\ell=1}^{S}\sum_{q=1}^{T}
                \theta^{\circ}_{\ell,q} Z_{i,\ell,q} + \gamma_{s,t}^{\circ} X_{i,s,t} + \epsilon_{i,s,t}
        \label{eq: posimplified}
    \end{equation}
Or equivalently,
    \begin{equation}
                Y_{i,s,t}(0) = \theta^{\circ}_{s,t} + \gamma^{\circ}_{s,t} X_{i,s,t} + \epsilon_{i,s,t}
        \label{eq: podoublesimplified}
    \end{equation}

\begin{assumption}[Stable Unit Treatment Value Assumption]
\label{assumption: sutva}
For each unit $i$, region $s$, and period $t$, the potential outcomes
$Y_{i,s,t}(1)$ and $Y_{i,s,t}(0)$ are well defined and depend only on the
treatment status of unit $i$'s region. In particular, there is no interference
across units or regions, and the observed outcome satisfies
\begin{equation}
\label{equation: sutva}
Y_{i,s,t}
=
D_{s,t}Y_{i,s,t}(1)
+
\left(1-D_{s,t}\right)Y_{i,s,t}(0),
\end{equation}
where $D_{s,t}\in\{0,1\}$ denotes the treatment status of region $s$ in
period $t$.
\end{assumption}

Assumption \eqref{assumption: sutva} implies that
\begin{equation}
Y_{i,s,t}
=
\begin{cases}
Y_{i,s,t}(1), & \text{if } D_{s,t}=1,\\
Y_{i,s,t}(0), & \text{if } D_{s,t}=0.
\end{cases}
\end{equation}

The ATT in terms of potential outcomes is given by:
    \begin{equation}
    \label{eq: rawatt}
        ATT := \mathbb{E}\biggr[ Y_t(1) - Y_t(0)|D_{s,t}=1 \biggr] 
    \end{equation}

In DID analyses, researchers include covariates to ensure that parallel trends are more plausible, as discussed in more detail in Section \ref{section: PT}. Despite recommendations in the DID literature, empirical researchers continue to use covariates, particularly time-varying ones, without considering the assumptions required to justify their inclusion. In this paper, we explicitly introduce yet another assumption researchers need to consider when including covariates in DID, which we call the \textbf{common causality of covariates (CCC)}. While all methods mentioned in Section \ref{sec:olddid} implicitly assume that the effects of the covariates are stable across regions and time, this assumption is not explicitly stated or analyzed directly.

\begin{assumption}[Two-way Common Causality of Covariate] 
\label{assumption: twowayCCC} 
The effect of the covariate is equal between regions and across all periods.
        \begin{equation*}
            \begin{gathered}
                \gamma^{\circ}_{s,t} = \gamma^{\circ}_{s',t'}.
            \end{gathered}
        \end{equation*}
\end{assumption}

Under Assumption \ref{assumption: twowayCCC}, the true coefficient of $X_{i,s,t}$ on the untreated potential outcome is assumed to be the same between regions and across periods. Therefore, we can re-write Equation \eqref{eq: posimplified} with a common $\gamma^{\circ}$, as shown in Equation \eqref{eq: poCCC}.
    \begin{equation}
                Y_{i,s,t}(0) = \sum_{\ell=1}^{S}\sum_{q=1}^{T}\theta^{\circ}_{s,t}Z_{i,s,t} + \gamma^{\circ} X_{i,s,t} + \epsilon_{i,s,t}
        \label{eq: poCCC}
    \end{equation}               

\begin{assumption}[Region-invariant Common Causality of Covariate]
\label{as2: regioninvariantCCC} The effect of the covariate is equal between regions, but can vary across periods.
            \begin{equation*}
                \begin{gathered}
                    \gamma^{\circ}_{s,t} = \gamma^{\circ}_{s',t} \quad \forall s,s',t.
                \end{gathered}    
            \end{equation*}
\end{assumption}

\begin{assumption}[Time-invariant Common Causality of Covariate]
\label{as2: timeinvariantCCC} The effect of the covariate is equal across all periods, but can vary between regions.
            \begin{equation*}
                \begin{gathered}
                    \gamma^{\circ}_{s,t} = \gamma^{\circ}_{s,t'} \quad \forall s,t,t'.
                \end{gathered}
            \end{equation*}
\end{assumption}

In a simple setting with two periods (1 and 2) and two regions (A and B), the \textbf{two-way CCC} implies $\gamma^{\circ}_{A,1} = \gamma^{\circ}_{A,2} = \gamma^{\circ}_{B,1} = \gamma^{\circ}_{B,2} = \gamma^{\circ}$. The \textbf{region-invariant CCC} implies $\gamma^{\circ}_{A,1} = \gamma^{\circ}_{B,1}$ and $ \gamma^{\circ}_{A,2} = \gamma^{\circ}_{B,2}$, and the \textbf{time-invariant CCC} implies $\gamma^{\circ}_{A,1} = \gamma^{\circ}_{A,2}$ and $\gamma^{\circ}_{B,1} = \gamma^{\circ}_{B,2}$. The two-way CCC assumption is more restrictive compared to Assumptions \eqref{as2: regioninvariantCCC} and \eqref{as2: timeinvariantCCC}, requiring that the effect of the covariates are the same across both regions and time. When the two-way CCC assumption holds, both the region-invariant and time-invariant CCC assumptions hold as well. However, if the two-way CCC is violated, either the region-invariant CCC, or the time-invariant CCC, or both may be violated. 

\subsection{Motivating Examples}

To motivate the practical importance of the CCC assumption we consider a number of empirically relevant situations.  Specifically, we consider examples where the coefficient is not the same across regions, across time, and across both regions and time.

\subsubsection{Region Varying Coefficients}

This example comes from \citet{bray2025bearspaw}. This paper examines the impact of water-use restrictions on water use following a water main break in the City of Calgary. The restrictions were in place during the summer and primarily applied to outdoor water use. The author intended to use urban neighborhoods as the control group and suburban neighborhoods as the treated group. This was motivated by the fact that the majority of outdoor water use is for watering lawns, which are concentrated in suburban neighborhoods.  

\begin{figure}[h!]
    \centering
    \begin{subfigure}[b]{0.31\textwidth}
        \centering
        \includegraphics[width=\textwidth]{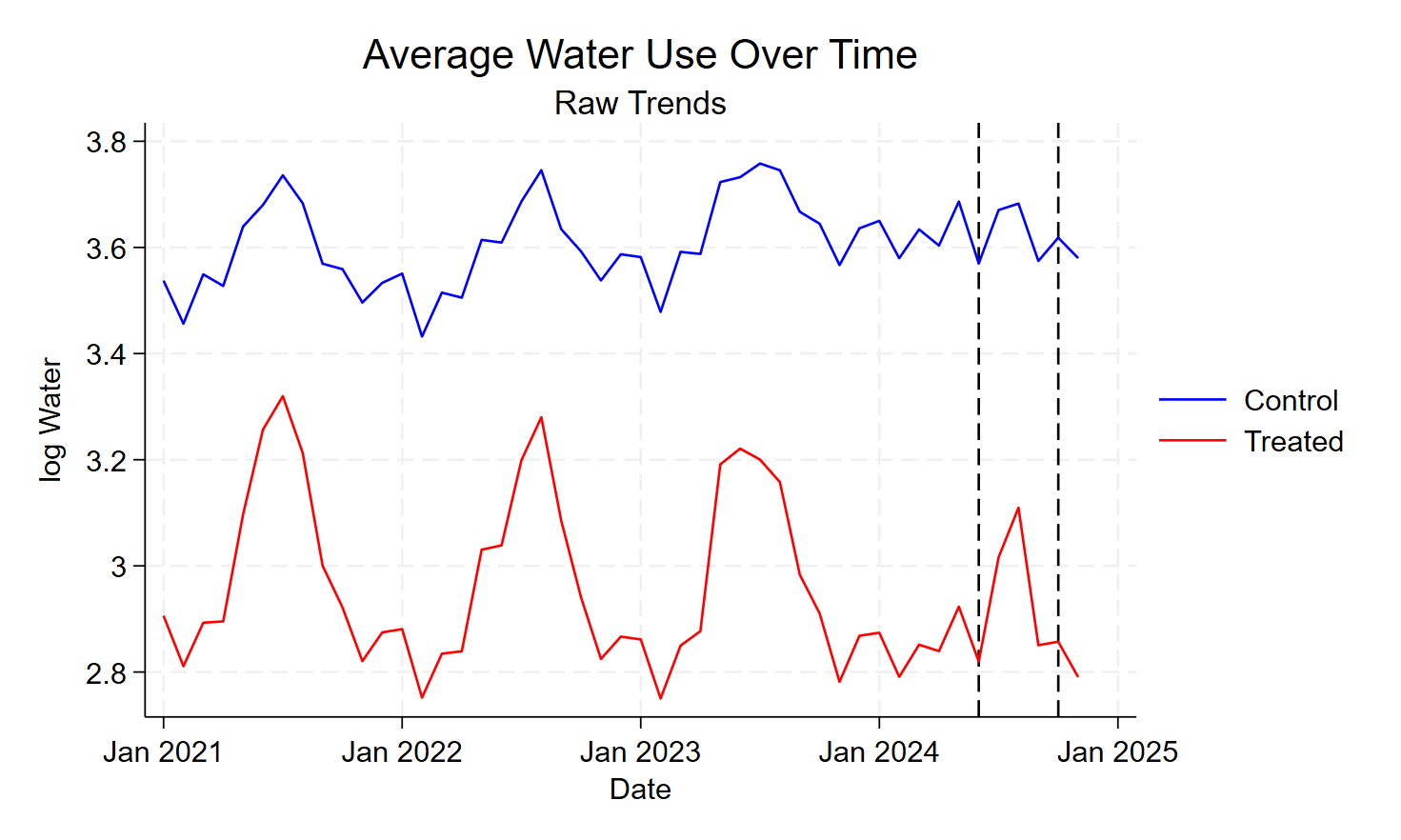}
        \caption{Raw trends}
        \label{fig:bray_raw}
    \end{subfigure}
    \hfill
    \begin{subfigure}[b]{0.31\textwidth}
        \centering
        \includegraphics[width=\textwidth]{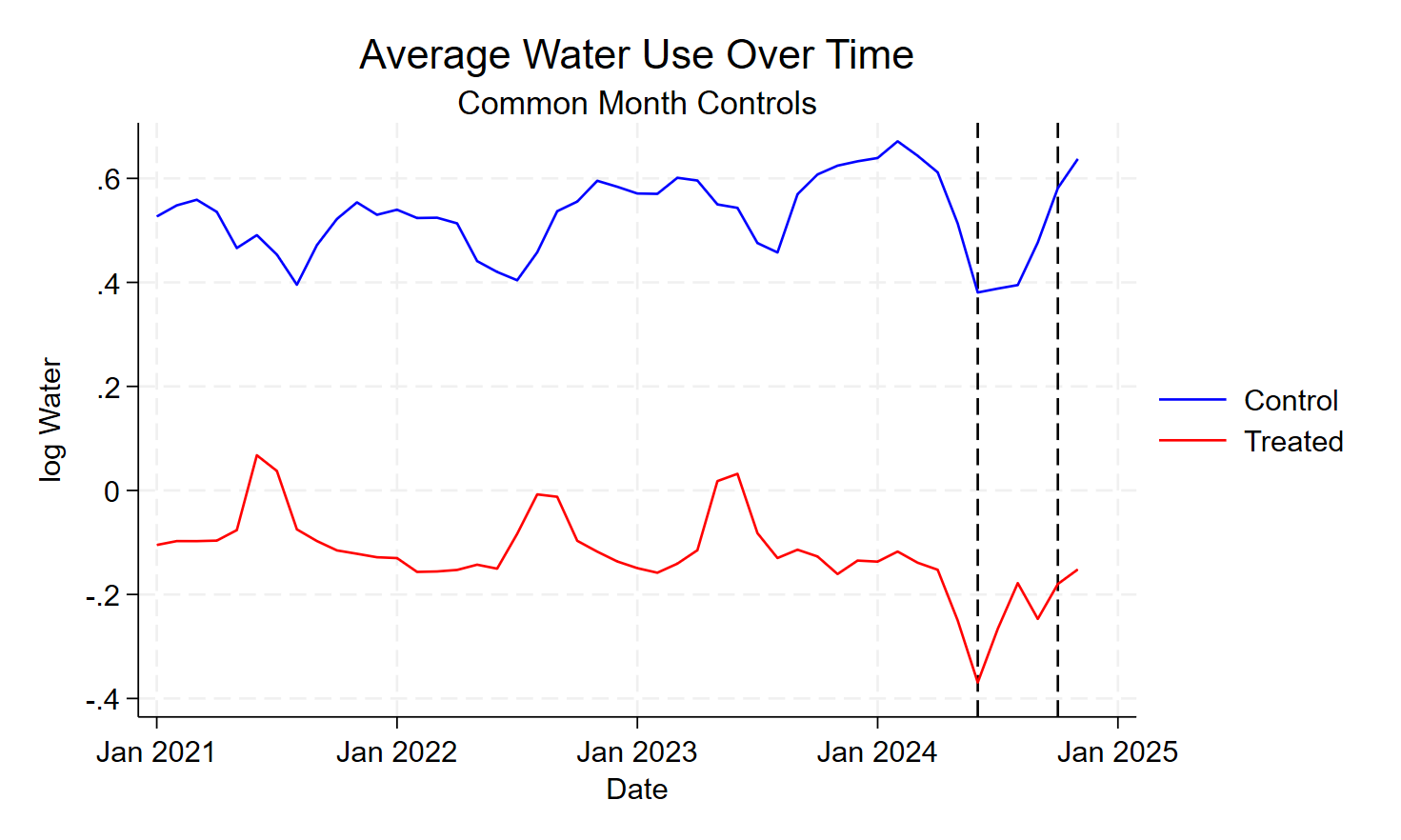}
        \caption{Common months}
        \label{fig:bray_common}
    \end{subfigure}
    \hfill
    \begin{subfigure}[b]{0.31\textwidth}
        \centering
        \includegraphics[width=\textwidth]{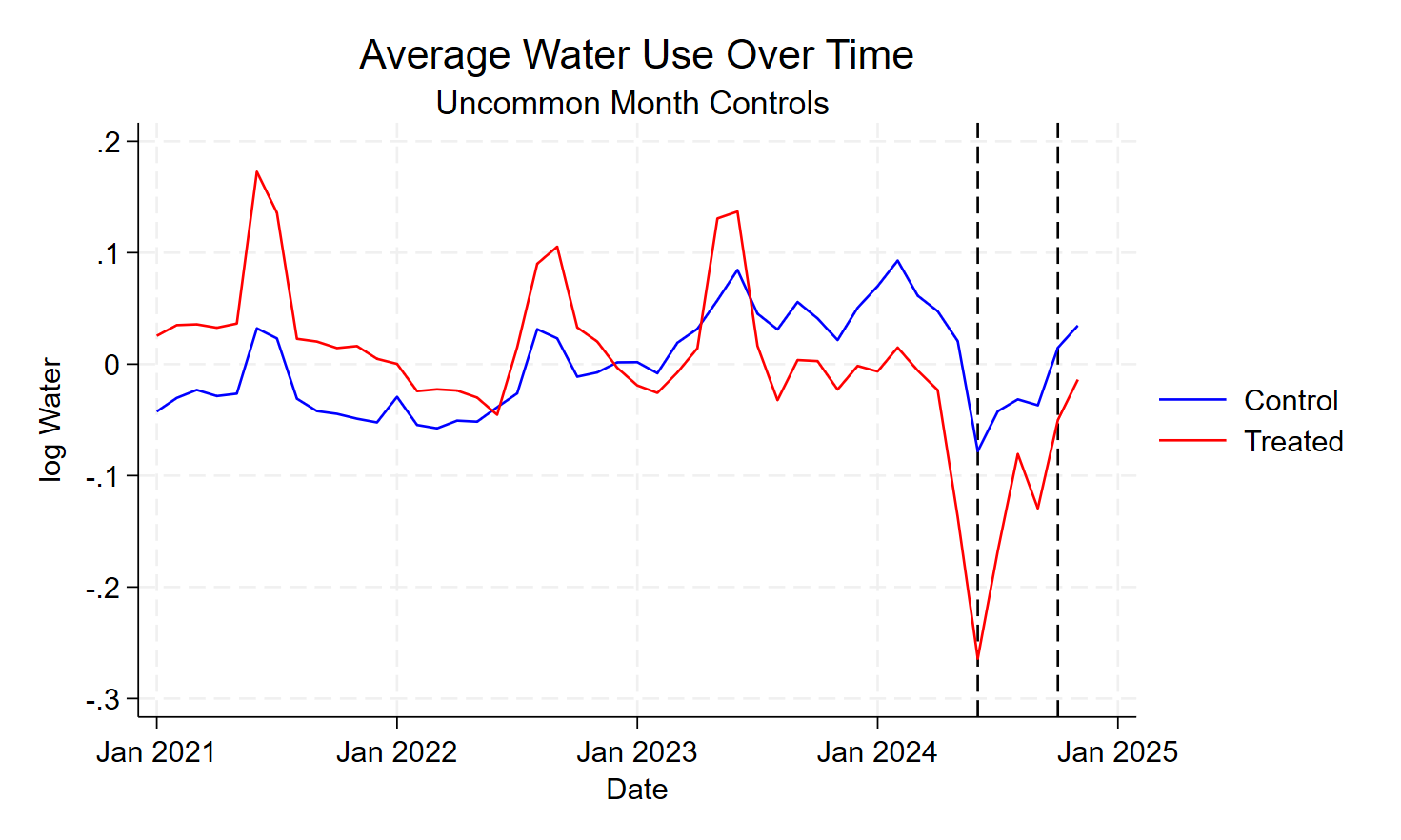}
        \caption{Uncommon months}
        \label{fig:bray_uncommon}
    \end{subfigure}
    \caption{Comparison of raw and adjusted trends.}
\end{figure}

Figure \ref{fig:bray_raw} shows the unconditional trends for the treated and control groups.  In this case, the pre-trends are clearly not parallel, as the seasonality is far more pronounced in the treatment group than in the control group. Given this, one might be tempted to model the seasonality, assuming a common covariate, by including month-of-year dummies.  Figure \ref{fig:bray_common} shows the conditional trends of water use, by treatment and control, after residualizing out the monthly dummies.  In this figure the pre-trends are clearly not parallel either.  In fact, the residualizing based on assuming common covariates has made things worse.  The estimates of the common monthly coefficients are an overestimate of the seasonality in the control group and an underestimate in the treated groups.  Accordingly, the residualized series show opposite seasonality across the treated and control groups, contrary to the raw data. At this point, many researchers would be tempted to conclude that CPT is likely implausible, given that the conditional pre-trends are not parallel.

Instead, we could relax the CCC assumption and estimate separate month-of-year dummies for both the treated and control groups.  Figure \ref{fig:bray_uncommon} plots these residuals.  The pre-trends in this figure, while not perfect, suggest that IPT assumption is plausible.  This example highlights how the new IPT assumption can expand the scope of settings in which valid ATTs can be estimated.

\subsubsection{Time Varying Covariates or Coefficients}

\begin{figure}[h!]
    \centering
    \includegraphics[width=0.6\textwidth]{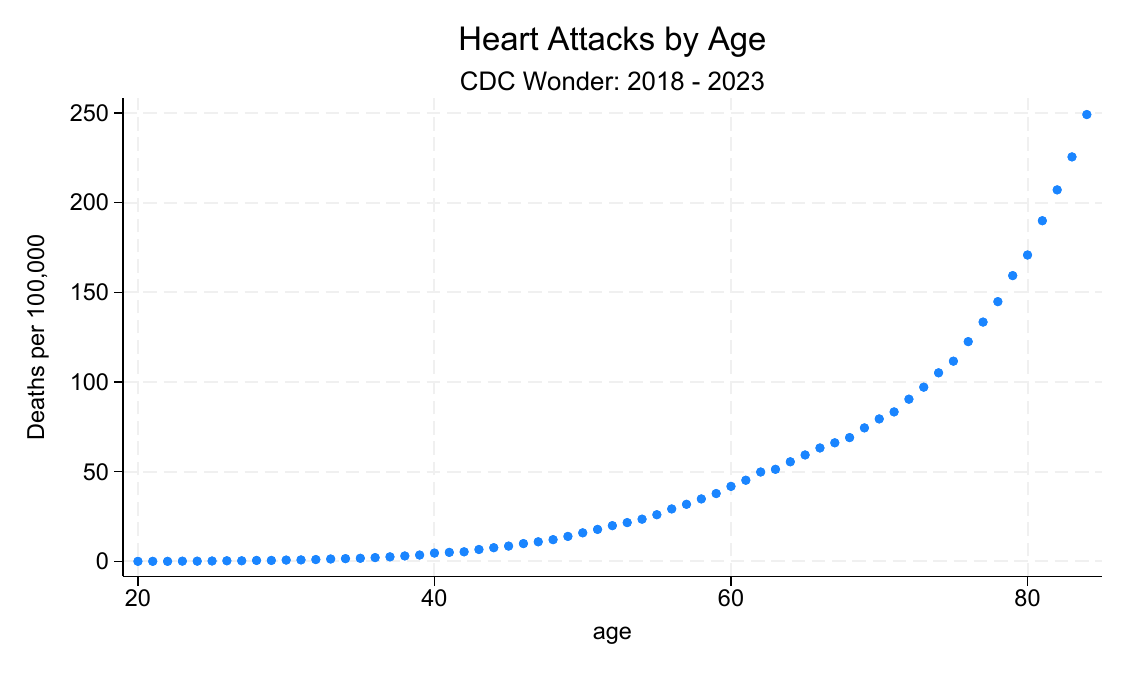}
    \caption{Heart attack deaths over time.}
    \label{fig:attacks}
\end{figure}

Consider a hypothetical intervention to reduce deaths from heart attacks. Further suppose we observe individuals over the period 2010--2020, where we have a relatively small panel of individuals in both a treated and a control group.    Figure \ref{fig:attacks} shows the relationship between heart attack deaths and age using Centre for Disease Control Data from 2018-2023.   Given the small sample, and the importance of age on the prevalence of heart attacks, a researcher would likely want to add age as a control variable. 

However, there are several problems with controlling for age in this example.  The first is that age is a time-varying covariate.  Conventionally, there are two ways around this: either recode age as the birth year, or use fixed pre-treatment values.  The birth-year approach changes a time-varying variable into a time-invariant one.  However, this directly assumes that the covariate is constant over time.  For example, consider a person born in 1940.  In 2010, that person is 70; in 2020, that person is 80. Given the rapid increase in the hazard rate shown in Figure \ref{fig:attacks}, the common causality of covariates assumption is clearly violated here.  

The other approach, following \cite{caetano2022timevarying} is to use pre-treatment values of covariates. Unfortunately, this approach suffers from the same problem.  The 1940 birth year person now gets a value for ``age" of 70 in both 2010 and 2020.  The estimate of the coefficient would then be an average of the relationship between age and heart attacks over the entire period, rather than the age-by-age relationship found in the data.  Again, the figure suggests that assuming a common coefficient for age (in 2010) on heart attacks across the period of 2010 to 2020 is likely to be violated.  

\subsubsection{Region and Time Varying Covariates}
\label{subsec: wageexample}

Consider an example where we want to estimate the ATT of a region-level policy on wage outcomes. Given the well-known link between education and wages \citep{mincer1958investment}, researchers may want to include a dummy variable for education (BA or higher) as a control. Since the 1990s, the number of college graduates has increased \citep{denning2022have}. This rise has lead to a decline in the relative value of a college degree, especially in jobs that require less cognitive effort \citep{horowitz2018relative}. As a result, the coefficient of the dummy variable may decrease over time, because of a shift in the educational attainment levels. This scenario suggests that the time-invariant CCC is unlikely to hold. Additionally, the coefficient of the dummy variable may be higher in regions with historically better higher education policies and higher enrollment rates \citep{fortin2006higher}. Regions with more capital-intensive industries may also show a higher coefficient compared to regions that rely more on hospitality, education, and health industries \citep{card2024industry}. This indicates that the region-invariant CCC is unlikely to hold. If we analyze the data for multiple regions and periods together, the two-way CCC may not hold.

To highlight that this assumption may be violated in actual datasets we consider a simple analysis using the Labour Force Survey (LFS) dataset for Canada.  This survey is used to calculate the official unemployment rate in Canada and surveys 100,000 people per month, following individuals for 6 months in total.  While we do not analyze an actual intervention here, we could imagine someone wanting to estimate the ATT of a set of policies which were thought to effect earnings.  Moreover, the anticipated ATT is small, so controlling for variation in earnings due to education is important. The sample is restricted to men between the ages of 35 to 55 who are employed from all 10 provinces between years 2008 to 2019.   The final sample has 1,573,585 observations. In line with the discussion above, we estimate the following regression: 
\begin{equation}
\label{equation:cccexampleemp}
    \text{earnings}_{i,p,t} = \alpha + \beta \text{college}_{i,p,t} + \epsilon_{i,p,t}.
\end{equation}
Here, $\text{earnings}_{i,p,t}$ is the dollar earnings for person $i$ in province $p$ in year $t$, and $\text{college}_{i,p,t}$ is a dummy variable which takes on a value of one if the person has a college degree (BA or higher). We
then re-estimate the model for each province and year separately. For each province and year pair, we record both the $\hat \beta_{p,t}$ coefficient estimate, but also the number of observations in that  province$\times$year pair that have a college degree. 

\begin{figure}[h!]
    \centering
    \includegraphics[width=0.8\textwidth]{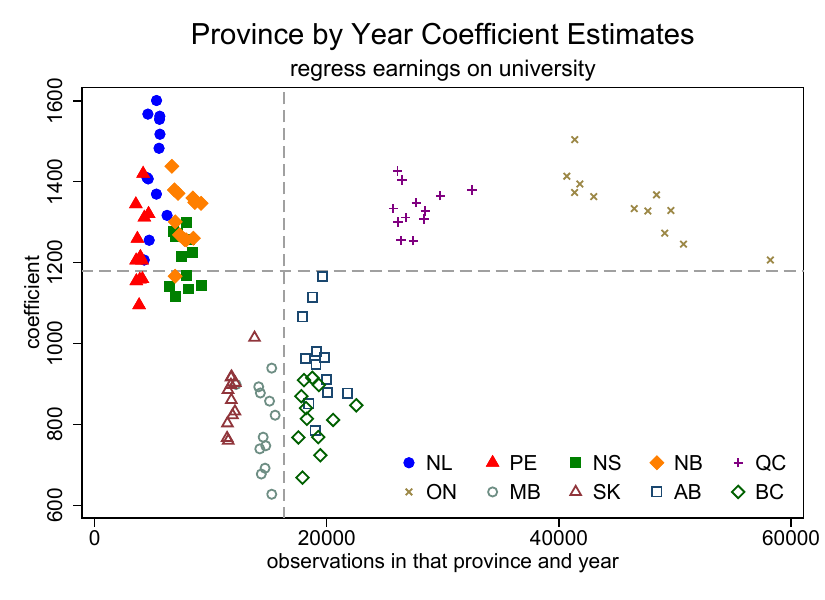}  
    \caption{Region$\times$Year coefficients of earnings on a dummy for university (BA or better)}
    \label{fig:CCCviolations}
\end{figure}

The results are shown in Figure \ref{fig:CCCviolations}. This plot shows that there is considerable variation in the coefficient estimates between provinces. There are also considerable variations in the coefficients across years for some regions, like Ontario and Quebec. The average number of observations per cell is around 16,332, and there is considerable variation in the number of observations ranging from 3,549 to 58,215. However, even the smallest counts represent a fairly large sample. This suggests that the variation in the coefficients is not coming from small sample sizes either.  Obviously, these are just estimates of the coefficients, and not the underlying causal parameters, but taken together this figure suggests 
that the assumption that the relationship between earnings and college being constant in all provinces and years is implausible.

Taken together, these three examples show why the assumptions of both region and/or invariant coefficients may be implausible.

\section{Nature of Covariates} \label{sec:nature}

In this section, we distinguish between 5 types of covariates in DID analysis, each based on the specific CCC assumption applied to them. We classify covariates for which the two-way CCC holds as \textbf{good controls}, the DAG for which is shown in Panel (a) of  Figure \eqref{fig:DAGs}. In other words, we assume \(\gamma^{\circ}_{s,t} = \gamma^{\circ}_{s',t'}\), implying that the effect of the covariate is the same across all regions and time periods. If the covariate is truly ``good" in the DGP, we can get consistent estimates of the ATT using TWFE and CS-DID, provided the assumptions for the respective estimators hold.

\begin{figure}[h!]
    \centering

    \begin{subfigure}[t]{0.45\textwidth}
        \centering
        \begin{tikzpicture}
            \node[draw, circle] (X) at (5,0) {$X$};
            \node[draw, circle] (D) at (3,2) {$D$};
            \node[draw, circle] (Y) at (7,2) {$Y$};
            \draw[->,>=stealth] (X) -- (Y);
            \draw[->,>=stealth] (D) -- (Y);
            \node at (6.25,0.75) {$\gamma^{\circ}$};
        \end{tikzpicture}
        \caption{Good controls}
        \label{fig:goodcontrols}
    \end{subfigure}
    \hfill
    \begin{subfigure}[t]{0.45\textwidth}
        \centering
        \begin{tikzpicture}
            \node[draw, circle] (X1) at (4,0) {$X_{A,1}$};
            \node[draw, circle] (X2) at (6,0) {$X_{A,2}$};
            \node[draw, circle] (X3) at (8,0) {$X_{B,1}$};
            \node[draw, circle] (X4) at (10,0) {$X_{B,2}$};
            \node[draw, circle] (D) at (3,2) {$D$};
            \node[draw, circle] (Y) at (7,2) {$Y$};
            \draw[->,>=stealth] (X1) -- (Y);
            \draw[->,>=stealth] (X2) -- (Y);
            \draw[->,>=stealth] (X3) -- (Y);
            \draw[->,>=stealth] (X4) -- (Y);
            \draw[->,>=stealth] (D) -- (Y);
            \node at (5,1.25) {$\gamma^{\circ}_{A,1}$};
            \node at (6.8,1) {$\gamma^{\circ}_{A,2}$};
            \node at (7.8,1) {$\gamma^{\circ}_{B,1}$};
            \node at (8.75,1.30) {$\gamma^{\circ}_{B,2}$};
        \end{tikzpicture}
        \caption{Temporally shifting good controls gone bad}
        \label{fig:goodcontrolsgonetwoway}
    \end{subfigure}

    \vspace{1em}

    \begin{subfigure}[t]{0.45\textwidth}
        \centering
        \begin{tikzpicture}
            \node[draw, circle] (X1) at (5,0) {$X_A$};
            \node[draw, circle] (X2) at (9,0) {$X_B$};
            \node[draw, circle] (D) at (3,2) {$D$};
            \node[draw, circle] (Y) at (7,2) {$Y$};
            \draw[->,>=stealth] (X1) -- (Y);
            \draw[->,>=stealth] (X2) -- (Y);
            \draw[->,>=stealth] (D) -- (Y);
            \node at (6.25,0.75) {$\gamma^{\circ}_A$};
            \node at (8.4,1.25) {$\gamma^{\circ}_B$};
        \end{tikzpicture}
        \caption{Good controls gone bad}
        \label{fig:goodcontrolsgonebad}
    \end{subfigure}
    \hfill
     \begin{subfigure}[t]{0.45\textwidth}
        \centering
        \begin{tikzpicture}
            \node[draw, circle] (X1) at (5,0) {$X_1$};
            \node[draw, circle] (X2) at (9,0) {$X_2$};
            \node[draw, circle] (D) at (3,2) {$D$};
            \node[draw, circle] (Y) at (7,2) {$Y$};
            \draw[->,>=stealth] (X1) -- (Y);
            \draw[->,>=stealth] (X2) -- (Y);
            \draw[->,>=stealth] (D) -- (Y);
            \node at (6.25,0.75) {$\gamma^{\circ}_1$};
            \node at (8.4,1.25) {$\gamma^{\circ}_2$};
        \end{tikzpicture}
        \caption{Temporally shifting controls}
        \label{fig:goodcontrolsgonetemporal}
    \end{subfigure}

    \vspace{1em}

    \begin{subfigure}[t]{0.45\textwidth}
        \centering
        \begin{tikzpicture}
            \node[draw, circle] (X) at (5,0) {$X$};
            \node[draw, circle] (D) at (3,2) {$D$};
            \node[draw, circle] (Y) at (7,2) {$Y$};
            \draw[->,>=stealth] (X) -- (Y);
            \draw[->,>=stealth] (D) -- (X);
            \draw[->,>=stealth] (D) -- (Y);
        \end{tikzpicture}
        \caption{Bad controls}
        \label{fig:badcontrol}
    \end{subfigure}

    \caption{Illustration of different types of covariates}
    \label{fig:DAGs}
\end{figure}

The second type of covariates, which we refer to as the \textbf{temporally shifting good controls gone bad}, include covariates that violate both the region-invariant and the time-invariant CCC (or the two-way CCC) assumptions. The DAG for this type of covariates is shown in Panel (b) of Figure \eqref{fig:DAGs}. In a simple case, with only two groups ($A$ and $B$) and two periods ($1$ and $2$), the effect of $X$ on $Y$ varies both across groups and over time. Therefore, \(\gamma^{\circ}_{A,1} \neq \gamma^{\circ}_{A,2} \neq \gamma^{\circ}_{B,1} \neq \gamma^{\circ}_{B,2}\).

The third type of covariates, which we refer to as \textbf{good controls gone bad}, are covariates for which the region-invariant CCC assumption is violated. The DAG for good controls gone bad is shown in Panel (c) of Figure \ref{fig:DAGs}. In a simple case where there are only two groups, $A$ and $B$, the effect of \( X \) on \( Y \) is different for $A$ compared to $B$. In other words, this violation occurs when  \(\gamma^{\circ}_A \neq \gamma^{\circ}_B\). However, the effect of the covariate remains the same across time. 

The fourth classification, \textbf{temporally shifting  controls}, refers to covariates that violate the time-invariant CCC assumption. The DAG for good controls gone temporal is shown in Panel (d) of Figure \eqref{fig:DAGs}. In this case, the effect of the control variable \( X \) on \( Y \) is the same across groups but changes over time. Consider two distinct periods 1 and 2. If the relationship between \( X \) and \( Y \) differs between these periods while remaining the same for each group, we observe a violation of time-invariant CCC. In this case, \(\gamma^{\circ}_1 \neq \gamma^{\circ}_2\). The DAG for \textbf{bad controls} are shown in Panel (e) of Figure \eqref{fig:DAGs}. Bad controls are controls which violate Assumption \ref{as: covariateunaffected}.

\section{Parallel Trends}
\label{section: PT}

In this section, we provide a detailed discussion of the parallel trends assumption, a key identifying assumption in DID. We begin by deriving the necessary and sufficient conditions for both strong and conditional parallel trends to hold. We show that the \textit{two-way common causality of covariates} assumption, while not necessary for conditional parallel trends, strengthens its plausibility. In addition, we demonstrate why incorporating time-varying covariates can complicate DID analysis and derive the necessary and sufficient conditions for conditional parallel trends to hold in the presence of such covariates. Finally, we introduce an alternative to the parallel trends assumption which we call \textbf{Integrated Parallel Trends (IPT)}. The integrated parallel trends assumption remains plausible when the two-way CCC assumption is violated, but it requires fewer conditions to hold compared to conditional parallel trends.

\begin{figure}
    \centering
    \includegraphics[width=0.5\textwidth]{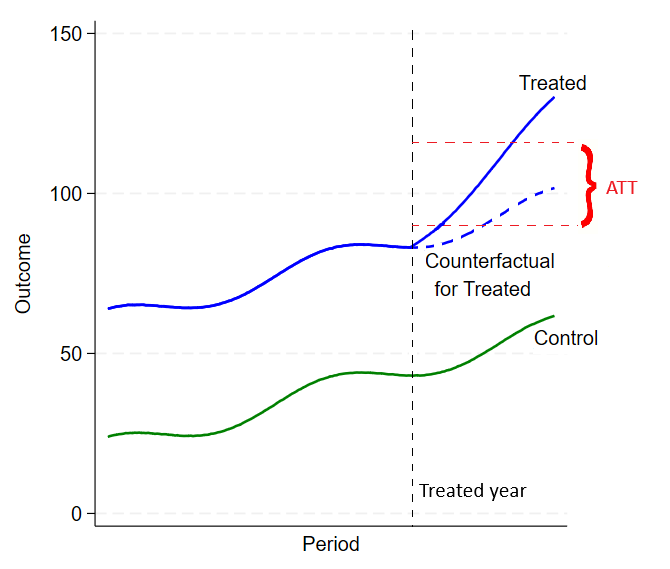}
    \caption{(Strong) Parallel Trends}
    \label{fig: SPT}
\end{figure}

\FloatBarrier

To formally illustrate the importance of the parallel trends assumption, consider a simple $2 \times 2$ setting with two periods (1 and 2) and two regions (A and B), where A is treated in period 2. In Figure \ref{fig: SPT}, the treated potential outcome is given by the solid blue line, which we observe. The untreated potential outcome is denoted by the dashed blue line, which we do not observe. Since the untreated potential outcome is unobserved, we cannot directly estimate the ATT using the sample analogues of Equation \eqref{eq: rawatt}. As a result, we require a control group and pre-treatment data for both the treated and control groups. We also rely on the \textit{parallel trends assumption}, which states that the treated and control groups would have followed the same trends in their outcomes in the absence of treatment. We use the trend of the control group to ``impute" the counterfactual outcomes of the treated group, which allows us to estimate the ATT. 

DID is commonly used to evaluate policies or interventions implemented in a region, where these policies are introduced in some regions (treated regions) and not in others (control regions). Regions could be provinces in Canada, states in the US, cities, municipalities, health authorities, etc. Since these regions may vary in demographic, economic, social and political characteristics, parallel trends without covariates is unlikely to hold. As a result, researchers use covariates to improve the plausibility of parallel trends, leading to the conditional parallel trends assumption (CPT). Under CPT, treated and control groups are assumed to follow similar trends in the absence of treatment, conditional on covariates.

\subsection{Parallel Trends without covariates}

In this subsection, we will derive the necessary and sufficient condition required for strong parallel trends to hold. In the absence of covariates, when $\gamma^{\circ}_{s,t}=0$ in Equation \eqref{eq: posimplified}, the strong parallel trends assumption in a $2 \times 2$ setup can be expressed as in Equation \eqref{equation: spt}.
\begin{assumption}[Strong Parallel Trends] 
    \label{as: spt} In the absence of treatment, the average untreated potential outcomes of the treated and control groups would evolve in parallel over time.
    \begin{equation}
    \label{equation: spt}
    \mathbb{E}[Y_{2}(0) \mid S_i = A] - \mathbb{E}[Y_{1}(0) \mid S_i = A] = \mathbb{E}[Y_{2}(0)\mid S_i = B] - \mathbb{E}[Y_{1}(0) \mid S_i = B].
\end{equation}
\end{assumption}

\begin{theorem}[Necessary and Sufficient condition for parallel trends without covariates] 
\label{theorem: ptnocovariates}
\begin{equation}
\label{eq: necessaryandsufficientspt}
    \mbox{Strong Parallel Trends} \Longleftrightarrow         \theta^{\circ}_{A,2} - \theta^{\circ}_{A,1} = \theta^{\circ}_{B,2} - \theta^{\circ}_{B,1}.
\end{equation}
\end{theorem}

The proof of Theorem \ref{theorem: ptnocovariates} can be found in Appendix \ref{appendix: ptproof}. In this paper, we do not aim to simplify \textbf{Theorem 1} further by imposing a particular functional form on $\theta^{\circ}_{s,t}$ . However, to intuitively explain what the equality on the right-hand side of Equation \eqref{eq: necessaryandsufficientspt} means, let us assume a two-way fixed effects type functional form for $\theta^{\circ}_{s,t}$ shown in Equation \eqref{eq: twfetypestructure}.
    \begin{equation}
    \label{eq: twfetypestructure}
        \theta^{\circ}_{s,t} \equiv \alpha^{\circ}_s + \delta^{\circ}_{t}
    \end{equation}
Here, $\alpha^{\circ}_s$ is the region fixed effect for region $s$, $\delta^{\circ}_{t}$ is the common period effect in period $t$. Plugging in \eqref{eq: twfetypestructure} into the left-hand side of Equation \eqref{eq: necessaryandsufficientspt} for region A only, we get:
    \begin{equation*}
        \alpha^{\circ}_A + \delta^{\circ}_{2} - \alpha^{\circ}_A - \delta^{\circ}_{1} = \delta^{\circ}_{2} - \delta^{\circ}_{1}
    \end{equation*}
Similarly, for region B we get:
    \begin{equation*}
        \alpha^{\circ}_B + \delta^{\circ}_{2} - \alpha^{\circ}_B - \delta^{\circ}_{1} = \delta^{\circ}_{2} - \delta^{\circ}_{1}
    \end{equation*}
    Within each region, the region fixed-effects cancel, but the remaining time fixed-effects remain. Since the TWFE specification imposes the same period effects across regions by default ($\delta_{A,1} = \delta_{B,1} = \delta_1$, $\delta_{A,2} = \delta_{B,2} = \delta_2$), both regions experience the same change $\delta^{\circ}_{2} - \delta^{\circ}_{1}$. Therefore, the necessary and sufficient condition for strong parallel trends is satisfied by construction under this additive representation. 
    
 This restriction is embedded directly in the TWFE regression, and may not accurately represent the underlying data-generating process. In practice, the time-related component of the untreated potential outcomes may vary across regions. Suppose that 
   \begin{equation}
    \label{eq: notreallytwfetypestructure}
        \theta^{\circ}_{s,t} \equiv \alpha^{\circ}_s + \delta^{\circ}_{s,t}
    \end{equation}
Here, $\delta^{\circ}_{s,t}$ is a region-specific time effect, which varies by region. Under this specification, strong parallel trend requires: 
    \begin{equation}
    \label{eq: sptfinal}
        \delta^{\circ}_{A,2} - \delta^{\circ}_{A,1} = \delta^{\circ}_{B,2} - \delta^{\circ}_{B,1}
    \end{equation}
    Equation \eqref{eq: sptfinal} implies that the region-specific time effects have to change by the same amount between periods in order for strong parallel trends to be satisfied. Consequently, when the true data-generating process contains region-specific period effects, the standard TWFE regression is misspecified and may attribute differences in untreated outcome trends to the treatment effect. Under the strong parallel trends and no anticipation assumptions, we can express the ATT shown in Equation \eqref{eq: rawatt} as:
    \begin{equation}
    \label{eq: SPTmaintext22}
        ATT = \mathbb{E} \biggr( \bigr[ Y_2 \mid S_i = A \bigr] - \mathbb{E} \bigr[ Y_1 \mid S_i = A\bigr] \biggr) - \biggr(\bigr[ Y_2 \mid S_i = B \bigr] - \mathbb{E} \bigr[ Y_1 \mid S_i = B \bigr]\biggr)
    \end{equation}

The no anticipation assumption is formally defined in Assumption \eqref{as: noanticipation}. Equation \eqref{eq: SPTmaintext22} follows directly from the same steps used in the proof presented in Appendix \ref{appendix: ATTcondcpt} and is therefore not derived separately.
   
    \begin{assumption}[No anticipation]
        \label{as: noanticipation}
    The treated potential outcome equals the untreated potential outcome for all units in the treated regions during all pre-intervention periods. 
    \begin{equation}
        \begin{gathered}
        \label{equation: noanticipation}
           Y_{i,s,t}(1) =  Y_{i,s,t}(0) \;\; \forall i \mbox{\quad\textit{a.s.} for all} \;\; t < t^s.
        \end{gathered}
    \end{equation}
    \end{assumption}

\subsection{Conditional Parallel Trends}
\label{subsection: cpt}

Due to differences in characteristics across regions, parallel trends without covariates are unlikely to hold. As a result, researchers include covariates to strengthen the credibility of parallel trends, leading to the conditional parallel trends assumption. When covariates matter in parallel trends, $\gamma^{\circ}_{s,t}$ in Equation \eqref{eq: posimplified} is no longer assumed to be 0. Conventional methods for assessing the plausibility of conditional parallel trends include regressing the outcome on covariates with a common coefficient across regions and periods. In this section, we derive the jointly necessary and sufficient conditions required for conditional parallel trends to hold.
\begin{assumption}[Conditional Parallel Trends] 
    \label{as:cpt}
    In the absence of treatment, the average untreated potential outcomes of the treated and control regions would evolve in parallel over time, conditional on covariates. 
    \begin{equation}
    \label{equation:cpt}
    \begin{aligned}
    &\mathbb{E}\left[Y_{2}(0) \mid S_i=A, X_{2}=x\right]
    - \mathbb{E}\left[Y_{1}(0) \mid S_i=A, X_{1}=x\right]
    \\
    = & \mathbb{E}\left[Y_{2}(0) \mid S_i=B, X_{2}=x\right]
    - \mathbb{E}\left[Y_{1}(0) \mid S_i=B, X_{1}=x\right],
    \qquad \forall x.
    \end{aligned}
    \end{equation}
\end{assumption}

A key aspect of the conditional parallel trends assumption is that it is a point-wise restriction. This means that the change in the untreated potential outcome between periods must be the same across the treated and control regions for each value of the covariates. To clarify this point, let us revisit the example discussed in Section \ref{subsec: wageexample}. Suppose education is included as a binary covariate that takes the value 1 if an individual has a BA degree or higher. Conditional parallel trends requires that the change in untreated potential outcomes be identical for treated and untreated regions within each education group. Specifically, the change in untreated potential outcomes should be identical across individuals with a college degree and separately across those without. Instead, when the covariate is continuous, assessing conditional parallel trends becomes much more challenging because the restriction must hold across every value of the covariate, not just within a finite set of categories.  

To highlight the point-wise nature of the conditional parallel trends, let the difference between the trends of the untreated potential outcome between the treated and the control regions at a given value of $x$ be $d(x)$. This is formally introduced in Equation \eqref{equation: dx}. 
\begin{equation}
\label{equation: dx}
\begin{aligned}
d(x) = &\mathbb{E}\left[Y_{2}(0) \mid S_i=A, X_{2}=x\right]
    - \mathbb{E}\left[Y_{1}(0) \mid S_i=A, X_{1}=x\right]
    \\
    - & \biggr[ \mathbb{E}\left[Y_{2}(0) \mid S_i=B, X_{2}=x\right]
    - \mathbb{E}\left[Y_{1}(0) \mid S_i=B, X_{1}=x\right] \biggr].
\end{aligned}
\end{equation}
Under the conditional parallel trends, we require that this difference is zero for every value of the covariates. Therefore, conditional parallel trends requires:
\begin{equation}
\label{equation: cptdx}
d(x)=0 \qquad \forall x.
\end{equation}
For the necessity result below, assume that the support of the covariate contains at least two distinct values.
\begin{theorem}[Necessary and sufficient condition for conditional parallel trends]
\label{theorem: sufficientconditioncptmaintext}
\begin{equation}
\label{eq: sufficientconditioncptmaintext}
\begin{aligned}
\text{Conditional Parallel Trends}
\Longleftrightarrow\;&
\left[
\theta^{\circ}_{A,2}-\theta^{\circ}_{A,1}
=
\theta^{\circ}_{B,2}-\theta^{\circ}_{B,1}
\right]
\\
&\land
\left[
\gamma^{\circ}_{A,2}-\gamma^{\circ}_{A,1}
=
\gamma^{\circ}_{B,2}-\gamma^{\circ}_{B,1}
\right].
\end{aligned}
\end{equation}
\end{theorem}
The two jointly necessary and sufficient conditions required for conditional parallel trends are:
   \begin{align}
    \begin{split}
    \label{equation: necessaryconditioncpt1maintext}
        \mbox{\textbf{Condition 1:}} \; \theta^{\circ}_{A,2} - \theta^{\circ}_{A,1} = \theta^{\circ}_{B,2} - \theta^{\circ}_{B,1}.
    \end{split}
    \end{align}
    \begin{align}
    \begin{split}
    \label{equation: necessaryconditioncpt2maintext}    
        \mbox{\textbf{Condition 2:}} \; \gamma^{\circ}_{A,2} - \gamma^{\circ}_{A,1} = \gamma^{\circ}_{B,2} - \gamma^{\circ}_{B,1}.
    \end{split}
    \end{align}

The proof of Theorem \ref{theorem: sufficientconditioncptmaintext} is provided in Appendix \ref{appendix: cptproof}. Conditional parallel trends hold only if the necessary and sufficient condition in Equation \eqref{eq: sufficientconditioncptmaintext} is satisfied, even if the working model is misspecified with a common coefficient. This implies that, even under two-way CCC violations, we may observe a parallel trend using pre-trend diagnostics. However, estimates of the ATT obtained by imposing a common coefficient with the conventional TWFE can be biased, as shown in Section \ref{sec:twfeproof}.

The second necessary and sufficient condition establishes a direct connection between CCC violations and conditional parallel trends. If either the region-invariant or the time-invariant CCC assumption holds, the second condition is satisfied automatically. Therefore, the three types of CCC assumptions are only sufficient for the covariate-coefficient component of this condition. Our findings show that, even when both assumptions are violated, conditional parallel trends can still hold provided that the change in the covariate effect between the pre- and post-intervention periods is the same across the treated and control regions. 

Since there is no statistical test for conditional parallel trends, the test proposed by \citet{karim2026ccc} provides an indirect diagnostic by assessing restrictions on the covariate coefficients. Using the test to find evidence in support of which CCC assumption is violated makes Condition 2 required for conditional parallel trends more credible. However,  CCC assumptions are not necessary for conditional parallel trends, and their validity alone does not confirm that the parallel trend assumption holds.  

Under the conditional parallel trends and no anticipation assumptions, the ATT shown in Equation \eqref{eq: rawatt} can be expressed in terms of observed outcomes, as shown in Equation \eqref{eq: CPTmaintext22}. 
    \begin{equation}
    \begin{aligned}
    \label{eq: CPTmaintext22}
        ATT = &\mathbb{E}\biggr[\mathbb{E} \biggr( \bigr[ Y_2 \mid S_i = A, X_t = x \bigr] - \mathbb{E} \bigr[ Y_1 \mid S_i = A, X_t = x\bigr] \biggr) \\ - & \biggr(\bigr[ Y_2 \mid S_i = B, X_t = x \bigr] - \mathbb{E} \bigr[ Y_1 \mid S_i = B, X_t = x \bigr]\biggr)\mid S_i = A \biggr].
    \end{aligned}
    \end{equation}
The expression within the outer expectation represents $ATT_x$, which is the average treatment effect on the treated, conditional on $X_{i,s,t} = x$. By taking the outer expectation over the covariate distribution of the treated distribution, we obtain the unconditional $ATT$. Under the conditional parallel trends and the no anticipation assumptions, the derivation of the conditional ATT as shown in Equation \eqref{eq: CPTcondmaintext22} is shown in Appendix \ref{appendix: ATTcondcpt}.
\begin{equation}
\label{eq: CPTcondmaintext22}
\begin{aligned}
    ATT_x = & \mathbb{E} \biggr( \bigr[ Y_2 \mid S_i = A, X_t = x \bigr] - \mathbb{E} \bigr[ Y_1 \mid S_i = A, X_t = x\bigr] \biggr) \\ - & \biggr(\bigr[ Y_2 \mid S_i = B, X_t = x \bigr] - \mathbb{E} \bigr[ Y_1 \mid S_i = B, X_t = x \bigr]\biggr)
\end{aligned}
\end{equation}

\subsection{Integrated Parallel Trends (IPT)}

In this subsection, we formally define \textit{Integrated Parallel Trends (IPT)}, which accommodates violations of the two-way CCC and does not require the two jointly necessary and sufficient conditions to hold for every value of $x$ to be valid. We also derive the necessary and sufficient condition for integrated parallel trends. Additionally, we demonstrate that integrated parallel trends hold when conditional parallel trends hold, but the converse is not true. 

\begin{assumption}[Integrated Parallel Trends] The average untreated potential outcomes between the treated and control regions would evolve in parallel over time after integrating over the covariate distribution of the treated region.
\begin{gather}
    \int \biggr[\mathbb{E}\bigr(Y_2(0) \mid S_i = A, X_2 = x \bigr) - \mathbb{E}\bigr(Y_1(0) \mid S_i = A, X_1 = x \bigr)\biggr] dF_{X_2 \mid S=A}(x) \nonumber \\ 
    = \int \biggr[\mathbb{E}\bigr(Y_2(0) \mid S_i = B, X_2 = x \bigr) -
    \mathbb{E}\bigr(Y_1(0) \mid S_i = B, X_1 = x \bigr)\biggr] dF_{X_2 \mid S=A}(x) \label{eq: integratedpt}.
\end{gather}    
\end{assumption}
One key feature of Integrated parallel trends is that it is no longer a point-wise restriction. To illustrate the distinction between integrated parallel trends and conditional parallel trends, we return to the binary education example introduced in Section \ref{subsec: wageexample}. Suppose the binary education dummy is zero for 25\% of the population in the treated region $A$, while the remaining 75\% have a value of one. Additionally, assume the difference between the expected post- and pre-period untreated potential outcomes in the treated region is 2 for those with a value of zero and 4 for those with a value of one. Let us also assume that the control region $B$ has a different distribution compared to region $A$, where 30\% of the population have a value of zero, while the remaining 70\% have a value of one. Moreover, assume the difference between the expected values of the post- and pre-period untreated potential outcomes in region $B$ is 5 for those with a value of zero and 3 for those with a value of one. These are summarized in the following table:
\begin{table}[htbp]
\centering
\caption{Differences in Expected Untreated Potential Outcomes}
\label{tab:ipt_example}
\begin{tabular}{lcc}
\toprule
Education Group
& Treated Region $A$
& Control Region $B$ \\
\midrule
No BA, $X=0$ 
& $2$ 
& $5$ \\
BA or higher, $X=1$
& $4$
& $3$ \\
\bottomrule
\end{tabular}
\end{table}

The left-hand side of Equation \eqref{eq: integratedpt} is:
\begin{gather}
    \text{For region A:} \quad 0.25(2) + 0.75(4) = 3.5.
\end{gather}
Since we evaluate the trend of the control region using the covariate distribution of the treated region, we apply the same weights to both regions, even though the control region has a different covariate distribution. Therefore, $X=0$ receives a weight of 0.25 and $X=1$ receives a weight of 0.75, corresponding to the proportions of the treated population at each value of $X$. So the right-hand side of Equation \eqref{eq: integratedpt} is:
\begin{gather}
    \text{For region B:} \quad 0.25(5) + 0.75(3) = 3.5.
\end{gather}
In this example, integrated parallel trends hold after averaging the conditional untreated trends over the covariate distribution of the treated region. In contrast, conditional parallel trends, which require the equality in Equation \eqref{equation:cpt} to hold for every value of $X$, do not hold.

In this case in particular, it holds for neither value of $X$.  The point-wise requirement for conditional parallel trends is relatively strong, particularly when the covariates are continuous. Integrated parallel trends relaxes this requirement by allowing untreated trends to differ at particular values of $X$, although the trends must be the same over the covariate distribution of the treated region. A second feature of the integrated parallel trends is that the conditional untreated outcomes of the control region are evaluated using the covariate distribution of the treated region. This provides a more relevant counterfactual for the treated region, since the objective is to recover how the treated region would have evolved in the absence of treatment.
\begin{theorem}[Necessary and sufficient condition for integrated parallel trends] \label{theorem: iptnscondition}
        \begin{gather}
        \text{Integrated Parallel Trends}
\Longleftrightarrow\ \bigr[ (\theta^{\circ}_{A,2} - \theta^{\circ}_{A,1}) - (\theta^{\circ}_{B,2} - \theta^{\circ}_{B,1}) \bigr] \nonumber \\ + \bigr[(\gamma^{\circ}_{A,2} - \gamma^{\circ}_{A,1}) -(\gamma^{\circ}_{B,2} - \gamma^{\circ}_{B,1})\bigr] \mathbb{E}[X_2 \mid S_i = A] = 0 \label{eq: nandsconditioniptmaintext}.
\end{gather}
\end{theorem}
A proof of Theorem \ref{theorem: iptnscondition} is shown in Appendix \ref{appendix: iptnscondition}. Now, we compare the expressions in Equations \eqref{eq: sufficientconditioncptmaintext} and \eqref{eq: nandsconditioniptmaintext} to highlight the difference between conditional and integrated parallel trends. Under conditional parallel trends, the jointly necessary and sufficient conditions need to hold for every value of $x$. However, under integrated parallel trends, these conditions need not hold for every value of $x$, as long as the necessary and sufficient condition in Equation \eqref{eq: nandsconditioniptmaintext} is satisfied.

\begin{theorem}[Nesting of Parallel Trends Assumptions] When conditional parallel trends hold, integrated parallel trends hold as well, but the converse does not hold. \label{theorem: ptnesting}
\begin{gather}
    \text{Conditional Parallel Trends} \Longrightarrow     \text{Integrated Parallel Trends}.
\end{gather}    
\end{theorem}
The proof of Theorem \ref{theorem: ptnesting} follows directly from the expressions in Equations \eqref{eq: sufficientconditioncptmaintext} and \eqref{eq: nandsconditioniptmaintext}. Under conditional parallel trends, the two jointly necessary and sufficient conditions can be rearranged as:
   \begin{align}
    \begin{split}
    \label{equation: theorem4c1}
        \bigr(\theta^{\circ}_{A,2} - \theta^{\circ}_{A,1}\bigr) - \bigr(\theta^{\circ}_{B,2} - \theta^{\circ}_{B,1} \bigr) = 0.
    \end{split}
    \end{align}
    \begin{align}
    \begin{split}
    \label{equation: theorem4c2} \bigr(\gamma^{\circ}_{A,2} - \gamma^{\circ}_{A,1} \bigr) - \bigr(\gamma^{\circ}_{B,2} - \gamma^{\circ}_{B,1}\bigr) = 0.
    \end{split}
    \end{align}
    Plugging in Equations \eqref{equation: theorem4c1} and \eqref{equation: theorem4c2} into the necessary and sufficient condition for integrated parallel trends in Equation \eqref{eq: nandsconditioniptmaintext} yields:
\begin{gather}
\underbrace{\bigr[ (\theta^{\circ}_{A,2} - \theta^{\circ}_{A,1}) - (\theta^{\circ}_{B,2} - \theta^{\circ}_{B,1}) \bigr]}_{=0} + \underbrace{\bigr[(\gamma^{\circ}_{A,2} - \gamma^{\circ}_{A,1}) -(\gamma^{\circ}_{B,2} - \gamma^{\circ}_{B,1})\bigr]}_{=0} \mathbb{E}[X_2 \mid S_i = A] = 0 \label{eq: nandsconditioniptmaintext2}.
\end{gather}
This shows that, when conditional parallel trends hold, the change in the untreated potential outcome is the same between the treated and the control at every value of $x$. Therefore, integrating these trends over the same treated covariate distribution must also yield equal integrated trends. Thus, integrated parallel trends hold when conditional parallel trends hold. However, the converse is not true. 
    
Under the integrated parallel trends and no anticipation assumptions, the ATT shown in Equation \eqref{eq: rawatt} can be expressed in terms of observed outcomes, as shown in Equation \eqref{eq: integratedATT}.

\begin{equation}
\label{eq: integratedATT}
\begin{aligned}
    ATT(A,2) = \int\biggr\{ & \biggr[\mathbb{E}[Y_{2} \mid S_i=A,X_{2}=x] - \mathbb{E}[Y_{1}\mid S_i=A,X_{1}=x] \biggr] \\
    - & \biggr[\mathbb{E}[Y_{2}\mid S_i=B,X_{2}=x] - \mathbb{E}[Y_{1}\mid S_i=B,X_{1}=x]\biggr]\biggr\}dF_{X_2\mid S=A}(x).
\end{aligned}
\end{equation}

\subsection{Pre-test Diagnostic for Integrated Parallel Trends}

Both conditional and integrated parallel trends are assumptions related to the untreated potential outcomes for treated and control regions. Since these are assumptions about the untreated potential outcomes for the treated region after the intervention, which are unobserved, these assumptions cannot be directly tested with observed data. Therefore, researchers usually evaluate the plausibility of conditional parallel trends by inspecting trends in pre-treatment outcomes. They often rely on visual checks to see if the treated and control regions display similar trends before the intervention, though this evaluation is inherently subjective. 

To check for conditional parallel trends, figures are plotted by regressing the observed outcome on covariates during the pre-intervention periods, then plotting the residuals from these regressions separately for treated and control regions on the same axis. There are two issues associated with verifying parallel trends through this method. Firstly, as the CCC assumption has not been formally introduced prior to this paper to the best of our knowledge, this adjustment is typically implemented by imposing a common coefficient on the covariates across different regions and periods. Secondly, there exists a mismatch between the conditional parallel trends assumption as a point-wise restriction concerning the covariates and the manner in which this assumption is assessed visually. Although the conditional parallel trends assumption requires that the condition hold for each individual value of the covariates, researchers generally do not verify pre-trends for every possible value of these covariates. Doing so may be more challenging if the covariates are continuous. 

For conditional parallel trends, we recommend that researchers adopt more flexible covariate specifications when assessing parallel trends to account for CCC violations. This involves interacting covariates with region dummies, time dummies, or both, and then regressing these interactions on the outcome to derive residuals. To select the correct functional form, we recommend using the CCC test proposed in \cite{karim2026ccc}. The CCC test uses a formal Wald-test-based method to assess whether the covariate effects vary across regions, over time, or both. Using more flexible functional forms of covariates can be empirically consequential, since pre-treatment trends that seem non-parallel under conventional common-coefficient adjustments might actually be plausibly parallel once CCC violations are considered. Therefore, examining pre-trends with more flexible covariate specifications can uncover trends which appear parallel even when conventional diagnostics fail. This may broaden the range of empirical applications where credible DID identification is feasible, rather than dismissing an analysis simply because parallel trends are not visible under overly restrictive covariate specifications. Researchers can alternatively rely on a check for integrated parallel trends, which can also accommodate CCC violations. 

Like conditional parallel trends, integrated parallel trends is an assumption about unobserved counterfactuals. Therefore, we can assess its plausibility using subjective pre-trends figures. In this paper, we propose a graphical assessment of integrated parallel trends based on period-to-period differences in expected outcomes, evaluated over the covariate distribution of the treated regions. The following steps describe how we construct the plot. 

To construct this plot, we first regress the outcome on the covariates separately for each region and period:
\begin{equation}
\label{eq: cptplot}
Y_{i,s,t} = \theta_{s,t} + \gamma_{s,t}X_{i,s,t} + u_{i,s,t} \quad \forall s,t.
\end{equation}
From this regression, we store the estimated coefficients $\widehat{\theta}_{s,t}$ and $\widehat{\gamma}_{s,t}$ for each region and period. Unlike the regression used to assess conditional parallel trends, which uses the full sample to estimate the residuals, these regressions are estimated separately for each region and period. As a result, they can account for violations of the CCC assumption. We use these coefficients to predict outcomes for both the treated and control regions using the covariate values observed in the treated region. This allows us to compare how predicted outcomes change from one period to the next in the treated and control regions while evaluating both regions over the same covariate distribution. So the change in the predicted values for the treated region $s$ evaluated at $\overline{X}_{s,t}$ is: 
\begin{equation}
    \Delta Y_{s,t}(\overline{X}_{s,t}) = \theta_{s,t} + \gamma_{s,t} \overline{X}_{s,t} - \theta_{s,t-1} - \gamma_{s,t-1} \overline{X}_{s,t}.
\end{equation}
Similarly, the change in the predicted values for the control region $s'$ evaluated at $\overline{X}_{s,t}$ is: 
\begin{equation}
    \Delta Y_{s',t}(\overline{X}_{s,t}) = \theta_{s',t} + \gamma_{s',t} \overline{X}_{s,t} - \theta_{s',t-1} - \gamma_{s',t-1} \overline{X}_{s,t}.
\end{equation}
We then plot these differences for each region separately on a common figure. This approach does not require a separate set of figures for each covariate specification, since Equation \eqref{eq: cptplot} can accommodate both cases in which the CCC assumption holds and cases in which it is violated. Since integrated parallel trends can hold when conditional parallel trends are violated, we may also be able to observe trends that appear inherently parallel, even when conditional parallel trends checks fail.

\section{Multiple Groups and Staggered Adoption} \label{sec: multgroups}

In the previous section, we have focused on a simple 2 $\times$ 2 case to derive the necessary and sufficient conditions for parallel trends, conditional parallel trends and integrated parallel trends. For the remainder of the paper, we focus on more general settings with $S$ regions and $T$ periods. Therefore, we restate the parallel trends, conditional parallel trends and integrated parallel trends assumptions in this more general setting, as shown below. For simplicity of notation, we continue to assume that there is a single time-varying covariate $X_{i,s,t}$.

\begin{assumption}[Parallel Trends]
    \label{as: pt}
The treated and control groups would have followed the same trends in their outcomes in the absence of treatment.
\begin{equation}
    \begin{gathered}
    \label{equation: ptgeneral}
        \mathbb{E}[Y_{t}(0) - Y_{t^s-1}(0) \mid S_i = s] = \sum_{s' \in \mathbf{s}^c}\omega_{s',s,t} \biggr[\mathbb{E}[Y_{t}(0) - Y_{t^s-1}(0) \mid S_i = s']\biggr]. 
    \end{gathered}
\end{equation}
\end{assumption}

\begin{assumption}[Conditional Parallel Trends]
    \label{as: cpt}
The treated and control groups would have followed the same trends in their outcomes in the absence of treatment, conditional on covariates.
\begin{equation}
    \begin{aligned}
    \label{equation: cptgeneral}
        & \mathbb{E}[Y_{t}(0) \mid S_i = s, X_{t}] - \mathbb{E}[Y_{t^s-1}(0) \mid S_i = s, X_{t^s-1}]
        \\ = \sum_{s' \in \mathbf{s}^c} \omega_{s',s,t}\biggr[ &  \mathbb{E}[Y_{t}(0) \mid S_i = s', X_{t}] - \mathbb{E}[Y_{t^s-1}(0) \mid  S_i = s', X_{t^s-1}]\biggr].
    \end{aligned}
\end{equation}
\end{assumption}

\begin{assumption}[Integrated Parallel Trends]
    \label{as: ipt}
        In the absence of treatment, the treated and control regions would evolve in parallel over time after integrating over the covariate distribution of the treated region.
        \begin{equation}
        \begin{aligned}
            & \int \biggr[ \mathbb{E}[Y_{t}(0) \mid S_i = s, X_{t}] - \mathbb{E}[Y_{t^s-1}(0) \mid S_i = s, X_{t^s-1}] \biggr] dF_{X_t\mid S = s^t}(x) 
            \\ = \sum_{s' \in \mathbf{s}^c}\omega_{s',s,t} & \biggr[\int \biggr[\mathbb{E}[Y_{t}(0) \mid S_i = s', X_{t}] - \mathbb{E}[Y_{t^s-1}(0) \mid S_i = s', X_{t^s-1}]\biggr]dF_{X_t\mid S = s}(x)\biggr] .
            \label{eq:ipt22}
        \end{aligned}
        \end{equation}
\end{assumption}

There may be multiple relevant control regions for treated region $s$ in period $t$. Let $\omega_{s',s,t}$ be the weight assigned to control region $s'$, such that these weights sum to one. With multiple control regions, we take a weighted average across the relevant control regions $s' \in \mathbf{s}^c$. With covariates and assuming conditional parallel trends and no anticipation, the ATT can be written as shown in Equation \eqref{eq: attgeneral}. In common treatment adoption settings, $t^s-1$ spans the entire pre-intervention period, which is different than in staggered adoption, where only the period right before treatment is the pre-intervention period.
\begin{equation}
    \begin{aligned}
    \label{eq: attgeneral}
       ATT = & \mathbb{E} \biggr[ \bigr(\mathbb{E}[Y_{t} \mid s \in \mathbf{s}^t, X_{t}] - \mathbb{E}[Y_{t^s-1} \mid s \in \mathbf{s}^t, X_{t^s-1}] \bigr) 
        \\ - & \biggr( \sum_{s' \in \mathbf{s}^c} \omega_{s',s,t} \bigr( \mathbb{E}[Y_{t} \mid s' \in \mathbf{s}^c, X_{t}] - \mathbb{E}[Y_{t^s-1} \mid  s' \in \mathbf{s}^c, X_{t^s-1}] \bigr) \biggr) \mid s \in \mathbf{s}^t \biggr].
    \end{aligned}
\end{equation}
In staggered treatment adoption designs, where different regions can implement treatment at different periods, the TWFE estimator is known to be inconsistent under heterogeneous treatment effects due to forbidden comparisons and negative weighting issues \citep{goodman2021difference}. To address the problem of “forbidden comparisons”, \cite{callaway2021difference} proposes a framework that decomposes staggered adoption designs into a series of simpler “2 × 2” comparisons, which we call \textit{stacks}. Each stack isolates one treated cohort and an appropriate control cohort, which can either be never treated or not yet treated. With covariates, the ATT for each stack, called $ATT(s,t)$ is written as:
\begin{equation}
    \begin{aligned}
    \label{eq: attstaggered}
       ATT(s,t) = & \mathbb{E} \biggr[ \bigr(\mathbb{E}[Y_{t} \mid S_i=s, X_{t}] - \mathbb{E}[Y_{t^s-1} \mid S_i=s, X_{t^s-1}] \bigr) 
        \\ - & \biggr( \sum_{s' \in \mathbf{s}^c} \omega_{s',s,t} \bigr( \mathbb{E}[Y_{t} \mid S_i=s', X_{t}] - \mathbb{E}[Y_{t^s-1} \mid S_i=s', X_{t^s-1}] \bigr) \biggr) \mid S_i=s \biggr].
    \end{aligned}
\end{equation}

In staggered adoption designs, the pre-intervention period $t^s-1$ is applied separately to each treated region $s$ \citep{callaway2021difference}. In this paper, we use regions to define cohorts rather than pooling regions by treatment timing, similar to \citep{callaway2021difference}. The benefits of this approach are explored in more detail in \citet*{karim2025policy}. Under IPT and the no-anticipation assumption, the ATT can be written as:
\begin{equation}
\label{eq: attgeneralipt}
\begin{aligned}
ATT(s,t) = & \int \biggr\{\biggr[\mathbb{E}[Y_t \mid S_i=s,X_t=x] - \mathbb{E}[Y_{t^s-1} \mid S_i=s,X_{t^s-1}=x]\biggr] \\
&- \sum_{s' \in \mathbf{s}^c} \omega_{s',s,t} \biggr[\mathbb{E}[Y_t \mid S_i=s',X_t=x] - \mathbb{E}[Y_{t^s-1} \mid S_i=s',X_{t^s-1}=x]\biggr]
\biggr\} dF_{X_t \mid S_i=s}(x).
\end{aligned}
\end{equation}

\section{Integrated Difference-in-Differences (DID-INT)} \label{sec: didint}

In this section, we formally develop the Integrated Difference-in-Differences (DID-INT), which can provide a consistent estimate of the ATT and is robust to the three types of CCC violations we discuss in this paper (described in detail in Section \ref{sec: ccc}). The ATT is estimated in five steps. In the first step, we employ the \textbf{CCC test} to determine the functional form of covariates $f(X^k_{i,s,t})$. Details of this test are presented in \cite{karim2026ccc}. In the second step, we propose running the following regression without a constant: 
\begin{equation}
\label{equation: DIDINTfirstreg}
    Y_{i,s,t} = \sum_{s} \sum_{t} \lambda_{s,t} I(s, t) + f(X_{i,s,t}) + \epsilon_{i,s,t},
\end{equation}
where, $I(s, t)$ is a dummy variable that takes a value of 1 if the observation is in region $s$ in period $t$, or the region$\times$time intersection. $f(X_{i,s,t})$ represents a function of covariates, which varies according to the specific CCC violations researchers intend to account for in their analysis. Depending on the function of $f(X_{i,s,t})$, we also generate two types of dummy variables: $I(s)$, which takes on a value of 1 if the observation is in region $s$; and $I(t)$, which takes on a value of 1 if the observation is from period $t$. 

In the third step, we store the differences in $\lambda_{s,t}$ in a matrix for each period after region $s$ is first treated, using $t^{s}-1$ as the pre-intervention period. In addition to this difference, we integrate the change in the covariate component over the covariate distribution of the treated region $s$ in period $t$. This integrated difference for region $s$ in period $t$ is denoted by $\widehat{\Delta}_{s,t}$. 
\begin{equation}
\label{equation: diffsteptreat}
    \widehat{\Delta}_{s,t} = (\widehat{\lambda}_{s,t} - \widehat{\lambda}_{s,t^{s}-1}) + \int (\widehat{\gamma}_{s,t} - \widehat{\gamma}_{s,t^s-1})x d\widehat{F}_{X_t \mid S=s}(x).
\end{equation}
Here, $t^s$ is the period when region $s$ is first treated. For the relevant control region $s^ \prime$ in period $t$, the covariate component is evaluated using the same covariate distribution of the treated region.  
\begin{equation}
\label{equation: diffstepcontrol}
    \widehat{\Delta}_{s',t} = (\widehat{\lambda}_{s',t} - \widehat{\lambda}_{s',t^{s}-1}) + \int (\widehat{\gamma}_{s',t} - \widehat{\gamma}_{s',t^s-1})x d\widehat{F}_{X_t \mid S=s}(x).
\end{equation}
In the fourth step, we estimate the ATT for region $s$ in period $t$, denoted by $\widehat{\tau}^{DT}_{s,t}$ as follows:
\begin{equation}
\label{equation: secondstep}
    \widehat{\tau}^{DT}_{s,t} =
    \widehat{\Delta}_{s,t} - \sum_{s' \in \mathbf{s}^c} \omega_{s',s,t}
    \widehat{\Delta}_{s',t}.
\end{equation}
We compute a weighted average of the integrated differences across all control regions in $\mathbf{s}^c$ because there may be multiple relevant control regions for each treated region $s$ in period $t$.  These are drawn from the matrix of differences in the third step. In the fifth step, we estimate the overall ATT by taking a weighted average of the $\widehat{\tau}^{DT}_{s,t} $ estimates obtained in the fourth step. The expression of the overall ATT is:
\begin{equation}
\label{equation: aggregation}
    \widehat{\tau}^{DT} = \sum_{s=1}^{S} \sum_{t=1}^{\mathcal{T}} 1\{t^s \leq t\} w_{s,t} \widehat{\tau}^{DT}_{s,t}.
\end{equation}

Here, $w_{s,t}$ are the region-period weights which are defined in \cite{callaway2021difference}. In the above expression, the forbidden comparisons highlighted by \cite{goodman2021difference} are excluded from the calculation. Cluster-robust inference on the ATT can be done on the ATT using a cluster jackknife and randomization inference, see \ref{sec:crve} for details.

Let the population regression model for Equation \eqref{equation: DIDINTfirstreg} be given by:
\begin{equation}
\label{equation: DIDINTpopulationregmain}
    Y_{i,s,t} = \sum_{s} \sum_{t} \lambda^*_{s,t} I(s, t) + \gamma^*_{s,t} I(s)I(t)X_{i,s,t} + \epsilon^*_{i,s,t},
\end{equation}
Here, $\lambda^*_{s,t}$ and $\gamma^*_{s,t}$ represent the probability limits of the intercept and the covariate coefficient in the regression shown in Equation \eqref{equation: DIDINTfirstreg}, respectively. $\epsilon^*_{i,s,t}$ denotes the population regression error. Consider a treated region $s$ in the post-intervention period $t > t^s - 1$. Under these definitions, the probability limit of $\widehat{\tau}^{DT}_{s,t}$ is given by:
\begin{equation}
\begin{aligned}
\tau^{{DT}^*}_{s,t} = & \bigr( \lambda^*_{s,t} - \lambda^*_{s,t^s-1} \bigr) - \sum_{s' \in \mathbf{s}^c} \omega_{s',s,t} \bigr(\lambda^*_{s',t} - \lambda^*_{s',t^s-1} \bigr) \\
& + \int \biggr[ \bigr( \gamma^*_{s,t} - \gamma^*_{s,t^s-1} \bigr) - \sum_{s' \in \mathbf{s}^c} \omega_{s',s,t} \bigr(\gamma^*_{s',t} - \gamma^*_{s',t^s-1} \bigr) \biggr] x dF_{X_t \mid S=s}(x).
\end{aligned}
\end{equation}

\begin{theorem}[Identification under DID-INT under either conditional parallel trends or integrated parallel trends.] Suppose Assumptions \ref{as: randomsampling}, \ref{as: linearity}, \ref{as: covariateunaffected}, \ref{assumption: sutva} and \ref{as: noanticipation} hold,  and either Assumption \ref{as: cpt} or \ref{as: ipt} holds. Under these assumptions,
\label{theorem: didintidentificationproof}
\begin{equation}
    \tau^{{DT}^*}_{s,t} = ATT(s,t). 
\end{equation}
\end{theorem}
A proof of the above Theorem is shown in Appendix \ref{appendix: didintconsistency}. The proof also shows that when the treatment effect is heterogeneous with respect to $X_{i,s,t}$, such that $ATT_x(s,t) \neq ATT(s,t)$ for at least some values of $x$, the result in Theorem \ref{theorem: didintidentificationproof} holds.

\subsection{Additional Implementation Details and Software}

When there is more than one control region, the fourth step of the DID-INT estimator is easiest to do in the form of a regression.  This allows for $\widehat{\Delta}_{(s)}$ terms from multiple regions to contribute to the estimate of $\widehat{\theta}_{s,t}$.  This regression based approach is identical to the one proposed in \cite{karim2024difference}, except for how the $\widehat{\Delta}_{(s,t)}$ are calculated, see Section 5.2 of that paper for details. To facilitate the implementation of the DID-INT estimator we have a package available in \texttt{Julia} which is available at \url{https://github.com/ebjamieson97/DIDInt.jl}.  We also have a wrapper for \texttt{Stata} which calls the \texttt{Julia} program to perform the calculations, using the approach in \cite{Roodman_Julia}. The \texttt{Stata} program is available at \url{https://github.com/ebjamieson97/didintjl}. A wrapper in \texttt{R} is forthcoming. The software package allows for cluster robust inference using both a cluster-jackknife and randomization inference.  The details of these routines and their finite sample performance are discussed in Section \ref{sec:crve} and the companion paper \cite{austin2026policy}.


\subsection{Variants of the DID-INT Estimator} \label{subsection: modelselectionalgorithm}

In this section, we will explore the four distinct ways to model covariates in DID-INT, depending on the type of CCC violations researchers want to account for. When the two-way CCC seems plausible, we recommend modeling the covariates as $f(X_{i,s,t}) = \gamma X_{i,s,t}$. This version of DID-INT will be referred to as the \textbf{homogeneous DID-INT}. If the time-invariant CCC assumption is plausible but the region-invariant CCC is not, we recommend that researchers interact the covariates with the $I(s)$ dummies and include the interacted terms as covariates in the model. Therefore, $f(X_{i,s,t}) = \sum_{s=1}^S \gamma_s I(s) X_{i,s,t}$, which adjusts for potential violations of the region-invariant CCC. This approach is referred to as the \textbf{region-varying DID-INT}. The third approach, referred to as the \textbf{time-varying DID-INT}, accounts for violations of the time-invariant CCC assumption when the region-invariant CCC assumption is plausible. Potential violations of time-invariant CCC are accounted for by interacting the covariates with the $I(t)$ dummy variables.
This implies: $f(X_{i,s,t}) = \sum_{t=1}^T \gamma_t I(t) X_{i,s,t}$. Lastly, the \textbf{two-way DID-INT} allows for two-way CCC violations, where \(f(X_{i,s,t}) = \sum_{t=1}^T \sum_{s=1}^S \gamma_{s,t} I(s) I(t) X_{i,s,t}\). Here, the covariates are interacted with both the $I(s)$ and the $I(t)$ dummy variables and included as covariates in the model. Figure \ref{fig:functionalform} provides a summary.  Here A and B are two groups, 1 and 2 are two time periods.  The true $\gamma$ terms, $\gamma^{\circ}$, are allowed to potentially vary either across groups, across periods, or across both groups and periods. 

\begin{figure}[ht] 
    \centering
    
    \begin{minipage}[t]{0.45\textwidth}
        \centering
        \caption*{I - Homogeneous:}
        \begin{tabular}{c|cc}
            & A & B \\
            \hline
            1 & $\gamma^0$ & $\gamma^0$ \\
            2 & $\gamma^0$ & $\gamma^0$
        \end{tabular}
        \vspace{0.05cm} 
        \begin{footnotesize}
            \[
            f(X_{i,s,t})  = \gamma X_{i,s,t}
            \]
        \end{footnotesize}
        
        \caption*{III - Year Variation:}
        \begin{tabular}{c|cc}
            & A & B \\
            \hline
            1 & $\gamma^{\circ}_1$ & $\gamma^{\circ}_1$ \\
            2 & $\gamma^{\circ}_2$ & $\gamma^{\circ}_2$
        \end{tabular}
        \vspace{0.05cm} 
        \begin{footnotesize}
            \[
            f(X_{i,s,t})  = \sum_{t=1}^{T} \gamma_t I(t) X_{i,s,t}
            \]
        \end{footnotesize}
    \end{minipage}
    \hfill

    \begin{minipage}[t]{0.45\textwidth}
        \centering
        \caption*{II - Region Variation:}
        \begin{tabular}{c|cc}
            & A & B \\
            \hline
            1 & $\gamma^{\circ}_A$ & $\gamma^{\circ}_B$ \\
            2 & $\gamma^{\circ}_A$ & $\gamma^{\circ}_B$
        \end{tabular}
        \vspace{0.05cm} 
        \begin{footnotesize}
            \[
            f(X_{i,s,t})  =  \sum_{s=1}^{S^T} \gamma_s I(s) X_{i,s,t}
            \]
        \end{footnotesize}
        
        \caption*{IV - Region \& Year (Two-way):}
        \begin{tabular}{c|cc}
            & A & B \\
            \hline
            1 & $\gamma^{\circ}_{A,1}$ & $\gamma^{\circ}_{B,1}$ \\
            2 & $\gamma^{\circ}_{A,2}$ & $\gamma^{\circ}_{B,2}$
        \end{tabular}
        \vspace{0.05cm} 
        \begin{footnotesize}
            \[
            f(X_{i,s,t})  =  \sum_{t=1}^{T} \sum_{s=1}^{S^T} \gamma_{s,t} I(s) I(t) X_{i,s,t}
            \]
        \end{footnotesize}
    \end{minipage}
    
    \caption{Modeling Covariates in DID-INT}
    \label{fig:functionalform}
\end{figure}

\FloatBarrier

 We have introduced four versions of DID-INT that researchers can employ based on the functional form of the covariates, $f(X_{i,s,t})$. In practice, the true functional form is unobserved. Therefore, we recommend using the CCC test introduced in \citep{karim2026ccc} to determine the appropriate functional form of the covariates in DID-INT. Depending on the restrictions supported by the test, covariates can be modeled using one of four specifications.


\section{The Two-way Fixed Effects (TWFE) Estimator} \label{sec:twfeproof}

In this section, we formally examine the identification properties of the standard and the flexible TWFE estimators under violations of the three CCC assumptions. The TWFE model can be written as:
    \begin{equation}
        \label{eq: twfeonecov}
            Y_{i,s,t} = \alpha_s + \delta_t + \beta_{D} D_{s,t} + f(X_{i,s,t}) + \epsilon_{i,s,t}
    \end{equation}

The key parameter of interest is $\beta_{D}$, which we derive in this section. For simplicity, we assume that there is a single time-varying covariate $X_{i,s,t}$ that is required for either \textbf{conditional parallel trends} or \textbf{integrated parallel trends} to hold.  One important feature of the fixed effects structure in Equation \eqref{eq: twfeonecov} is that the region and period effects are restricted to be additive, as shown in Equation \eqref{eq: additivefe}. 
\begin{equation}
\label{eq: additivefe}
    \theta^{\circ}_{s,t} = \alpha^{\circ}_s + \delta^{\circ}_t
\end{equation}
Under the correct specification of the population regression, $\alpha^{\circ}_s = \alpha^{*}_s$ and $\delta^{\circ}_t = \delta^{*}_t$. When the restriction shown in Equation \eqref{eq: additivefe} is violated, the TWFE estimator can suffer from a misspecification bias. The DID-INT does not impose such a restriction on the region-period intercepts, and is therefore not susceptible to this bias. 
    
Similar to DID-INT, the TWFE regression can take four distinct functional forms depending on the type of CCC violation. When the two-way CCC assumption is plausible, we model the covariates as $f(X_{i,s,t}) = \gamma X_{i,s,t}$. We refer to this specification as the \textbf{standard TWFE}. If the time-invariant CCC assumption is plausible but the region-invariant CCC assumption is violated, we interact the covariates with the $I(s)$ dummies and include the interacted terms as covariates in the model. Therefore, $f(X_{i,s,t}) = \sum_{s=1}^S \gamma_s I(s) X_{i,s,t}$, which accounts for potential violations of the region-invariant CCC. We refer to this specification as the \textbf{region-varying flexible TWFE}. The third specification, referred to as the \textbf{time-varying flexible TWFE}, accounts for violations of the time-invariant CCC when the region-invariant CCC assumption is plausible. Potential violations of the time-invariant CCC are accounted for by interacting the covariates with the $I(t)$ dummy variables. This implies $f(X_{i,s,t}) = \sum_{t=1}^T \gamma_t I(t) X_{i,s,t}$. Lastly, the \textbf{two-way flexible TWFE} allows for two-way CCC violations, where \(f(X_{i,s,t}) = \sum_{t=1}^T \sum_{s=1}^S \gamma_{s,t} I(s) I(t) X_{i,s,t}\). Here, the covariates are interacted with both the $I(s)$ and the $I(t)$ dummy variables and included as covariates in the model. While the homogeneous specification corresponds to the standard TWFE, the remaining three specifications are flexible TWFE estimators that accommodate different types of CCC violations.

Regardless of the specification used, both the standard and the flexible TWFE estimators can be biased in staggered treatment adoption designs under heterogeneous treatment effects because of forbidden comparisons and negative weighting issues \citep{goodman2021difference, de2020twott}. Our objective in this section is to isolate an additional source of bias arising from CCC violations. In particular, CCC violations can cause conditional parallel trends to be violated, even when integrated parallel trends continue to hold. To separate this bias from the bias associated with staggered adoption, we impose the homogeneous treatment effects assumption throughout this section. 
\begin{assumption}[Homogeneous Treatment Effect] The average treatment effect on the treated is constant across both regions and time.
\label{as: homte}
\begin{equation}
    \mathbb{E}\bigr[ Y_{t}(1) - Y_{t}(0) \mid S_i = s  \bigr] = \mathbb{E}\bigr[ Y_{t'}(1) - Y_{t'}(0) \mid S_i = s'  \bigr] \quad \forall s,s',t,t' \text{such that} D_{s,t} = D_{s',t'} = 1.  
\end{equation}
\end{assumption}

Although Assumption \ref{as: homte} imposes homogeneous treatment effects across regions and periods, it does not require the treatment effect to be homogeneous across values of the covariate. In particular, we allow the conditional treatment effect to vary with $x$, where:
\begin{equation}
    \tau_{s,t}(x)=\mathbb{E}\bigr[Y_t(1)-Y_t(0)\mid S_i=s,X_t=x\bigr].
\end{equation}
Thus, Assumption \ref{as: homte} requires the average treatment effect on the treated to be constant across regions and periods, while allowing $\tau_{s,t}(x)$ to vary with $x$.

\begin{theorem}[Flexible TWFE under integrated parallel trends] Suppose that Assumptions \ref{as: randomsampling}, \ref{as: linearity}, \ref{as: covariateunaffected}, \ref{assumption: sutva}, \ref{as: noanticipation} and \ref{as: homte} hold, and that either Assumption \ref{assumption: twowayCCC}, \ref{as2: regioninvariantCCC} or  \ref{as2: timeinvariantCCC} is violated. Also suppose that $f(X_{i,s,t})$ correctly accounts for the relevant CCC violations, and the additive fixed-effects restriction in Equation \eqref{eq: additivefe} holds. Furthermore, suppose Assumption \ref{as: ipt} holds, and that the treatment effect is homogeneous with respect to $x$, such that $ATT_x(s,t)=ATT(s,t)$ for all $x$. Under these conditions, integrated parallel trends are sufficient for the flexible TWFE to identify the ATT. 
\label{theorem: twfeipt}
\begin{equation}
\label{eq: modtwfeatt}
    \beta^*_{D} = ATT.
\end{equation}
\end{theorem}

\begin{theorem}[Flexible TWFE under integrated parallel trends and treatment-effect heterogeneity with respect to covariates.] Suppose that Assumptions \ref{as: randomsampling}, \ref{as: linearity}, \ref{as: covariateunaffected}, \ref{assumption: sutva}, \ref{as: noanticipation} and \ref{as: homte} hold, and that either Assumption \ref{assumption: twowayCCC}, \ref{as2: regioninvariantCCC} or  \ref{as2: timeinvariantCCC} is violated. Also suppose that $f(X_{i,s,t})$ correctly accounts for the relevant CCC violations, and the additive fixed-effects restriction in Equation \eqref{eq: additivefe} holds. Furthermore, suppose Assumption \ref{as: ipt} holds, and that the treatment effect is heterogeneous with respect to $x$, such that $ATT_x(s,t)\neq ATT(s,t)$ for at least some values of $x$. In the non-trivial case where the integrated covariate slope component of the flexible TWFE is non-zero, integrated parallel trends are not sufficient for the flexible TWFE to identify the ATT.
\label{theorem: twfeiptheterogeneous}
\begin{equation}
\label{eq: modtwfeattheterogeneous}
    \beta^*_{D} \neq ATT.
\end{equation}
\end{theorem}

The proofs of Theorems \ref{theorem: twfeipt} and \ref{theorem: twfeiptheterogeneous} are provided in Appendix \ref{appendix: twfeconsistencymod}. Since conditional parallel trends imply integrated parallel trends, the identification result in Theorem \ref{theorem: twfeipt} also holds under conditional parallel trends. The result in Theorem \ref{theorem: twfeipt} relies on several strong assumptions. There are two main reasons why the flexible TWFE may fail to identify the ATT when these assumptions are relaxed. First, the additive fixed-effects restriction in Equation \eqref{eq: additivefe} might not apply to real-world data, but the flexible TWFE imposes this restriction. This could result in misspecification bias. Second, when Assumption \ref{as: homte} is violated, the flexible TWFE estimator can be inconsistent even when integrated parallel trends hold because of negative weighting and forbidden comparisons. Therefore, the result in Theorem \ref{theorem: twfeipt} is no longer valid. 

 In practice, neither the conditional nor the integrated parallel trends can be statistically verified because both assumptions concern the untreated potential outcome in the post-treatment period, which we do not observe. Therefore, researchers examine pre-intervention trends in the treated and control groups to assess the plausibility of parallel trends after the intervention. The conventional approach to evaluating conditional parallel trends begins by regressing the observed outcomes on covariates in the pre-intervention period, including a constant. In this regression, the covariates are assumed to share a common coefficient. In the second step, the residuals are plotted for each region separately in a common figure.  The residuals are shown in Equation \eqref{eq: incorrectresiduals}.
\begin{equation}
\label{eq: incorrectresiduals}
    \tilde{Y}^{CPT}_{i,s,t} = Y_{i,s,t} - \widehat{\gamma} X_{i,s,t}.
\end{equation}

Once the pre-trends are plotted in this way, researchers typically assess whether the trends appear parallel through visual inspection. The conclusions from these visual inspections are inherently subjective and may lead researchers to incorrectly conclude that the trends are parallel under a homogeneous covariate specification, although there may be CCC violations. Consequently, they may end up estimating the standard TWFE model, even though the CCC assumptions may be violated in the data. 

\begin{theorem}[Standard TWFE estimator under CCC violations] Suppose that Assumptions \ref{as: randomsampling}, \ref{as: linearity}, \ref{as: covariateunaffected}, \ref{assumption: sutva}, \ref{as: noanticipation} and \ref{as: homte} hold, and that either Assumption \ref{assumption: twowayCCC}, \ref{as2: regioninvariantCCC} or  \ref{as2: timeinvariantCCC} is violated. Also suppose that the additive fixed-effects restriction in Equation \eqref{eq: additivefe} holds. The standard TWFE with a common covariate coefficient can be biased even when Assumption \ref{as: cpt} holds.
\label{theorem: twfeunflexible}
    \begin{align}
    \begin{split}
    \label{equation: finalresults}
       \beta_{D} = ATT + \text{Bias}_D
    \end{split}
    \end{align}
\end{theorem}
A proof of Theorem \ref{theorem: twfeunflexible} is shown in Appendix \ref{appendix: twfeconsistency}.


\section{Other Difference-in-Difference Estimators} \label{sec:other}

In this section we discuss three alternative difference-in-differences estimators.  Specifically, we discuss: the widely used \cite{callaway2021difference}
estimator for staggered adoption in Section \ref{sec:CS-DID}; the imputation estimator from \cite{borusyak2024revisiting} in Section \ref{sec:imputation}; and the new FLEX estimator from \cite{deb2024flexible} in Section \ref{sec:flex}.

\subsection{Callaway and Sant'Anna (2021) DID estimator} \label{sec:CS-DID}

In this section, we will explore the potential inconsistency that arises in the \cite{callaway2021difference} DID estimator (CS-DID) when integrated parallel trends hold. We show that, when the CS-DID is conducted with the outcome regression (OR), it can identify the ATT under integrated parallel trends, heterogeneity of the treatment effect with respect to covariates, and the no compositional changes assumption. However, when the compositional changes assumption is violated, the regression adjustment fails. The inverse probability weighting (IPW) approach is also problematic under the interacted covariate structures considered in Figure \ref{fig:functionalform}. In particular, region-specific covariate interactions can generate a lack of overlap between the treated and comparison groups within the relevant group-time comparisons, preventing the propensity score from being well defined. This issue also limits the applicability of doubly robust DID (DR-DID) estimators that rely on propensity-score weighting in combination with OR. 

The CS-DID is a semi-parametric method that estimates the ATT without forbidden comparisons, as demonstrated by \cite{goodman2021difference} and \cite{de2020twott}. The estimation of the ATT involves two steps. In the first step, the dataset is decomposed into several ``2x2 comparison'' blocks, each containing a treated group and a control (or not yet treated) group. The pre-intervention period is the period right before the treated group is treated. Without covariates, the ATT of each of the ``2x2 comparison" blocks, known as $ATT(g,t)$, is estimated non-parametrically as follows:
        \begin{align}
            \begin{split}
                ATT(g,t) = \biggr( \mathbb{E} \bigr[ Y_{i,g,t} \bigr] - \mathbb{E} \bigr[ Y_{i,g,g-1} \bigr] \biggr) - \biggr( \mathbb{E} \bigr[Y_{i,g',t}\bigr] - \mathbb{E}\bigr[Y_{i,g',g-1}\bigr] \biggr) 
            \end{split}
        \end{align}

Here, groups or cohorts are determined by the period they were first treated, $g$, which differs from DID-INT. The differences in estimating the jurisdiction and period ATT at different levels are detailed in  \citet*{austin2026policy}. The period $g-1$ is the period right before treatment for cohort $g$. For treated cohort $g$, $g'$ is a relevant control group for cohort $g$, which is either not yet treated or never treated in period $t$. The second step involves taking a weighted average of all the $ATT(g,t)$'s to get an overall estimate of the ATT: 
    \begin{equation}
        ATT = \sum_{g=2}^{G} \sum_{t=2}^\mathcal{T} 1\{g \leq t\} w_{g,t} ATT(g,t)
    \end{equation}  

The above avoids all the forbidden comparisons demonstrated by \cite{goodman2021difference} and \cite{de2020twott}. With covariates, the first step is estimated using the Doubly Robust DID (DR-DID) approach \citep{sant2020doubly} by default. The DR-DID approach combines the IPW approach proposed by \cite{abadie2005semiparametric} and the OR approach proposed by \cite{heckman1997matching} to derive a doubly robust estimator. This estimator is robust to misspecification, provided either the propensity score model or the outcome regression model is correctly specified. The CS-DID can also estimate the $ATT(g,t)$'s using only the IPW or OR \citep*{RePEc:boc:bocode:s458976}. However, using the DR-DID can be advantageous if the propensity score and outcome regressions depend on time-varying covariates in both periods, due to the property of double robustness \citep*{caetano2022timevarying}. 

Estimators such as IPW, OR, and DR-DID have two disadvantages. The first disadvantage concerns interacted covariates. Although the OR can incorporate interacted covariates directly, propensity score-based approaches such as IPW and DR-DID may encounter some problems if these interacted covariates lack overlap, formally defined in Assumption \ref{as: overlap}.  

\begin{assumption}[Overlap] For each $t \in \{2,\ldots,T\}$ and $g$, there exists some $\varepsilon > 0$ such that \[\mathbb{P}(G_g = 1) > \varepsilon\]
and \[p_{g,t}(X) < 1-\varepsilon\qquad\text{a.s.}
\]
\label{as: overlap}
\end{assumption}

We demonstrate this problem using the two-way specification. The same argument applies to specifications containing region-specific covariate interactions, including the region-varying and two-way specifications. Since the region-varying and two-way interacted covariate structures can violate the required overlap condition, propensity-score-based approaches such as IPW and DR-DID are not applicable with these specifications. Pure time-varying interactions do not generate this problem solely by construction.

\begin{gather}
f(X_{i,s,t}) = \Big[X_{i,s,t} I(Treat) I(t^s-1), X_{i,s,t} I(Treat) I(t), \nonumber
\\ X_{i,s,t} I(Control) I(t^s-1), X_{i,s,t} I(Control) I(t) \Big].
\end{gather}

For treated observations in the pre-treatment period,
\begin{equation}
f(X_{i,Treat,t^s-1}) = \left[ X_{i,Treat,t^s-1}, \; 0, \; 0, \; 0 \right].
\end{equation}

For treated observations in the post-treatment period,
\begin{equation}
f(X_{i,Treat,t}) = \left[0, \; X_{i,Treat,t}, \; 0, \; 0 \right].
\end{equation}

For control observations in the pre-treatment period,
\begin{equation}
f(X_{i,Control,t^s-1}) = \left[ 0, \; 0, \; X_{i,Control,t^s-1}, \; 0 \right].
\end{equation}

For control observations in the post-treatment period,
\begin{equation}
f(X_{i,Control,t}) = \left[0, \; 0, \; 0, \; X_{i,Control,t} \right].
\end{equation}

Except when the relevant value of $X_{i,s,t}$ is zero, each group-period cell therefore occupies a distinct component of the support of $f(X_{i,s,t})$. In particular, the interacted covariates distinguish treated observations from control observations within each period. Consequently, for points in the treated support,
\begin{equation}
\Pr\left(S_i=Treat \mid f(X_{i,s,t})\right) = 1,
\end{equation}
whereas for points in the control support,
\begin{equation}
\Pr\left( S_i=Treat \mid f(X_{i,s,t}) \right) = 0.
\end{equation}
Hence, the overlap condition $0 < \Pr\left( S_i=Treat \mid f(X_{i,s,t}) \right) < 1$ is violated.

Since the interacted covariate structures violate the required overlap condition, propensity score-based approaches such as IPW and DR-DID are not applicable with these specifications. For this reason, we do not consider CS-DID specifications with interacted covariates when implementing IPW or DR-DID.

The second disadvantage of IPW, OR, and DR-DID is that, with repeated cross-sectional data, these estimators require a no compositional changes assumption to identify the ATT. We formally introduce this assumption below.

\begin{assumption}[No Compositional Changes]
\label{as: compositional}
Within each region, the distribution of the covariates is independent of the time period. In our notation, this condition can be written as
\begin{equation}
    X_{i,s,t} \perp T_t \mid S_i=s.
\end{equation}
Accordingly, the distribution of the covariates within each region remains unchanged between periods $t^s-1$ and $t$. That is,
\begin{equation}
    F_{X_{i,s,t}\mid S_i=s}(x)
    =
    F_{X_{i,s,t^s-1}\mid S_i=s}(x),
    \qquad
    s\in\{Treat,Control\},
    \quad \forall x.
\end{equation}
\end{assumption}

Assumption \ref{as: compositional} allows the realized values of the covariates to differ across observations and periods, but requires their distribution within each region to remain unchanged over time. It is weaker than the no compositional changes assumption imposed by \cite{hong2013measuring}, which requires
\begin{equation}
    T_t \perp \left(X_{i,s,t},S_i\right).
\end{equation}
The condition in \cite{hong2013measuring} implies both independence between the covariates and time conditional on region, $X_{i,s,t} \perp T_t \mid S_i=s$, and independence between regional assignment and time. Assumption \ref{as: compositional} requires only the former. Although this assumption is not explicitly stated in \cite{sant2020doubly}, Assumption 1(b) in \citet[p.g. 103]{sant2020doubly} requires the joint distribution of treatment assignment and the covariates to remain invariant over time, thereby ruling out compositional changes.

Since propensity score-based implementations of CS-DID, such as IPW and DR-DID, become problematic with the interacted covariate structures considered above due to violations of the overlap condition, we focus on the case in which CS-DID is implemented using OR. Since CS-DID estimates the ATT of several ``2x2" comparison blocks by default, let us analyze the OR estimator in a two-region, two-period framework. The OR estimand for repeated cross-sectional data is given by
\begin{equation}
    \tau^{or,rc} = \mathbb{E}\left[      \omega_1^{rc}(D_s,T_t) OR_{i,s,t} \right],
\end{equation}
where
\begin{equation}
    \omega_1^{rc}(D_s,T_t) = \left(    \frac{D_sT_t}{\mathbb{E}[D_sT_t]} - \frac{D_s(1-T_t)}{\mathbb{E}[D_s(1-T_t)]}
    \right), 
\end{equation}
and 
\begin{equation}
\begin{aligned}
OR_{i,s,t} = & T_t \left[ Y_{i,s,t} - \mathbb{E}^*\left[Y_{i,s',t} \mid S_i=s', T_t=1,
X_{i,s',t}=x \right] \right] \\
& + (1-T_t) \left[ Y_{i,s,t^s-1} - \mathbb{E}^*\left[ Y_{i,s',t^s-1} \mid S_i=s', T_t=0, X_{i,s',t^s-1}=x \right] \right].
\end{aligned}
\end{equation}

Here, $D_s$ is the treatment-group indicator, taking a value of 1 if observation $i$ belongs to the treated region $s$ and 0 if it belongs to the control region $s'$. $T_t$ is the period indicator, taking a value of 1 if observation $i$ is observed in the post-intervention period $t$ and 0 if it is observed in the pre-intervention period $t^s-1$. Finally,
$\mathbb{E}^*[Y_{i,s',t}\mid D_s=d,T_t=t,X_{i,s',t}]$
denotes the model-implied conditional expectation, which may differ from the true population conditional expectation under model misspecification.

\begin{theorem}[OR identifies the ATT under integrated parallel trends and no compositional changes]
\label{theorem: orcompositional}
Suppose Assumptions \ref{as: randomsampling}, \ref{as: linearity}, \ref{as: covariateunaffected}, \ref{assumption: sutva}, \ref{as: noanticipation}, \ref{as: ipt}, and \ref{as: compositional} hold, and the outcome regression is correctly specified. Then, with repeated cross-sectional data,
\begin{equation}
    \tau^{or,rc} = ATT(s,t).
\end{equation}
\end{theorem}

\begin{theorem}[Outcome regression is generally biased under compositional changes]
\label{theorem: ornoncompositional}
Suppose Assumptions \ref{as: randomsampling}, \ref{as: linearity}, \ref{as: covariateunaffected}, \ref{assumption: sutva}, \ref{as: noanticipation}, and \ref{as: ipt} hold, and the outcome regression is correctly specified, but Assumption \ref{as: compositional} is violated. Then, with repeated 
cross-sectional data, in the non-trivial case,
\begin{equation}
    \tau^{or,rc} \neq ATT(s,t).
\end{equation}
\end{theorem}

The proofs of Theorems \ref{theorem: orcompositional} and \ref{theorem: ornoncompositional} are shown in Appendix \ref{Appendix: DRDID}. Therefore, with repeated cross-sectional data, CS-DID requires the no compositional changes assumption to identify the ATT. In contrast, DID-INT does not require this assumption and can accommodate changes in the distribution of covariates over time.

\subsection{Imputation Estimator} \label{sec:imputation}

In this section we compare the (static) imputation estimator proposed by \citet*{borusyak2024revisiting}, also designed to estimate the ATT under staggered treatment adoption. The imputation estimator is done in four steps. In the first step, the unit ($\alpha_i$) and period ($\beta_t$) fixed effects, and coefficients of covariates ($\gamma$) are fitted by using a TWFE model on the untreated units only. The regression is shown in Equation \eqref{equation: imp1}.
\begin{equation}
\label{equation: imp1}
    Y_{i,s,t} = \alpha_i + \beta_t + \tau^{static} D_{i,s,t} + \gamma X_{i,s,t} + \epsilon_{i,s,t}
\end{equation}
Note, for all observations in the control group, $D_{i,s,t} = 0$. In the second step, the estimated fixed effects from this model are used to impute the untreated potential outcomes of the treated group, shown in Equation \eqref{equation: imp2}.
\begin{equation}
\label{equation: imp2}
    \widehat{Y}_{i,s,t}(0) = \widehat{\alpha}_i + \widehat{\beta}_t  + \widehat{\gamma} X_{i,s,t} + \epsilon_{i,s,t}
\end{equation}
In the third step, the individual treatment effects are estimated by taking a difference of the observed outcome of each treated unit and the imputed counterfactuals. This is shown in Equation \eqref{equation: imp3}. 
\begin{equation}
\label{equation: imp3}
    \widehat{\tau}_{i,s,t} = Y_{i,s,t} - \widehat{\alpha}_i + \widehat{\beta}_t - \widehat{\gamma} X_{i,s,t} \; \forall i \in S^T
\end{equation}
In the final step, the overall ATT is estimated by taking a weighted average of $\tau_{i,s,t}$ across all treated units and periods. Although the imputation estimator provides a viable method for estimating the ATT in the presence of time-varying covariates and staggered treatment designs, it is not robust to CCC violations. Specifically, step 2 relies on the assumption that the coefficients of the covariates are identical for the treated and control groups in order to impute the untreated potential outcomes of the treated observations. As a result, the overall estimate of the ATT obtained in step 4 may be inconsistent under CCC violations.  

This means that the imputation estimator does not provide a consistent estimate of the ATT when the CCC assumption is violated.  Unfortunately, it is also not possible to easily modify the estimator as the current version fits all parameters using regressions with only untreated units. 

\subsection{The FLEX model} \label{sec:flex}

In this section, we compare the DID-INT to the flexible linear model or FLEX proposed by \citet*{deb2024flexible}. The FLEX model also interacts the covariates with a region dummy and a time dummy. However, FLEX model generates three types of variables: one where the covariates are interacted with the group dummies ($I(g) X^k_{i,g,t}$); one where the covariates are interacted with the time dummies ($I(t) X^k_{i,g,t}$); and the third where the covariates are interacted with both time and group dummies ($\sum_{g \neq \infty} \sum_{t \geq r} \sum_{k} \beta_{gtk} I(g) I(t) X_{gtk}$).  Importantly, these ``interacted'' dummies are only for the treated units in either the post period, or all periods, depending on whether or not the `leads' option is specified. These covariates are then included in the FLEX model in an additive way: $\sum_{g \neq \infty} \sum_{t \geq t^*} \sum_{k} \beta_{gtk} I(g) I(t) X_{gtk} + \sum_{g} \sum_{k}  \beta_{gk} I(g) X_{gk} + \sum_{t} \sum_{k} \beta_{tk} I(t) X_{tk}$. The regression for the FLEX model is shown below:

\begin{equation}
\begin{aligned}
    y_{gt} = & \sum_{g \neq \infty} \sum_{t \geq t^*} \tau_{gt} I(g) I(t)
    + \sum_{g \neq \infty} \sum_{t \geq t^*} \sum_{k} \beta_{gtk} I(g) I(t) X_{gtk} \\
    & + \sum_{g} \sum_{k} \beta_{gk} I(g) X_{gk} 
    + \sum_{t} \sum_{k} \beta_{tk} I(t) X_{tk} \\
    & + \sum_{k} \beta_k X_k 
    + \sum_{t} \phi_t I(t) + \sum_{g} \psi_g I(g)     + \epsilon_{gt}.
\end{aligned}
\end{equation}
The second step involves taking a weighted average of the estimated treatment effect similar to the DID-INT. We highlight a few key differences between the FLEX and the two-way DID-INT. First, the FLEX model includes the three types of interacted covariates in the regression specification shown above, in addition to non-interacted covariates. In contrast, the DID-INT only includes the covariates interacted with the time or group dummies, or both. Second, the FLEX model includes two-way interactions of covariates for only a subset of the `interactions'.  As noted, these are estimated only for the treated regions.  They are either estimated only for the post-intervention period when `leads' is not specified, and both pre-intervention and post-intervention periods when it is not.
This implies that the DID-INT can capture the variations across time and group in both treatment and control groups.
Whether DID-INT or FLEX is estimating more parameters depends on the number of groups, the number of time periods, and the number of covariates.  Finally, FLEX is based on the TWFE model and tries to model the untreated outcomes with group and time fixed effects.  

\section{Monte Carlo Simulation Study}

\label{section: MC}

In this section, we introduce the design of our Monte Carlo Simulation Study.  We use data from the Current Population Survey (CPS) for these simulations.  In Section \ref{sec:mc-att} we compare the distribution of ATT estimates from DID-INT with several other DID estimators. We then consider cluster-robust inference in Section \ref{sec:crve}.

\subsection{ATT Estimation} \label{sec:mc-att}

To keep the constructed dataset as realistic as possible, we use data from the Current Population Survey (CPS) covering the years 2000 to 2013. The CPS is a repeated cross-sectional dataset that includes information on employment status, earnings, education, and demographic trends of individuals. Similar to \citet*{bertrand2004much}, we restrict our sample to women between the ages of 24 and 55 in their fourth interview month. We keep only observations with positive earnings and non-missing education, resulting in a sample of 346,589 individuals.

\begin{figure}
	\centering
	\includegraphics[width=1\linewidth]{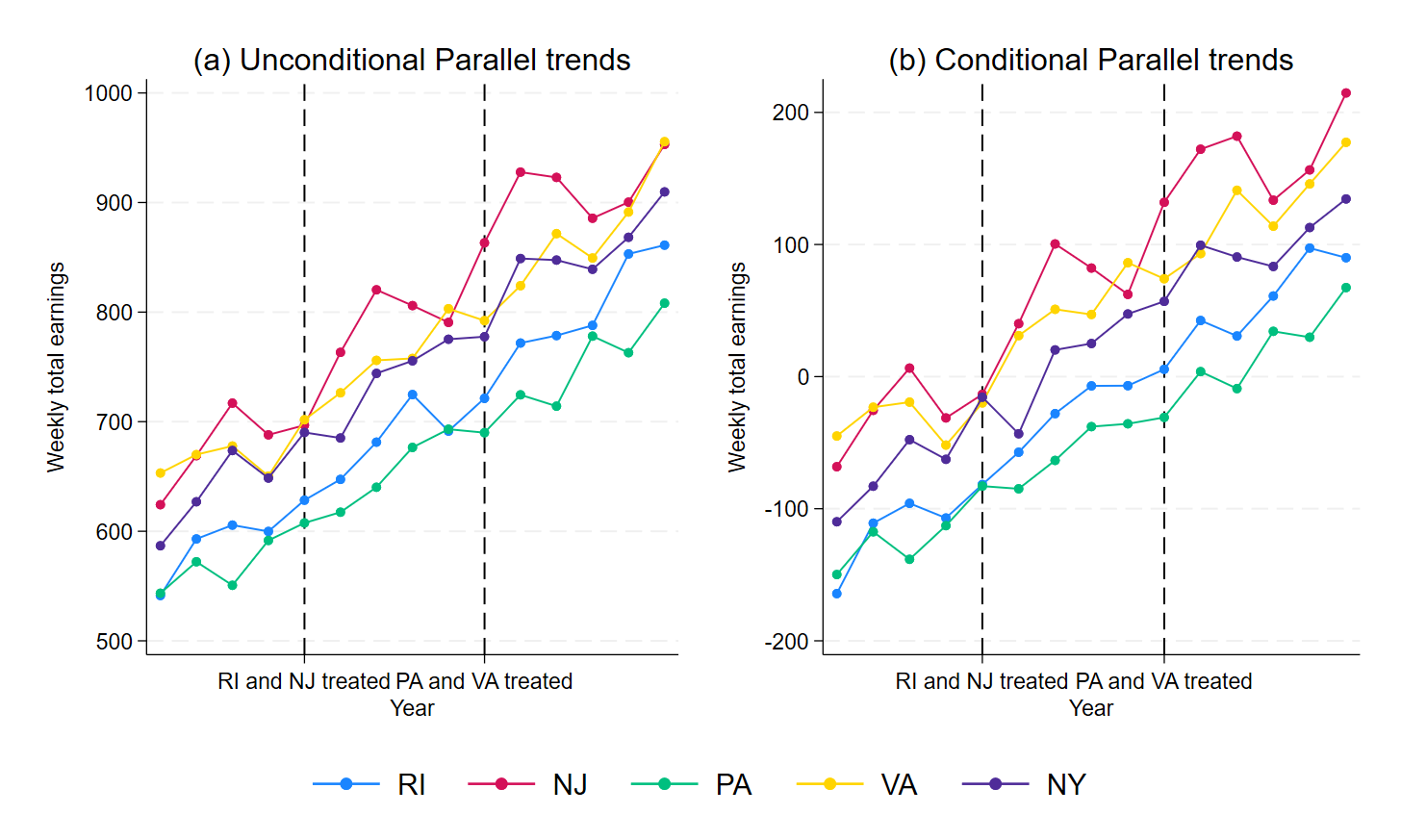}
	\caption{Parallel trends for weekly earnings}
	\label{figure: weeklypt}
\end{figure}

To generate our constructed outcome, we start by estimating coefficient for selected covariates based on individual's weekly earnings. We limit our analysis to Rhode Island, New Jersey, Pennsylvania, Virginia, and New York, where parallel trends seem plausible. An appendix example uses just data from New York and New Jersey. The parallel trends figures are shown in Figure \ref{figure: weeklypt}. The chosen covariates include age, race, education and marital status, which are known to influence weekly wages. Race, education, and marital status are transformed into binary variables, while age remains continuous.  For individual $i$ in state $s$ and time period $t$ let $X_{i,s,t}$ be the vector of these covariates.  In these simulations RI and NJ are treated in 2004, and PA and VA are treated in 2009, and NY is never treated.

To investigate the various difficulties for empirical research, we consider three problematic issues.  The first is that covariates can change in composition (violations of Assumption \ref{as: compositional}), or that the covariates are not time-invariant.  The second is that there can be a covariate-specific treatment effect (or the treatment effects are heterogeneous with respect to $X_{i,s,t}$).  The third is CCC violations.  The distributions of the covariates are matched to the CPS, but treatment effects are imposed.

 All Monte Carlo designs  are special cases of
\begin{equation}
	Y_{i,s,t}^{(c)}
	=
	a_{s,t}(c)
	+cX_{i,s,t}'\beta_{s,t}
	+D_{s,t}\tau(X_{i,s,t})
	+\epsilon_{i,s,t},
	\qquad
	\epsilon_{i,s,t}\overset{\mathrm{iid}}{\sim}N(0,\sigma^2).
	\label{eq:mc-outcome}
\end{equation}

Here, $Y_{i,s,t}^{(c)}$ is the outcome variable. The state--year intercepts are $a_{s,t}(c)$, this differ according to the various DGPs.  
The vector $\beta_{s,t}$ allows for covariates to vary across either states, years, neither, or both.  This variation is a primary difference between DGPs.  The scalar $c$ controls the importance of 
the covariates, in the main DGP $c=1$ so the covariates are as important as in the original CPS.  In some sensitivity 
experiments $c>1$.  For most of the DGPs, the treatment effect $\tau=\tau(X_{i,s,t})=0$. $D_{s,t}$ is a binary treatment indicator.  In some 
DGPs the intercepts are adjusted so that IPT holds, to match the identifying assumptions.

The conditional mean of the untreated outcome is therefore
\begin{equation}
	m_{s,t}^{(c)}(x)
	=
	a_{s,t}(c)+cx'\beta_{s,t}.
	\label{eq:mc-untreated-mean}
\end{equation}

For all simulations the variance of the error terms was calibrated to match the residuals from the CPS, so $\sigma=466.64$.  
We use 2,500 Monte Carlo replications in each design.  Importantly, there is no anticipation, and treatment does not change the distribution of the covariates.  

\noindent\begin{minipage}{\linewidth}
	Let $t^s$ denote adoption, $b_s=t^s-1$, and $F_{s,t}$ the covariate
	distribution in state--year cell $(s,t)$. State--year and aggregate
	treatment effects are
	\begin{equation}
		\begin{aligned}
			ATT(s,t)
			&=\int\tau(x)\,dF_{s,t}(x),\\
			ATT
			&=\sum_{s,t:D_{s,t}=1}w_{s,t}ATT(s,t),
			\qquad
			w_{s,t}
			=\frac{N_{s,t}}{\sum_{s',t':D_{s',t'}=1}N_{s',t'}}.
		\end{aligned}
		\label{eq:mc-target}
	\end{equation}
\end{minipage}\par

The population ATT does not depend on $c$.  DID-INT estimates ATT(s,t) terms for each of the four 
treated states before estimating an ATT.  Similarly, CS-DID estimates ATT(g,t) terms for the treatment
cohorts before estimating an ATT through simple (population weight) averaging. 
In addition, we consider direct ATT estimates from each of FLEX, 
BJS, TWFE, and flexible TWFE.  The latter is just conventional TWFE with the covariate slopes estimated 
separately for each state--year cell.  Table ~\ref{tab:mc-summary} reports summary statistics for 
each estimator for each DGP with $c=1$, and Table \ref{tab:mc-sensitivity} reports them for $c>1$.
  Specifically, the tables report the Monte Carlo Mean Error, Standard Deviation, 
and Root Mean Squared Error (RMSE).  We now discuss the DGPs in turn.

\subsection{Changing the CCC Assumption}

We consider six different designs.  The first has $\beta_{s,t}=\mathbf{0}$ so the covariates do 
not enter the DGP, the second has homogeneous  $\beta_{s,t}$ in the DGP, 
the third uses state varying  $\beta_{s,t}$ and increased the education share by 10 percentage points
 in 2006, the fourth uses state--time varying  $\beta_{s,t}$.
These $\beta_{s,t}$ parameters are estimated from the CPS data. The fifth varies the treatment effects by education level, 
the sixth changes the share of educated individuals and the return to education for the control group.
The appendix considers a DGP with just two states. To highlight the differences between estimators 
when covariates matter more, several DGPs consider two values of $c$, both $c=1$ and $c>1$. 
For each design, \(c>1\) is chosen to double its within-state--year covariate $R^2$. For example, in the
 composition-shift design, $R^2$ rises from 0.1676 to 0.3353 with $c=1.582$.

For the three DGPs $a_{s,t}(c) = \alpha_s+\delta_t$ and  $\tau(x)=0$. In the fourth DGP, the 
intercepts need to be offset for IPT to hold. Using the CPS sample, let the separately estimated
 state--year slopes be  $\beta^{\mathrm{CPS}}_{s,t}$.
Set $\beta_{s,t}=\beta^{\mathrm{CPS}}_{s,t}$ and $\tau(x)=0$ in \eqref{eq:mc-outcome}, and define
	\begin{align}
		d_{s,t}
		&=
		(\beta_{s,t}-\beta_{s,b_s})
		-
		(\beta_{\mathrm{NY},t}-\beta_{\mathrm{NY},b_s}),
		\notag\\
		a_{s,t}(c)
		&=
		a_{s,b_s}(c)
		+a_{\mathrm{NY},t}(c)
		-a_{\mathrm{NY},b_s}(c)
		-cd_{s,t}'\bar X_{s,t},
		\qquad t\geq t^s.
		\label{eq:mc-two-way}
	\end{align}

Before adoption and in New York, intercepts are calibrated from
CPS state earnings levels net of a common linear trend, with
that trend then added back. These intercepts remain fixed as
$c$ varies. The treated-post intercepts in
\eqref{eq:mc-two-way} offset the integrated slope contrast so IPT holds.
At either signal strength, the conditional untreated-trend
discrepancy is
\[
q_{s,t}^{(c)}(x)
=
cd_{s,t}'(x-\bar X_{s,t}),
\qquad
\int q_{s,t}^{(c)}(x)\,dF_{s,t}(x)=0.
\]
CPT and two-way CCC fail, while IPT holds for every
treated-post comparison. The covariate slopes and composition
are empirically calibrated; the intercept restriction is
imposed by construction.

\begin{figure}[!htbp]
	\centering
	\caption{No Covariate Effect}
	\label{fig:mc-no-x}
	\includegraphics[width=\textwidth]{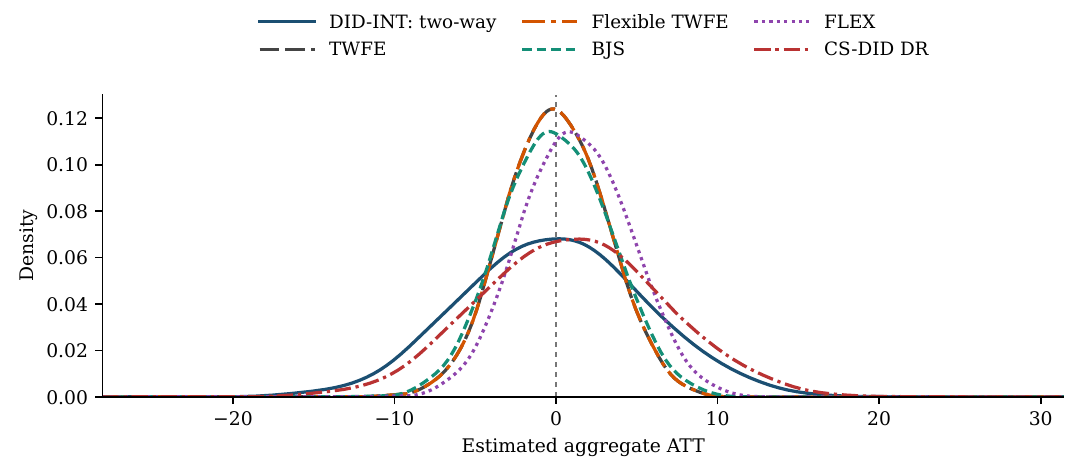}
	\par\smallskip
	{\footnotesize\raggedright
		\textit{Notes:}
		In \eqref{eq:mc-outcome}, $a_{s,t}(1)=\alpha_s+\delta_t$,
		$\beta_{s,t}=\mathbf{0}$, and $\tau(x)=0$. State-level PT holds,
		but changing state shares can violate cohort-level PT. All
		estimators omit covariates: ordinary and flexible TWFE coincide.
		 The reference line marks zero ATT.
		\par}
\end{figure}

Figure \ref{fig:mc-no-x} shows the estimates from the no covariate DGP.  In this
DGP both conventional and flexible TWFE coincide and are the most efficient.  
BJS is unbiased and slightly less efficient, DID-INT is also unbiased, but less 
efficient.  Both FLEX and CS-DID are slightly biased, while CS-DID is less efficient
than FLEX.

\begin{figure}[!htbp]
	\centering
	\caption{Homogeneous Covariate Effects}
	\label{fig:mc-hom}
	\includegraphics[width=\textwidth]{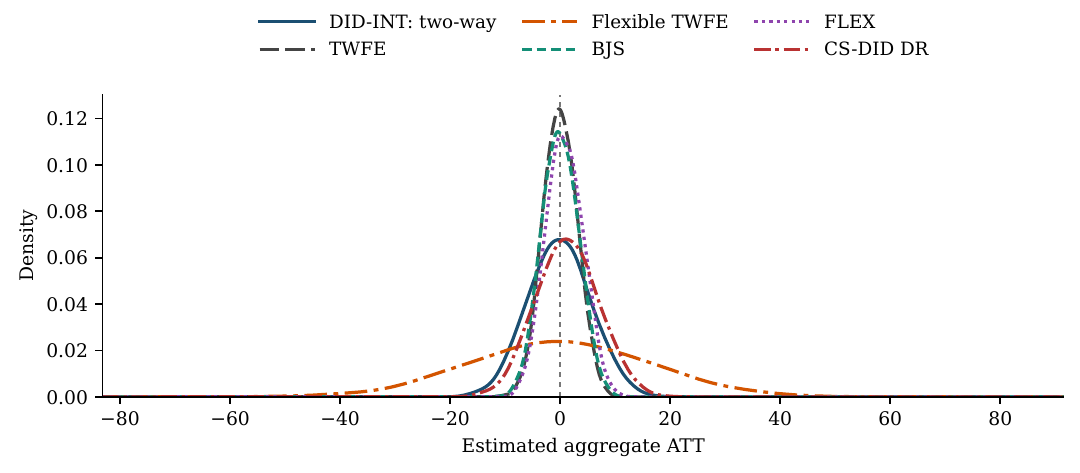}
	\par\smallskip
	{\footnotesize\raggedright
		\textit{Notes:}
		In \eqref{eq:mc-outcome}, $a_{s,t}(c)=\alpha_s+\delta_t$,
		$\beta_{s,t}=\gamma$, and $\tau(x)=0$. The four slopes equal
		their pooled CPS estimates, and observed CPS covariate evolution
		is retained. State-level CPT, IPT, and two-way CCC hold;
		unconditional PT need not hold. The reference line marks zero ATT.
		\par}
\end{figure}

Figure \ref{fig:mc-hom} shows the estimates from the homogeneous coefficients DGP.
In this DGP most of the estimators are unbiased, but FLEX and CS-DID are slightly
biased above zero.  TWFE, BJS, and FLEX are the most efficient, while DID-INT and
CS-DID are less efficient, flexible TWFE is quite inefficient.  

\begin{figure}[!htbp]
	\centering
	\caption{State-Varying Covariate Effects and Composition Shift}
	\label{fig:mc-drift}
	\includegraphics[width=\textwidth]{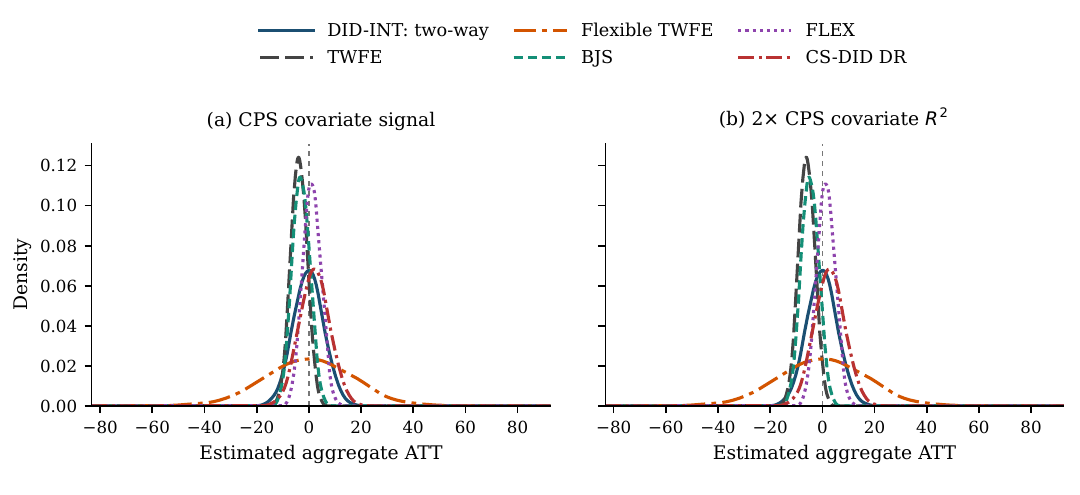}
	\par\smallskip
	{\footnotesize\raggedright
		\textit{Notes:}
		In \eqref{eq:mc-outcome}, $a_{s,t}(c)=\alpha_s+\delta_t$,
		$\beta_{s,t}=\gamma_s$, and $\tau(x)=0$. Share of education=1 increased
		by 10 percentage points for all states in 2006.
		Panel (a) uses $c=1$; panel (b) uses $c=1.582$.
		Intercepts and covariate distributions are unchanged.
		State-level CPT, IPT, and time-invariant CCC hold; unconditional
		PT and region-invariant CCC fail. The reference line marks zero ATT.
		\par}
\end{figure}

Figure \ref{fig:mc-drift} shows the estimates from the state varying coefficients DGP where 
the fraction of education=1 is also increased by 10 percentage points for all states in the 
year 2006 and beyond.  In this design both DID-INT and flexible TWFE are unbiased at either
value of c.  Both BJS and TWFE are negatively biased, while FLEX and CS-DID are positively biased.
The pattern of efficiency is the same as in Figure \ref{fig:mc-hom}.  As c increases, the extent
of bias increases but the efficiency is unchanged.

\begin{figure}[!htbp]
	\centering
	\caption{Two-Way Covariate-Coefficient Variation}
	\label{fig:mc-two-way}
	\includegraphics[width=\textwidth]{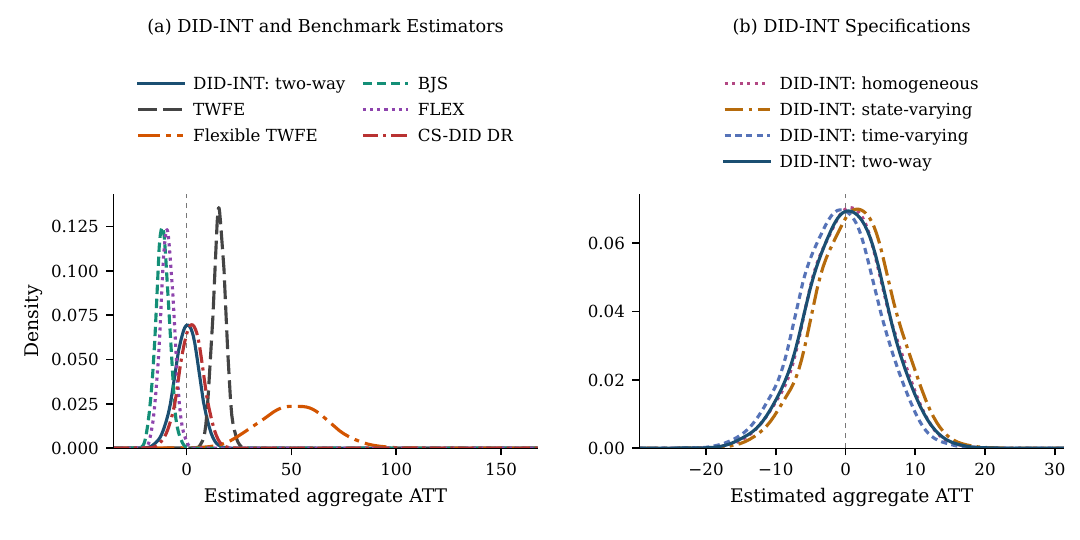}
	\par\smallskip
	{\footnotesize\raggedright
		\textit{Notes:}
		Equation~\eqref{eq:mc-outcome} uses
		$\beta_{s,t}=\beta^{\mathrm{CPS}}_{s,t}$, $\tau(x)=0$,
		and intercepts satisfying~\eqref{eq:mc-two-way}. CPT and
		two-way CCC fail; cell-level IPT holds. Panel (a) compares
		two-way DID-INT with TWFE, flexible TWFE, BJS, FLEX, and
		CS-DID DR; panel (b) compares the four DID-INT
		specifications. The reference line marks zero ATT.
		Horizontal scales differ across panels to display their
		full distributions.
		\par}
\end{figure}

Figure \ref{fig:mc-two-way} shows the estimates from the state--time varying coefficients DGP.
In this figure, panel (a) compares two-way DID-INT with the other estimators as usual.  
In this DGP, CS-DID is slightly biased as 2.17, while the other estimators are more severely
biased.  Specifically, BJS has a bias of -11.46, FLEX -9.19, TWFE of 15.68, and flexible TWFE 
of 51.38.  These biases are amplified when c is increased, as shown in Table \ref{tab:mc-sensitivity}. 
Panel (b) compares the four DID-INT specifications, interestingly while two-way DID-INT is the least biased, 
the homogeneous version of DID-INT is the second least, while time-varying DID-INT has a bias of -1.72, and 
state-varying has a bias of 1.97.

\subsection{Treatment effects vary by education}

The treatment effect heterogeneity DGP allows for variation by education. In this DGP, 
keep the same  $a_{s,t}(1)=\alpha_s+\delta_t$,$\beta_{s,t}=\gamma$, and $c=1$ from the 
homogeneous effects model, but replace $\tau=0$ with 
	\begin{equation}
		\tau(x)=25+25e,
		\qquad
		ATT(s,t)=25+25p_{s,t},
		\qquad
		p_{s,t}=\int e\,dF_{s,t}(x),
		\label{eq:mc-education-tau}
	\end{equation}

where $e$ is the education covariate.  In this DGP, treatment increases earnings
by \$25 for individuals with $E=0$ and by \$50 for observations with $E=1$.
The aggregate ATT is 38.48.  State-level CPT, IPT, and two-way CCC continue 
to hold for untreated outcomes.

\begin{figure}[!htbp]
	\centering
	\caption{Treatment Effects Vary by Education}
	\label{fig:mc-education-tau}
	\includegraphics[width=\textwidth]{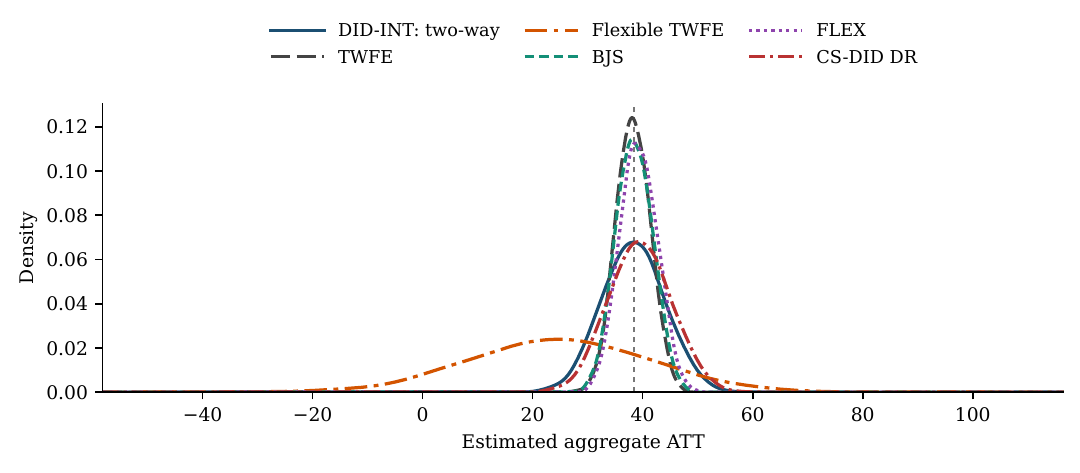}
	\par\smallskip
	{\footnotesize\raggedright
		\textit{Notes:}
		Equation~\eqref{eq:mc-outcome} uses
		$a_{s,t}(1)=\alpha_s+\delta_t$, $\beta_{s,t}=\gamma$,
		$c=1$, and $\tau(x)=25+25e$. State-level CPT, IPT, and
		two-way CCC hold for untreated outcomes. The reference line marks aggregate ATT
		38.47.
		\par}
\end{figure}

Figure \ref{fig:mc-education-tau} shows the estimates from this DGP.  In this figure, BJS, DID-INT, and
TWFE are unbiased, while CS-DID and FLEX are slightly biased at 1.01 and .78 respectively. Flexible TWFE
is quite biased at -13.58. The normal efficiency pattern holds.

\FloatBarrier
\subsection{Changing education returns and aggregation}

The final DGP highlights what happens if the CCC assumption is violated in a particular
way.  Specifically, if one group experiences a change in a coefficient over time.  For
this DGP, we increase the returns to education over time in New York, the control state, 
but not in the treated states.  To make the effect more dramatic, we also increase the share
of college educated individuals to 65\% in New York.  

Unlike the previous designs which use the state--year specific covariate distributions we 
use the state--specific distributions from the year 2000 for all years.  All other slopes
are estimated from the state-specific year 2000 values, except New York's education slope.
We adjust the $a_{s,t}(c)$ so that the integrated trend discrepancies cancel in the aggregate ATT.
Without this offset, the increase in returns to education in New York would generate systematic
differences in untreated trends. One can think of this as shifting untreated earnings growth 
toward more-educated workers within New York, while offsetting the resulting average trend 
difference for the treated population in aggregate.

This design therefore imposes
identification of the aggregate ATT, rather than state-level
IPT or identification of every dynamic effect. Its frozen
populations and compensating intercept trend make it more
stylized than the CPS-composition benchmarks.

\begin{figure}[!htbp]
	\centering
	\caption{New York Education Premium with Higher New York Education Share}
	\label{fig:mc-ny65}
	\includegraphics[width=\textwidth]{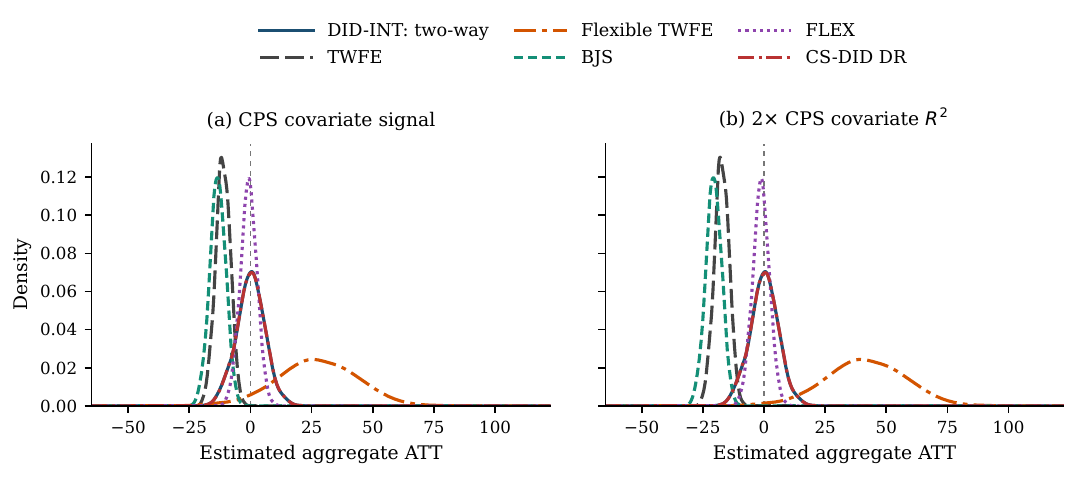}
	\par\smallskip
	{\footnotesize\raggedright
		\textit{Notes:}
        New York's education share is 0.65 and its return to education rises over time. Panel (a) uses the CPS covariate level $R^2$; panel (b) doubles within-cell $R^2$. Integrated discrepancies cancel only in aggregate, not state by state. Both panels use 2,500 replications; the reference line marks zero ATT.
		\par}
\end{figure}

DID-INT and CS-DID are unbiased at both signal
strengths (Figure~\ref{fig:mc-ny65}). Increasing $c$ changes
TWFE's bias from $-11.27$ to $-17.28$, BJS's from
$-13.38$ to $-20.54$, and flexible TWFE's from
27.15 to 41.31. FLEX has bias $-1.19$ in the higher $R^2$ case
but is more efficient than DID-INT. Aggregate identification does not imply a general
efficiency advantage. Appendix Figures~\ref{fig:mc-nj-trends}
and~\ref{fig:mc-nj-density} distinguish PT, CPT, and IPT in
the two-state subset without cross-state cancellation.

\subsection{New York and New Jersey - trends}

In the preceding example the intercept terms in the DGP were offset so that the IPT 
assumption holds for aggregate ATT.  In this DGP we consider a two state version in 
which the intercepts are offset so that IPT holds for each post treatment year.  Specifically, 
we restrict the sample to New York and New Jersey, and New Jersey is again treated 
in 2004.  There are 202,714 observations in this sample.  Again the education=1 share is 
increased to 65\% in New York, and the returns to education grow over time in New York, 
but not in New Jersey.  

\begin{figure}[!htbp]
	\centering
	\caption{Parallel Trends, Conditional Parallel Trends, and Integrated Parallel Trends}
	\label{fig:mc-nj-trends}
	\includegraphics[width=\textwidth]{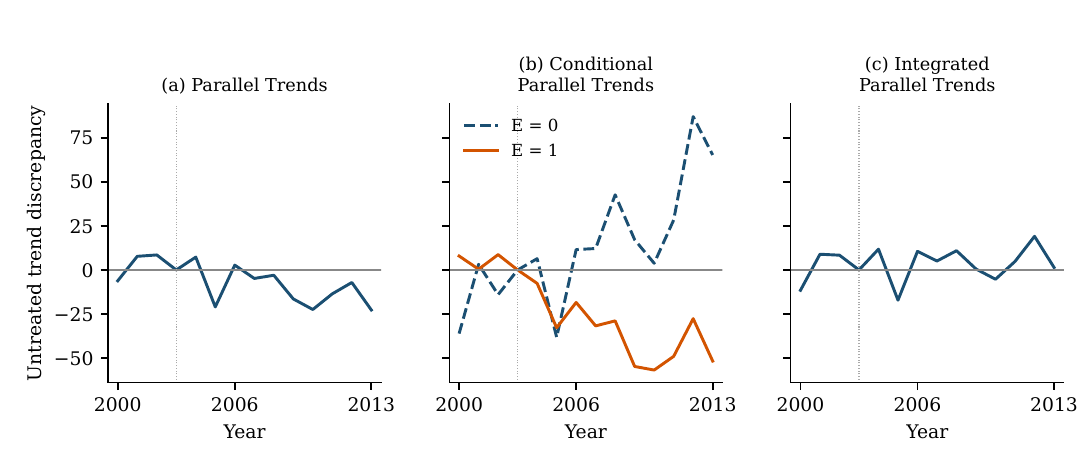}
	\par\smallskip
	{\footnotesize\raggedright\textit{Notes:} One DGP realization, shared across panels, from the NY and NJ DGP. Each curve is the New Jersey minus New York change relative to 2003. Panel (a) uses sample means under each state's own frozen composition. Panel (b) uses unrestricted state--year conditional-mean regressions at $E=0,1$ for a while, married, 38 year old. Panel (c) integrates those same four fitted means over New Jersey's baseline distribution, retaining all estimated slope contrasts. Population IPT holds year by year; the estimated discrepancies need not be zero. Post-adoption untreated outcomes are available only because this is a simulation. The draw is replication 1 of the same 2,500-replication density sample.\par}
\end{figure}

Figure \ref{fig:mc-nj-trends} compares parallel trends type figures for parallel trends,
 conditional parallel, and integrated parallel trends. It does so by subtracting the New
 York value from the New Jersey one.  When the relevant assumption holds, these difference
 curve should be zero.  In the left of the figure we can 
 see that parallel trends fail to hold here as the wages in New York have increased on
 average with the increasing returns to education.  The middle of the figure highlights 
 an important feature of conditional parallel trends, where we assume that parallel trends
 hold for all values of the covariates, not just for the residualized outcome. In this 
 panel we plot the outcomes conditional on education=1 and education=0 for both states.
 Recall in the DGP that the increasing return to education reflects a redistribution of 
 wages away from non-college educated workers towards college educated ones.  As a result, 
 the non-college educated individuals in New Jersey are comparatively better off than their
 counterparts in New York. Conversely, the college educated workers in New Jersey are worse
 off compared to those in New York.  As a result, conditional parallel trends fails for both
 education groups. Finally, the rightmost figure plots the integrated parallel trends.  By design, 
 this holds in this DGP.  Appendix \ref{app:mc-two-state} analyses the ATT estimates for this DGP.
In this DGP DID-INT, CS-DID, and FLEX are all unbiased, while BJS, TWFE, and flexible TWFE are 
all quite biased.

\FloatBarrier

\input{main_table_CPS.tex}
\FloatBarrier
\input{main_table_R2.tex}

\FloatBarrier

\subsection{Cluster Robust Inference} \label{sec:crve}
 
We evaluate the reliability of two different cluster robust procedures for inference on ATT estimation using DID-INT.  We use
a similar design from \citet*{karim2026improvedinferencecsdidusing}  and \citet*{bertrand2004much}.  Specifically, we use the Merged Outgoing Rotation Group (MORG)
from the 1979-1999 years in the CPS.  As in \citet*{karim2024difference}, we consider the log of women's weekly earnings as the outcome 
variable.  We exclude observations with weekly earnings below \$20.  We include the control variables of age, age-squared, and education 
categorical variables.  The total sample, from which we subsample, contains 547,818 observations.

Rather than a fixed set of states and a fixed number of treated states as in \citet*{bertrand2004much}, we vary both.  Specifically, we consider 
different numbers of total states $S$, and we vary $S\in\{8,16,24,32,36,40\}$. For each value of $S$, we also choose the number 
or early adopting states $J$ and late adopting state $L$.  For convenience we just set $J=L$.  We also draw $T=8$ consecutive years from the 1979-1999 panel.  We vary $J=L\in\{1,\ldots,S/2-1\}$, with the early adopters treated in period 5 and the late adopters treated in period 7.  We use 50,000 replications across each of these $S,J,L$ designs.  

\citet*{karim2026improvedinferencecsdidusing} provides more details on the specific clustering procedures.  In these experiments we consider both the cluster jackknife and randomization inference.  
The cluster jackknife deletes one jurisdiction at a time and estimates variance from the dispersion of the leave-one-state-out estimates:
\begin{equation}
	\label{eq:jk_var}
	\widehat{\mathrm{Var}}_{\mathrm{JK}}(\widehat{ATT})
	=
	\frac{S-1}{S}\sum_{s=1}^{S}\left(\widehat{ATT}^{(s)}-\bar{ATT}^{(\cdot)}\right)^2,
	\qquad
	\bar{ATT}^{(\cdot)}=\frac{1}{S}\sum_{s=1}^{S}\widehat{ATT}^{(s)}.
\end{equation}

\citet*{karim2026improvedinferencecsdidusing} proposes a similar variance estimator for inference with CSDID.  Randomization inference is done analogously to the RI-$\beta$ procedure
discusssed in \cite{MW-RI}, which is based on un-studentized estimates.  Across re-randomizations we hold the values of $J$ and $L$ fixed, along with the respective timings.  The re-randomizations can be thought of as permutations of the state specific labels as control, late treated, or early treated.  Let $M$ be the total number of possible permutations.  When $M<99$ we enumerate all of the permutations and use the approach in \cite{racine2007inference} which results in a kernel-smoothed $p$-value.  When $99 \leq M < 999$,  we enumerate all the permutations and compute the $p$-value as

\[
p=\frac{1+\sum_{m=1}^{M}
	\mathbf{1}\{|\widehat{ATT}^{(m)}|\geq|\widehat{ATT}^{obs}|\}}
{1+M}.
\]

If $M \geq 999$, we randomly pick 999 permutations and use the same formula.  Both the jackknife and randomization inference procedures are assessed at the five-percent level.

\begin{center}
	\begin{minipage}{\textwidth}
		\centering
		\input{final_didint_att_rejection}
		\captionsetup{skip=2pt}
		\captionof{figure}{Finite-Sample Size of DID-INT Inference Procedures.}
		\label{fig:didint_att_rejection}
	\end{minipage}
\end{center}

Figure \ref{fig:didint_att_rejection} reports the rejection frequencies.  The jackknife follows a familiar U-shape in the value of $J=L$.  This
was first reported in the common adoption literature by \citet*{MW-JAE}.  The same pattern exists for staggered adoption in \citet*{karim2026improvedinferencecsdidusing} and \citet*{karim2024difference}.  When each treatment-timing group contains only one treated state, rejection frequencies range from 8.2 to 16.1 percent across the displayed values of $S$; with two treated states per group, they range from 4.4 to 10.0 percent.  The rejection frequencies generally fall as additional treated states are added until half are treated, and then begin to rise again.  The jackknife thus requires both a non-small number of treated and control states.

Randomization inference seems quite reliable across all values of $S$ and $J,L$, with rejection frequencies between 4.3 and 5.2 percent. Inference appears to be reliable with both few treated states and few control states. These simulations average across both large treated and small treated groups.  \cite{MW-RI} find that the un-studentized version of RI can over-reject when small groups are treated, and under-reject when 
large ones are treated.  We suspect this also applies in the staggered adoption setting.  Additionally, randomization inference tests a sharp null of no treatment effects, while the jackknife tests the conventional null that the att is zero. \citet*{karim2026improvedinferencecsdidusing} considers inference for disaggregate ATTs.  

\section{Empirical Example} \label{sec:empex}

As a brief empirical example we consider the impact of the legalization of medical marijuana on BMI, as considered in \citet*{sabia2017effect}.  We use data from the Behavioral Risk Factor Surveillance System (BRFSS).  There are 10 always treated states which are removed from the analysis, which leaves 41 states in the sample.  The District of Columbia is counted as a state and is treated.  There are seven states which legalize medical marijuana in the period from 2004 to 2011.  The adoption schedule is: RI in 2006; NM in 2007; MI in 2008; AZ, DC, and NJ in 2010, and DE in 2011.  

The outcomes of interest are body mass index (BMI), and severe obesity.  We restrict the sample to individuals 18 and above, and are left with a sample of 2,380,588 individuals.  We include  categorical control variables for sex, age, and race.  For both outcome variables we estimate ATTs for  the four variants of DID-INT and both the regession adjustment (RA) and inverse-probability weighting (IPW) versions of CSDID.  

\begin{table}[!ht]
	\centering
	\small
	\setstretch{1}
	\setlength{\tabcolsep}{4pt}
	\renewcommand{\arraystretch}{1.12}
	\caption{Fitted DID-INT and CSDID: corrected treatment timing}
	\label{tab:medical_fitted_didint}
	\begin{tabular}{@{}lrrr@{\hspace{14pt}}rrr@{}}
		\hline
		& \multicolumn{3}{c}{BMI} & \multicolumn{3}{c}{Severe obesity} \\
		\cline{2-4}\cline{5-7}
		Estimator & ATT & $p_{\mathrm{RIF}}$ & $p_{\mathrm{JK}}$
		& ATT & $p_{\mathrm{RIF}}$ & $p_{\mathrm{JK}}$ \\
		\hline
		DID-INT: homogeneous & $-0.02848$ & --- & 0.4698 & $-0.001074$ & --- & 0.3327 \\
		DID-INT: state-varying       & $-0.03493$ & --- & 0.3286 & $-0.001638$ & --- & 0.0957 \\
		DID-INT: year-varying        & $-0.03479$ & --- & 0.4272 & $-0.001288$ & --- & 0.3375 \\
		DID-INT: two-way & $-0.03326$ & --- & 0.3082 & $-0.001004$ & --- & 0.4050 \\
		CSDID-RA             & $-0.08143$ & 0.0184 & 0.0610 & $-0.004484$ & 0.0020 & 0.0134 \\
		CSDID-IPW            & $-0.06634$ & 0.0783 & 0.1617 & $-0.002675$ & 0.0866 & 0.1612 \\
		\hline
	\end{tabular}
\end{table}

Table \ref{tab:medical_fitted_didint} presents these ATT estimates along with jackknife $p$-values for both CS-DID and DID-INT
estimates, and native recentered influence function (RIF) $p$-values for CS-DID.  For BMI the CS-DID estimate is -0.08143 for RA, and 
significant at the 5\% level using RIF $p$-values and the 10\% using jackknife ones.  The IPW estimate is -0.06634, which is significant at
the 10\% level using RIF $p$-values and not significant at the 10\% using jackknife ones.  Conversely, the DID-INT estimates range from 
$-0.035$ to $-0.028$ and all the $p$-values exceed 0.30. 

For severe obesity, there is a broadly similar pattern where the CS-DID estimates are larger and more significant.  Specifically, the RA estimate is 
-0.004484 and is significant at the 1\% level using RIF and 5\% level using the jackknife.  The IPW estimate is -0.002675 and is significant at the 10\% level using RIF and not at the 10\% level using the jackknife.  The DID-INT estimates for two-way, homogeneous, and year range from -0.001004 to -0.001288 and all the $p$-values exceed 0.33.  Notably, the state varying DID-INT estimate is more extreme at -0.001638 and is significant at the 10\% level.

\section{Conclusion}
\label{sec:conclusion}

Difference-in-differences (DID) is widely used to estimate treatment effects for policies implemented at the jurisdictional level. However, existing DID methods require careful covariate selection to recover a consistent estimate of the ATT. The literature recommends using either time-invariant covariates or pre-treatment covariates when the covariates change with time. Nonetheless, researchers may still want to include time-varying covariates in DID analysis, even though they are not necessary for parallel trends to be plausible. We begin by introducing a new assumption called the common causality of covariates (CCC) assumption, which has been implied in the existing literature. In particular, we introduce three types of CCC assumptions: the region-invariant CCC, the time-invariant CCC, and the two-way CCC. The region-invariant CCC assumes that the effects of the covariates are the same between regions, while the time-invariant CCC assumes that these effects remain stable across time. The two-way CCC combines both, implying that covariate effects remain constant across regions and time. When the two-way CCC holds, both region-invariant CCC and time-invariant CCC hold as well. In this paper, we highlight issues that may arise when researchers use conventional DID methods, such as TWFE, or more modern methods, like CS-DID, to identify the ATT under CCC violations. We also contribute to the literature by providing researchers with a tool to obtain a consistent estimate of the ATT under CCC violations, called the Integrated Difference-in-Differences (DID-INT).

We consider four versions of DID-INT depending on the assumptions we make on the covariates. The homogeneous DID-INT imposes a common covariate coefficient across regions and periods. The region-varying DID-INT accounts for region-invariant CCC violations by interacting time-varying covariates with region dummies. Conversely, the time-varying DID-INT accounts for time-invariant CCC violations by interacting covariates with time dummies. Finally, the two-way DID-INT adjusts for two-way CCC violations by interacting the covariates with both region and time dummies. DID-INT relies on integrated parallel trends, which requires the untreated outcome trends to be parallel after integrating the conditional trends over the covariate distribution of the treated region. Integrated parallel trends can therefore hold even when conditional parallel trends is violated.

To illustrate the problems with both conventional and modern DID methods, we first derive the necessary and sufficient conditions for conditional parallel trends. The derivation shows that conditional parallel trends can still hold even when the two-way CCC assumption is violated, provided the covariate coefficients satisfy a strict restriction across regions and time periods. We then propose integrated parallel trends, which does not require untreated outcome trends to be parallel at each covariate level. Rather, deviations in untreated trends across covariate values can cancel out after integrating over the covariate distribution in the treated region. We show that standard common-coefficient adjustments can result in misleading pre-trend estimates and lead to misspecification bias when using the TWFE approach. We also show that the ability of the CS-DID to identify the ATT under CCC violations and under integrated parallel trends depends on implementation. In the case of repeated cross-sectional data, the outcome-regression approach further requires that the distribution of the covariates within each region remains constant over time. DID-INT can identify the ATT under integrated parallel trends while allowing CCC violations, treatment-effect heterogeneity with respect to covariates, and changes in covariate composition. It is also robust to forbidden comparisons and to negative weighting in staggered treatment-adoption situations.

Specifically, in the data-generating processes involving CCC violations that we examine in our simulation, CS-DID, the imputation estimator, and the FLEX estimator are all biased. Moreover, we consider data-generating processes in which the covariate distribution changes over time or treatment effects depend on covariates. If the covariate distribution changes over time and covariate effects vary by region, DID-INT remains unbiased, while TWFE, the imputation estimator, FLEX, and CS-DID are biased. However, when treatment effects depend on covariates, DID-INT, TWFE, and the imputation estimator remain unbiased, whereas flexible TWFE is biased and CS-DID and FLEX show small biases.

Overall, our results show that the role of covariates in DID is more complicated than simply deciding which ones to include; \textit{how} covariates affect untreated potential outcomes across regions and time periods can also have important implications for identification. DID-INT offers an alternative in these situations because it lets covariate effects vary across regions and periods and relies on a weaker parallel trends assumption called integrated parallel trends.

\bibliography{mybib.bib}

\clearpage

\appendix 

\setcounter{section}{0} 
\setcounter{equation}{0}  
\setcounter{assumption}{0}  

\appendix 
\renewcommand{\theHequation}{appendix.\arabic{equation}}

\setcounter{section}{0} 

\section{Proof of the necessary and sufficient condition for strong parallel trends without covariates}
\label{appendix: ptproof}


The parallel trends assumption can be written as:
\begin{equation}
\label{eq: appendixAspt}
    \mathbb{E}[Y_{2}(0) \mid S_i = A] - \mathbb{E}[Y_{1}(0) \mid S_i = A] = \mathbb{E}[Y_{2}(0)\mid S_i = B] - \mathbb{E}[Y_{1}(0) \mid S_i = B].
\end{equation}  

In the absence of covariates, the coefficient on the covariate is zero. Therefore, when $\gamma^{\circ}_{s,t}=0$, Equation \eqref{eq: posimplified} can be written as: 
    \begin{equation}
        Y_{i,s,t}(0) = \sum_{\ell=1}^{S}
        \sum_{q=1}^{T} \theta^{\circ}_{\ell,q} Z_{i,\ell,q} + \epsilon_{i,s,t}.
    \end{equation}
Or equivalently,
    \begin{equation}
        Y_{i,s,t}(0) = \theta^{\circ}_{s,t} + \epsilon_{i,s,t}.
    \end{equation}

Assuming that $\mathbb{E}[\epsilon_{i,s,t}|S_i = s] = 0$ within each region-period cell, we can write the expected value of the untreated potential outcome in region $s$ in period $t$ as:
    \begin{equation}
    \label{eq: appendixApotentialoutcome}
        \mathbb{E}[Y_{t}(0)|S_i = s] = \theta^{\circ}_{s,t}.
    \end{equation}
The change in the expected value of the untreated potential outcome for region $A$ between periods $2$ and $1$ can be written as:
    \begin{equation}
    \label{eq: appendixAdiffA}
        \mathbb{E}[Y_{2}(0)|S_i = A] - \mathbb{E}[Y_{1}(0)|S_i = A] = \theta^{\circ}_{A,2} - \theta^{\circ}_{A,1}.
    \end{equation}
Similarly, the change in the expected value of the untreated potential outcome for region $B$ between periods $2$ and $1$ can be written as:
    \begin{equation}
    \label{eq: appendixAdiffB}
        \mathbb{E}[Y_{2}(0)|S_i = B] - \mathbb{E}[Y_{1}(0)|S_i = B] = \theta^{\circ}_{B,2} - \theta^{\circ}_{B,1}.
    \end{equation}
Plugging in Equations \eqref{eq: appendixAdiffA} and \eqref{eq: appendixAdiffB} into Equation \eqref{eq: appendixAspt}, we get:
    \begin{equation}
    \label{eq: necessaryconditionspt}
        \theta^{\circ}_{A,2} - \theta^{\circ}_{A,1} = \theta^{\circ}_{B,2} - \theta^{\circ}_{B,1}.
    \end{equation}
This shows that Equation \eqref{eq: necessaryconditionspt} is a necessary condition for parallel trends.
Now, assume that the equality shown in Equation \eqref{eq: necessaryconditionspt} holds. Using Equation \eqref{eq: appendixApotentialoutcome}, we can also show that Equation \eqref{eq: appendixAspt} holds as well. Therefore, Equation \eqref{eq: necessaryconditionspt} is also a sufficient condition for strong parallel trends.
\begin{equation}
    \mbox{Strong Parallel Trends} \Longleftrightarrow         \theta^{\circ}_{A,2} - \theta^{\circ}_{A,1} = \theta^{\circ}_{B,2} - \theta^{\circ}_{B,1}.
\end{equation}


\section{Proof of the Necessary and Sufficient Condition for Conditional Parallel Trends}
\label{appendix: cptproof}

The conditional parallel trends assumption can be written as:
\begin{equation}
\begin{aligned}
\label{eq: appendixBcptnonresidualized}
    & \mathbb{E}[Y_{2}(0) \mid S_i = A, X_{2} = x] - \mathbb{E}[Y_{1}(0) \mid S_i = A, X_{1} = x]
    \\ = & \mathbb{E}[Y_{2}(0) \mid S_i = B, X_{2} = x] - \mathbb{E}[Y_{1}(0) \mid S_i = B, X_{1} = x] \qquad \forall x.    
\end{aligned}
\end{equation}
The true potential outcome model shown in Equation \eqref{eq: posimplified} can be written as: 
    \begin{equation}
    \label{eq: fauxcpt3}
        Y_{i,s,t}(0) = \sum_{\ell=1}^{S}
        \sum_{q=1}^{T} \theta^{\circ}_{\ell,q} Z_{i,\ell,q} + \gamma^{\circ}_{s,t} X_{i,s,t} +  \epsilon_{i,s,t}
    \end{equation}
Or equivalently,
    \begin{equation}
        Y_{i,s,t}(0) = \theta^{\circ}_{s,t} + \gamma^{\circ}_{s,t} X_{i,s,t} +  \epsilon_{i,s,t}.
    \end{equation}
Let us now analyze conditional parallel trends at a specific value of $X_{i,s,t}= x$. Assuming that $\mathbb{E}[\epsilon_{i,s,t}|S_i = s, X_{t}= x] = 0$ within each region-period cell, we can write the expected value of the untreated potential outcome in region $s$ in period $t$ as:
    \begin{equation}
    \label{eq: appendixBpotentialoutcome}
        \mathbb{E}[Y_{t}(0)|S_i = s, X_{t} = x] = \theta^{\circ}_{s,t}  + \gamma^{\circ}_{s,t} x.
    \end{equation}
The change in the expected value of the untreated potential outcome for region $A$ between periods $2$ and $1$ can be written as:
    \begin{equation}
    \label{eq: appendixBdiffA}
        \mathbb{E}[Y_{2}(0)|S_i = A, X_{2} = x] - \mathbb{E}[Y_{1}(0)|S_i = A, X_{1} = x] = \theta^{\circ}_{A,2} - \theta^{\circ}_{A,1} + (\gamma^{\circ}_{A,2} - \gamma^{\circ}_{A,1})x.
    \end{equation}
Similarly, the change in the expected value of the untreated potential outcome for region $B$ between periods $2$ and $1$ can be written as:
    \begin{equation}
    \label{eq: appendixBdiffB}
        \mathbb{E}[Y_{2}(0)|S_i = B, X_{2} = x] - \mathbb{E}[Y_{1}(0)|S_i = B, X_{1} = x] = \theta^{\circ}_{B,2} - \theta^{\circ}_{B,1} + (\gamma^{\circ}_{B,2} - \gamma^{\circ}_{B,1})x.
    \end{equation}
Plugging in Equations \eqref{eq: appendixBdiffA} and \eqref{eq: appendixBdiffB} into Equation \eqref{eq: appendixBcptnonresidualized}, we get:
    \begin{equation}
    \label{eq: necessaryconditioncpt}
        \theta^{\circ}_{A,2} - \theta^{\circ}_{A,1} +(\gamma^{\circ}_{A,2} - \gamma^{\circ}_{A,1})x = \theta^{\circ}_{B,2} - \theta^{\circ}_{B,1} +(\gamma^{\circ}_{B,2} - \gamma^{\circ}_{B,1})x.
    \end{equation}

This shows that Equation \eqref{eq: necessaryconditioncpt} is a necessary condition for conditional parallel trends for a given value of $X_{i,s,t} = x$.  Since conditional parallel trends needs to hold for every value of $x$, let us assume that the condition shown in Equation \eqref{eq: necessaryconditioncpt} holds for two distinct values of $X_{i,s,t}$, $x_1$ and $x_2$. Re-arranging Equation \eqref{eq: necessaryconditioncpt}, we get
    \begin{equation}
    \label{eq: necessaryconditioncptrearranged}
        \bigr(\theta^{\circ}_{A,2} - \theta^{\circ}_{A,1}\bigr) - \bigr(\theta^{\circ}_{B,2} - \theta^{\circ}_{B,1} \bigr) +\biggr[(\gamma^{\circ}_{A,2} - \gamma^{\circ}_{A,1}) -(\gamma^{\circ}_{B,2} - \gamma^{\circ}_{B,1})\biggr]x = 0.
    \end{equation}
Evaluating Equation \eqref{eq: necessaryconditioncptrearranged} at two distinct values, $x_1$ and $x_2$, we get
\begin{equation}
\label{eq: equalityofx1}
\begin{aligned}
&\bigl(\theta^{\circ}_{A,2} - \theta^{\circ}_{A,1}\bigr)
- \bigl(\theta^{\circ}_{B,2} - \theta^{\circ}_{B,1}\bigr)+ \biggl[(\gamma^{\circ}_{A,2} - \gamma^{\circ}_{A,1})
- (\gamma^{\circ}_{B,2} - \gamma^{\circ}_{B,1})\biggr]x_1
=0.
\end{aligned}
\end{equation}

\begin{equation}
\label{eq: equalityofx2}
\begin{aligned}
&\bigl(\theta^{\circ}_{A,2} - \theta^{\circ}_{A,1}\bigr)
- \bigl(\theta^{\circ}_{B,2} - \theta^{\circ}_{B,1}\bigr)+ \biggl[(\gamma^{\circ}_{A,2} - \gamma^{\circ}_{A,1})
- (\gamma^{\circ}_{B,2} - \gamma^{\circ}_{B,1})\biggr]x_2
=0.
\end{aligned}
\end{equation}
Taking a difference of Equations \eqref{eq: equalityofx1} and \eqref{eq: equalityofx2}, we get:
\begin{equation}
\bigr((\gamma^{\circ}_{A,2} - \gamma^{\circ}_{A,1})
- (\gamma^{\circ}_{B,2} - \gamma^{\circ}_{B,1})\bigr)(x_1 - x_2) = 0.
\end{equation}
Since $x_1 \neq x_2$, we get:
\begin{equation}
\label{equation: condition2proved}
\gamma^{\circ}_{A,2} - \gamma^{\circ}_{A,1}
= \gamma^{\circ}_{B,2} - \gamma^{\circ}_{B,1}.
\end{equation}
Substituting Equation \eqref{equation: condition2proved} into either Equation \eqref{eq: equalityofx1} or \eqref{eq: equalityofx2}, we get:
   \begin{align}
    \begin{split}
        \theta^{\circ}_{A,2} - \theta^{\circ}_{A,1} = \theta^{\circ}_{B,2} - \theta^{\circ}_{B,1}.
    \end{split}
    \end{align}
Therefore, in order for conditional parallel trends to hold, the following two conditions are jointly necessary:
   \begin{align}
    \begin{split}
    \label{equation: necessaryconditioncpt1app}
        \mbox{\textbf{Condition 1:}} \; \theta^{\circ}_{A,2} - \theta^{\circ}_{A,1} = \theta^{\circ}_{B,2} - \theta^{\circ}_{B,1}.
    \end{split}
    \end{align}
    \begin{align}
    \begin{split}
    \label{equation: necessaryconditioncpt2app}    
        \mbox{\textbf{Condition 2:}} \; \gamma^{\circ}_{A,2} - \gamma^{\circ}_{A,1} = \gamma^{\circ}_{B,2} - \gamma^{\circ}_{B,1}.
    \end{split}
    \end{align}
We now show that these two conditions are also jointly sufficient. Multiplying both sides of Condition 2 by an arbitrary value of $x$ and rearranging, we get
\begin{equation}
\label{eq: derivationsufficiencycpt}
    \biggr[(\gamma^{\circ}_{A,2} - \gamma^{\circ}_{A,1}) -(\gamma^{\circ}_{B,2} - \gamma^{\circ}_{B,1})\biggr]x = 0.
\end{equation}
Condition 1 can be equivalently written as $\bigr(\theta^{\circ}_{A,2} - \theta^{\circ}_{A,1}\bigr) - \bigr(\theta^{\circ}_{B,2} - \theta^{\circ}_{B,1} \bigr) = 0$. Since this term equals zero, adding it to the left-hand side of Equation \eqref{eq: derivationsufficiencycpt} preserves the equality. Re-arranging, we get: 
    \begin{equation}
    \label{eq: necessaryconditioncptrearrangedagain}
        \bigr(\theta^{\circ}_{A,2} - \theta^{\circ}_{A,1}\bigr) - \bigr(\theta^{\circ}_{B,2} - \theta^{\circ}_{B,1} \bigr) +\biggr[(\gamma^{\circ}_{A,2} - \gamma^{\circ}_{A,1}) -(\gamma^{\circ}_{B,2} - \gamma^{\circ}_{B,1})\biggr]x = 0.
    \end{equation}
This is precisely the condition required for conditional parallel trends, as shown in Equation \eqref{eq: necessaryconditioncpt}. Since the value of $x$ was arbitrarily chosen, this equality holds for every value of $x$. This proves that both conditions are jointly sufficient for conditional parallel trends. Therefore, 
\begin{equation}
\label{eq: necessaryandsufficientcpt}
\begin{aligned}
\mbox{Conditional Parallel Trends}
\Longleftrightarrow\;&
\left[
\theta^{\circ}_{A,2}-\theta^{\circ}_{A,1}
=
\theta^{\circ}_{B,2}-\theta^{\circ}_{B,1}
\right]
\\
&\land
\left[
\gamma^{\circ}_{A,2}-\gamma^{\circ}_{A,1}
=
\gamma^{\circ}_{B,2}-\gamma^{\circ}_{B,1}
\right].
\end{aligned}
\end{equation}


\section{Proof of Necessary and Sufficient Condition for Integrated Parallel Trends With Covariates}
\label{appendix: iptnscondition}

The integrated parallel trends assumption can be written as:
\begin{gather}
    \int \biggr[\mathbb{E}\bigr(Y_2(0) \mid S_i = A, X_2 = x \bigr) - \mathbb{E}\bigr(Y_1(0) \mid S_i = A, X_1 = x \bigr)\biggr] dF_{X_2 \mid S=A}(x) \nonumber \\ 
    = \int \biggr[\mathbb{E}\bigr(Y_2(0) \mid S_i = B, X_2 = x \bigr) -
    \mathbb{E}\bigr(Y_1(0) \mid S_i = B, X_1 = x \bigr)\biggr] dF_{X_2 \mid S=A}(x) \label{eq: integratedptapp}.
\end{gather}  
Under Assumption \ref{as: linearity}, the true unobserved potential outcome model can be written as:
\begin{gather}
    Y_{i,s,t}(0) = \theta^{\circ}_{s,t} + \gamma^{\circ}_{s,t} X_{i,s,t} + \epsilon_{i,s,t}
\end{gather}
Let us now analyze integrated parallel trends at a specific value of $X_{i,s,t} = x$. Assuming that $\mathbb{E}[\epsilon_{i,s,t} \mid S_i = s, X_t = x] = 0$ within each region-period cell, we can write the expected value of the untreated potential outcome in region $s$ in period $t$ as:
    \begin{equation}
    \label{eq: appendixCpotentialoutcome}
        \mathbb{E}[Y_{t}(0)|S_i = s, X_{t} = x] = \theta^{\circ}_{s,t}  + \gamma^{\circ}_{s,t} x.
    \end{equation}
The change in the expected value of the untreated potential outcome for region $A$ between periods $2$ and $1$ can be written as:
    \begin{equation}
    \label{eq: appendixCdiffA}
        \mathbb{E}[Y_{2}(0)|S_i = A, X_{2} = x] - \mathbb{E}[Y_{1}(0)|S_i = A, X_{1} = x] = \theta^{\circ}_{A,2} - \theta^{\circ}_{A,1} + (\gamma^{\circ}_{A,2} - \gamma^{\circ}_{A,1})x.
    \end{equation}
Integrating this expression over the treated covariate distribution and simplifying yields:
    \begin{gather}
    \int \biggr[\mathbb{E}\bigr(Y_2(0) \mid S_i = A, X_2 = x \bigr) - \mathbb{E}\bigr(Y_1(0) \mid S_i = A, X_1 = x \bigr)\biggr] dF_{X_2 \mid S=A}(x) \nonumber \\ = \int \biggr[(\theta^{\circ}_{A,2} - \theta^{\circ}_{A,1}) + (\gamma^{\circ}_{A,2} - \gamma^{\circ}_{A,1})x \biggr] dF_{X_2 \mid S=A}(x) \nonumber \\ 
    = (\theta^{\circ}_{A,2} - \theta^{\circ}_{A,1}) +  (\gamma^{\circ}_{A,2} - \gamma^{\circ}_{A,1}) \int x dF_{X_2 \mid S=A}(x) \nonumber \\ 
    = (\theta^{\circ}_{A,2} - \theta^{\circ}_{A,1}) +  (\gamma^{\circ}_{A,2} - \gamma^{\circ}_{A,1}) \mathbb{E}[X_2 \mid S_i = A] \label{eq: appendixCintA}.
    \end{gather} 
Similarly, the change in the expected value of the untreated potential outcome for region $B$ between periods $2$ and $1$ can be written as:
    \begin{equation}
    \label{eq: appendixCdiffB}
        \mathbb{E}[Y_{2}(0)|S_i = B, X_{2} = x] - \mathbb{E}[Y_{1}(0)|S_i = B, X_{1} = x] = \theta^{\circ}_{B,2} - \theta^{\circ}_{B,1} + (\gamma^{\circ}_{B,2} - \gamma^{\circ}_{B,1})x.
    \end{equation}
Integrating this expression over the treated covariate distribution and simplifying yields:
    \begin{gather}
    \int \biggr[\mathbb{E}\bigr(Y_2(0) \mid S_i = B, X_2 = x \bigr) - \mathbb{E}\bigr(Y_1(0) \mid S_i = B, X_1 = x \bigr)\biggr] dF_{X_2 \mid S=A}(x) \nonumber \\ = \int \biggr[(\theta^{\circ}_{B,2} - \theta^{\circ}_{B,1}) + (\gamma^{\circ}_{B,2} - \gamma^{\circ}_{B,1})x \biggr] dF_{X_2 \mid S=A}(x) \nonumber \\ 
    = (\theta^{\circ}_{B,2} - \theta^{\circ}_{B,1}) +  (\gamma^{\circ}_{B,2} - \gamma^{\circ}_{B,1}) \int x dF_{X_2 \mid S=A}(x) \nonumber \\ 
    = (\theta^{\circ}_{B,2} - \theta^{\circ}_{B,1}) +  (\gamma^{\circ}_{B,2} - \gamma^{\circ}_{B,1}) \mathbb{E}[X_2 \mid S_i = A] \label{eq: appendixCintB}.
    \end{gather}
Plugging in Equations \eqref{eq: appendixCintA} and \eqref{eq: appendixCintB} into Equation \eqref{eq: integratedptapp}, we get:
    \begin{gather}
        \text{Integrated Parallel Trends} \Longrightarrow \bigr[ (\theta^{\circ}_{A,2} - \theta^{\circ}_{A,1}) - (\theta^{\circ}_{B,2} - \theta^{\circ}_{B,1}) \bigr] \nonumber \\ + \bigr[(\gamma^{\circ}_{A,2} - \gamma^{\circ}_{A,1}) -(\gamma^{\circ}_{B,2} - \gamma^{\circ}_{B,1})\bigr] \mathbb{E}[X_2 \mid S_i = A] = 0. \label{eq: necessaryconditionipt}
    \end{gather}
This shows that Equation \eqref{eq: necessaryconditionipt} is a necessary condition for integrated parallel trends. Next, we show that this condition is also sufficient. Assume that the equality shown in Equation \eqref{eq: necessaryconditionipt} holds. We expand the expression as follows:
\begin{gather}
    \bigr[ (\theta^{\circ}_{A,2} - \theta^{\circ}_{A,1}) - (\theta^{\circ}_{B,2} - \theta^{\circ}_{B,1}) \bigr] \nonumber + \bigr[(\gamma^{\circ}_{A,2} - \gamma^{\circ}_{A,1}) -(\gamma^{\circ}_{B,2} - \gamma^{\circ}_{B,1})\bigr] \mathbb{E}[X_2 \mid S_i = A] \\ 
    =  \bigr[ (\theta^{\circ}_{A,2} - \theta^{\circ}_{A,1}) - (\theta^{\circ}_{B,2} - \theta^{\circ}_{B,1}) \bigr] +    \bigr[(\gamma^{\circ}_{A,2} - \gamma^{\circ}_{A,1}) -(\gamma^{\circ}_{B,2} - \gamma^{\circ}_{B,1})\bigr] \int x dF_{X_2 \mid S=A}(x) \nonumber \\
    = \int \biggr[(\theta^{\circ}_{A,2} - \theta^{\circ}_{A,1}) + (\gamma^{\circ}_{A,2} - \gamma^{\circ}_{A,1})x \biggr] dF_{X_2 \mid S=A}(x) 
    - \int \biggr[(\theta^{\circ}_{B,2} - \theta^{\circ}_{B,1}) + (\gamma^{\circ}_{B,2} - \gamma^{\circ}_{B,1})x \biggr] dF_{X_2 \mid S=A}(x) = 0. \label{eq: sufficientconditionintermediate}
\end{gather}
Under Assumption \ref{as: linearity}, the expression in Equation \eqref{eq: sufficientconditionintermediate} can be expressed as:
\begin{gather}
    \int \biggr[\mathbb{E}\bigr(Y_2(0) \mid S_i = A, X_2 = x \bigr) - \mathbb{E}\bigr(Y_1(0) \mid S_i = A, X_1 = x \bigr)\biggr] dF_{X_2 \mid S=A}(x) \nonumber \\ 
    = \int \biggr[\mathbb{E}\bigr(Y_2(0) \mid S_i = B, X_2 = x \bigr) -
    \mathbb{E}\bigr(Y_1(0) \mid S_i = B, X_1 = x \bigr)\biggr] dF_{X_2 \mid S=A}(x) \label{eq: integratedptapp2}.
\end{gather}
Therefore,
    \begin{gather}
        \text{Integrated Parallel Trends} \Longleftrightarrow \bigr[ (\theta^{\circ}_{A,2} - \theta^{\circ}_{A,1}) - (\theta^{\circ}_{B,2} - \theta^{\circ}_{B,1}) \bigr] \nonumber \\ + \bigr[(\gamma^{\circ}_{A,2} - \gamma^{\circ}_{A,1}) -(\gamma^{\circ}_{B,2} - \gamma^{\circ}_{B,1})\bigr] \mathbb{E}[X_2 \mid S_i = A] = 0. \label{eq: nandsconditionipt}
    \end{gather}


\section{Proof of ATT Under Conditional Parallel Trends and No Anticipation}
\label{appendix: ATTcondcpt}

In this section, we show that under the conditional parallel trends, no anticipation and SUTVA assumptions, the conditional ATT for treated region $A$ in period $2$, denoted by $ATT_x(A,2)$, can be expressed as: 
\begin{equation}
\begin{aligned}
    ATT_x(A,2) = & \biggr[\mathbb{E}[Y_{2} \mid S_i = A, X_{2} = x] - \mathbb{E}[Y_{1} \mid S_i = A, X_{1} = x] \biggr] \\ - & \biggr[\mathbb{E}[Y_{2} \mid S_i = B, X_{2} = x] - \mathbb{E}[Y_{1} \mid S_i = B, X_{1} = x]\biggr].
\end{aligned}
\end{equation}
The overall ATT for region $s$ in period $t$, denoted by $ATT(s,t)$ can then be obtained by averaging the $ATT_x(s,t)$ over all values of $X_{i,s,t}$ under treatment effect heterogeneity with respect to observed covariates. Therefore, the $ATT$ for region $A$ in period 2 is given by: 
\begin{equation}
\begin{aligned}
    ATT(A,2) = \mathbb{E}\bigr[ATT_x(A,2) \mid S_i = A \bigr] = \int ATT_x(A,2) dF_{X \mid S_i = A, T = 2}(x)
\end{aligned}
\end{equation}
By definition, $ATT_x(A,2)$ represents the expected difference between the treated and untreated potential outcomes for region $A$, given that $X_{i,s,t} = x$.
\begin{equation}
\label{eq: AppendixDpo}
    ATT_x(A,2) = \mathbb{E}[Y_{2}(1)\mid S_i = A, X_{2}= x] - \mathbb{E}[Y_{2}(0) \mid S_i = A, X_{2} = x].
\end{equation}
In Equation \eqref{eq: AppendixDpo}, we do not observe the untreated potential outcome of the treated group in the post-intervention period $t$, denoted by $Y_{i,A,2}(0)$. For any given value of $X_{i,s,t}$, the conditional parallel trends assumption between treated region $A$ and control region $B$, considering the post-intervention period $2$ and the pre-intervention period $1$, can be expressed as:
\begin{equation}
\begin{aligned}
\label{eq: appendixDcpt}
    & \mathbb{E}[Y_{2}(0) \mid S_i = A, X_{2} = x] - \mathbb{E}[Y_{1}(0) \mid S_i = A, X_{1}= x]
    \\ = & \mathbb{E}[Y_{2}(0) \mid S_i = B, X_{2} = x] - \mathbb{E}[Y_{1}(0) \mid S_i = B, X_{1}= x].    
\end{aligned}
\end{equation}
Re-arranging Equation \eqref{eq: appendixDcpt}, we get:
\begin{equation}
\begin{aligned}
\label{eq: appendixDcptrearranged}
    & \mathbb{E}[Y_{2}(0) \mid S_i = A, X_{2} = x] = \mathbb{E}[Y_{1}(0) \mid S_i = A, X_{1}= x]
    \\ + & \mathbb{E}[Y_{2}(0) \mid S_i = B, X_{2} = x] - \mathbb{E}[Y_{1}(0) \mid S_i = B, X_{1}= x].    
\end{aligned}
\end{equation}
Plugging Equation \eqref{eq: appendixDcptrearranged} into \eqref{eq: AppendixDpo}, we get:
\begin{equation}
\label{eq: AppendixDpointermediate}
\begin{aligned}
       ATT_x(A,2) = & \mathbb{E}[Y_{2}(1)\mid S_i = A, X_{2}= x] - \mathbb{E}[Y_{1}(0) \mid S_i = A, X_{1}= x]
    \\ - & \mathbb{E}[Y_{2}(0) \mid S_i = B, X_{2} = x] + \mathbb{E}[Y_{1}(0) \mid S_i = B, X_{1}= x]. 
\end{aligned}
\end{equation}
Under the no anticipation assumption $Y_{i,A,1}(0) = Y_{i,A,1}(1)$. Therefore, Equation \eqref{eq: AppendixDpointermediate} can be written as:
\begin{equation}
\label{eq: AppendixDpointermediate22}
\begin{aligned}
       ATT_x(A,2) = & \biggr[\mathbb{E}[Y_{2}(1)\mid S_i = A, X_{2}= x] - \mathbb{E}[Y_{1}(1) \mid S_i = A, X_{1}= x] \biggr]
    \\ - & \biggr[ \mathbb{E}[Y_{2}(0) \mid S_i = B, X_{2} = x] - \mathbb{E}[Y_{1}(0) \mid S_i = B, X_{1}= x] \biggr]. 
\end{aligned}
\end{equation}
By Assumption \eqref{assumption: sutva}, the observed outcome can be expressed in terms of the relevant potential outcome according to the treatment status of each region-period cell. Therefore, we can re-write Equation \eqref{eq: AppendixDpointermediate22} as: 
\begin{equation}
\label{eq: AppendixDpointermediate2}
\begin{aligned}
       ATT_x(A,2) = & \biggr[\mathbb{E}[Y_{2}\mid S_i = A, X_{2}= x] - \mathbb{E}[Y_{1} \mid S_i = A, X_{1}= x] \biggr]
    \\ - & \biggr[ \mathbb{E}[Y_{2} \mid S_i = B, X_{2} = x] - \mathbb{E}[Y_{1} \mid S_i = B, X_{1}= x] \biggr]. 
\end{aligned}
\end{equation}


\section{Proof of ATT Under Integrated Parallel Trends and No Anticipation}
\label{appendix: ATTintegratedpt}

In this section, we show that under the integrated parallel trends, no anticipation and SUTVA assumptions, the ATT for treated region $A$ in period $2$, denoted by $ATT(A,2)$, can be expressed as:
\begin{equation}
\begin{aligned}
    ATT(A,2) = \int \biggr\{ & \biggr[\mathbb{E}[Y_{2} \mid S_i = A, X_{2} = x] - \mathbb{E}[Y_{1} \mid S_i = A, X_{1} = x] \biggr] \\
    - & \biggr[\mathbb{E}[Y_{2} \mid S_i = B, X_{2} = x] - \mathbb{E}[Y_{1} \mid S_i = B, X_{1} = x] \biggr] \biggr\} dF_{X_2 \mid S=A}(x).
\end{aligned}
\end{equation}
By definition, $ATT(A,2)$ represents the expected difference between the treated and untreated potential outcomes for region $A$ in period $2$:
\begin{equation}
\label{eq: integratedATTpo}
    ATT(A,2) = \mathbb{E} [Y_2(1)-Y_2(0) \mid S_i=A].
\end{equation}
Using the law of iterated expectations over the covariate distribution of the treated region in period $2$, Equation \eqref{eq: integratedATTpo} can be written as:
\begin{equation}
\label{eq: integratedATTpo2}
\begin{aligned}
    ATT(A,2) = \int \bigr[ & \mathbb{E}[Y_2(1)\mid S_i=A,X_2=x]
    - \mathbb{E}[Y_2(0)\mid S_i=A,X_2=x] \bigr] dF_{X_2\mid S=A}(x).
\end{aligned}
\end{equation}

In Equation \eqref{eq: integratedATTpo2}, we do not observe the untreated potential outcome of the treated group in the post-intervention period $2$, denoted by $Y_{i,A,2}(0)$. The integrated parallel trends assumption between treated region $A$ and control region $B$, considering the post-intervention period $2$ and the pre-intervention period $1$, can be expressed as:
\begin{equation}
\label{eq: appendixIntegratedPT2}
\begin{aligned}
    & \int
    \biggr[\mathbb{E}[Y_2(0)\mid S_i=A,X_2=x] - \mathbb{E}[Y_1(0)\mid S_i=A,X_1=x] \biggr] dF_{X_2\mid S=A}(x) \\
    = & \int \biggr[\mathbb{E}[Y_2(0)\mid S_i=B,X_2=x] - \mathbb{E}[Y_1(0)\mid S_i=B,X_1=x]\biggr]dF_{X_2\mid S=A}(x).
\end{aligned}
\end{equation}

Re-arranging Equation \eqref{eq: appendixIntegratedPT2}, we get:
\begin{equation}
\begin{aligned}
    & \int \mathbb{E}[Y_2(0)\mid S_i=A,X_2=x] dF_{X_2\mid S=A}(x) = \int \mathbb{E}[Y_1(0)\mid S_i=A,X_1=x] dF_{X_2\mid S=A}(x) \\
    & + \int \mathbb{E}[Y_2(0)\mid S_i=B,X_2=x]dF_{X_2\mid S=A}(x) - \int \mathbb{E}[Y_1(0)\mid S_i=B,X_1=x] dF_{X_2\mid S=A}(x).
\end{aligned}
\label{eq: appendixIntegratedPTrearranged}
\end{equation}

Plugging Equation \eqref{eq: appendixIntegratedPTrearranged} into Equation \eqref{eq: integratedATTpo2}, we get:
\begin{equation}
\label{eq: integratedATTintermediate}
\begin{aligned}
    ATT(A,2) = & \int \mathbb{E}[Y_2(1)\mid S_i=A,X_2=x] dF_{X_2\mid S=A}(x) - \int \mathbb{E}[Y_1(0)\mid S_i=A,X_1=x]dF_{X_2\mid S=A}(x) \\ - & \int \mathbb{E}[Y_2(0)\mid S_i=B,X_2=x] dF_{X_2\mid S=A}(x) + \int \mathbb{E}[Y_1(0) \mid S_i=B,X_1=x]dF_{X_2\mid S=A}(x).
\end{aligned}
\end{equation}
Since all terms are integrated over the same covariate distribution of the treated region, Equation \eqref{eq: integratedATTintermediate} can be written as:
\begin{equation}
\label{eq: integratedATTintermediate2}
\begin{aligned}
    ATT(A,2) = \int \biggr\{ & \biggr[\mathbb{E}[Y_2(1)\mid S_i=A,X_2=x] - \mathbb{E}[Y_1(0)\mid S_i=A,X_1=x] \biggr] \\
    - & \biggr[ \mathbb{E}[Y_2(0)\mid S_i=B,X_2=x] - \mathbb{E}[Y_1(0)\mid S_i=B,X_1=x] \biggr]\biggr\} dF_{X_2\mid S=A}(x).
\end{aligned}
\end{equation}

Under the no anticipation assumption, $Y_{i,A,1}(0)=Y_{i,A,1}(1)$. Therefore, Equation \eqref{eq: integratedATTintermediate2} can be written as:
\begin{equation}
\label{eq: integratedATTintermediate3}
\begin{aligned}
    ATT(A,2) = \int \biggr\{ & \biggr[\mathbb{E}[Y_2(1)\mid S_i=A,X_2=x] - \mathbb{E}[Y_1(1)\mid S_i=A,X_1=x]\biggr]
    \\ - & \biggr[\mathbb{E}[Y_2(0)\mid S_i=B,X_2=x] - \mathbb{E}[Y_1(0)\mid S_i=B,X_1=x] \biggr] \biggr\} dF_{X_2\mid S=A}(x).
\end{aligned}
\end{equation}

By Assumption \eqref{assumption: sutva}, the observed outcome can be expressed in terms of the relevant potential outcome according to the treatment status of each region-period cell. Therefore, we can re-write Equation \eqref{eq: integratedATTintermediate3} in terms of observed outcomes as:
\begin{equation}
\label{eq: integratedATTappendix}
\begin{aligned}
    ATT(A,2) = \int\biggr\{ & \biggr[\mathbb{E}[Y_{2} \mid S_i=A,X_{2}=x] - \mathbb{E}[Y_{1}\mid S_i=A,X_{1}=x] \biggr] \\
    - & \biggr[\mathbb{E}[Y_{2}\mid S_i=B,X_{2}=x] - \mathbb{E}[Y_{1}\mid S_i=B,X_{1}=x]\biggr]\biggr\}dF_{X_2\mid S=A}(x).
\end{aligned}
\end{equation}


\section{Identification under Integrated Difference-in-Differences}
\label{appendix: didintconsistency}

For simplicity, we analyze the two-way DID-INT in this proof. The same argument applies to the region-varying and time-varying DID-INT specifications when they are correctly specified. In the second step of the two-way Integrated Difference-in-Differences, we run the following regression:
\begin{equation}
\label{equation: DIDINTfirstregapp}
    Y_{i,s,t} = \sum_{s} \sum_{t} \lambda_{s,t} I(s, t) + \gamma_{s,t} I(s)I(t)X_{i,s,t} + \epsilon_{i,s,t},
\end{equation}
For simplicity, we continue to assume that there is a single time-varying covariate $X_{i,s,t}$ required for integrated parallel trends. We allow the treatment effect to vary linearly with the covariate. Let the conditional ATT be defined as:
\begin{equation}
\label{eq: attxindidint}
    ATT_x(s,t) = \mathbb{E}\bigr[ Y_t(1) - Y_t(0) \mid S_i = s, X_t = x \bigr]. 
\end{equation}

Let the population regression model for Equation \eqref{equation: DIDINTfirstregapp} be given by:
\begin{equation}
\label{equation: DIDINTpopulationregapp}
    Y_{i,s,t} = \sum_{s} \sum_{t} \lambda^*_{s,t} I(s, t) + \gamma^*_{s,t} I(s)I(t)X_{i,s,t} + \epsilon^*_{i,s,t},
\end{equation}
Here, $\lambda^*_{s,t}$ and $\gamma^*_{s,t}$ represent the probability limits of the intercept and the covariate coefficient in the regression shown in Equation \eqref{equation: DIDINTfirstregapp}, respectively. $\epsilon^*_{i,s,t}$ denotes the population regression error. Let $s$ denote a treated region in period $t$, with $t > t^s - 1$, where $t^s - 1$ is the period immediately before treatment begins in region $s$. Additionally, let $s' \in \mathbf{s}^c$ represent a relevant control region that has not yet received treatment in period $t$, and $\mathbf{s}^c$ is the set of relevant control regions for treated region $s$ in period $t$. Under these definitions, the probability limit of $\widehat{\tau}^{DT}_{s,t}$ is given by:
\begin{equation}
\begin{aligned}
\tau^{{DT}^*}_{s,t} = & \bigr( \lambda^*_{s,t} - \lambda^*_{s,t^s-1} \bigr) - \sum_{s' \in \mathbf{s}^c}
\omega_{s',s,t} \bigr( \lambda^*_{s',t} - \lambda^*_{s',t^s-1} \bigr) \\
& + \int \biggr[ \bigr( \gamma^*_{s,t} - \gamma^*_{s,t^s-1} \bigr) - \sum_{s' \in \mathbf{s}^c} \omega_{s',s,t} \bigr( \gamma^*_{s',t} - \gamma^*_{s',t^s-1} \bigr) \biggr] x dF_{X_t \mid S=s}(x).
\label{eq: tauiptexpanded}
\end{aligned}
\end{equation}
Taking the conditional expectation of the population regression model shown in Equation \eqref{equation: DIDINTpopulationregapp} for the treated region $S_i = s$ for the post-intervention period $t$, we get:
    \begin{align}
    \label{eq: lambda1}
         \begin{split}
           E[Y_{t} \mid S_i = s, X_{t} = x] =  \lambda^*_{s,t} + \gamma^*_{s,t}x.
        \end{split}
    \end{align}
In the above equation, we assume that $\mathbb{E}[\epsilon^*_{t} \mid S_i = s, X_{t} = x] = 0$ for all $s$,$t$ and $x$. The indicators $I(s,t)$, $I(s)$ and $I(t)$ are excluded because they take the value of one for all observations in the sample under consideration. Moreover, $E[X_{t} \mid S_i = s, X_{t}= x] = x$ since $X_{i,s,t}$ is conditioned on itself. For a relevant control region $S_i = s' \in \mathbf{s}^c$ in period $t$: 
    \begin{align}
        \label{eq: lambda2}
         \begin{split}
           E[Y_{t} \mid S_i = s', X_{t} = x] =  \lambda^*_{s',t} + \gamma^*_{s',t}x.
        \end{split}
    \end{align}
For the treated region $S_i = s$ in the pre-intervention period $t^s-1$: 
    \begin{align}
        \label{eq: lambda3}
         \begin{split}
           E[Y_{t^s-1} \mid S_i = s, X_{t^s-1} = x] =  \lambda^*_{s,t^s-1} + \gamma^*_{s,t^s-1}x.
        \end{split}
    \end{align}
For a relevant control region $S_i = s' \in \mathbf{s}^c$ in the pre-intervention period $t^s-1$: 
    \begin{align}
        \label{eq: lambda4}
         \begin{split}
           E[Y_{t^s-1} \mid S_i = s', X_{t^s-1} = x] =  \lambda^*_{s',t^s-1} + \gamma^*_{s',t^s-1}x.
        \end{split}
    \end{align}
Under linearity, the model for the untreated potential outcome is:
\begin{equation}
    Y_{i,s,t}(0) = \theta^{\circ}_{s,t} + \gamma^{\circ}_{s,t} X_{i,s,t} + \epsilon_{i,s,t}.
\end{equation}
The conditional expectation of the untreated potential outcome for a given value of $X_{i,s,t} = x$ for region $S_i = s$ is:
\begin{equation}
\label{eq: poexpecationdidint}
    \mathbb{E} \bigr[Y_{t}(0) \mid S_i = s, X_{t} = x \bigr] = \theta^{\circ}_{s,t} + \gamma^{\circ}_{s,t}x.
\end{equation}

Under SUTVA, Equation \eqref{eq: lambda1} can be expressed as:
    \begin{align}
    \label{eq: lambda1sutva}
         \begin{split}
           E[Y_{t} \mid S_i = s, X_{t} = x] = E[Y_{t}(1) \mid S_i = s, X_{t} = x] =  \lambda^*_{s,t} + \gamma^*_{s,t}x.
        \end{split}
    \end{align}
Adding and subtracting $E[Y_{t}(0) \mid S_i = s, X_{t} = x]$ in Equation \eqref{eq: lambda1sutva} we get,
    \begin{align}
    \label{eq: lambda1sutva2}
         \begin{split}
           & E[Y_{t} \mid S_i = s, X_{t} = x] = E[Y_{t}(1) \mid S_i = s, X_{t} = x] \\ - & E[Y_{t}(0) \mid S_i = s, X_{t} = x] + E[Y_{t}(0) \mid S_i = s, X_{t} = x] \\ = &  \lambda^*_{s,t} + \gamma^*_{s,t}x.
        \end{split}
    \end{align}
By the definition of $ATT_x(s,t)$ in Equation \eqref{eq: attxindidint}, and the conditional expectation of the untreated potential outcome model shown in Equation \eqref{eq: poexpecationdidint} yields:
    \begin{align}
    \label{eq: lambda1sutvafinal}
         \begin{split}
           & E[Y_{t} \mid S_i = s, X_{t} = x] = ATT_x(s,t) + \theta^{\circ}_{s,t} + \gamma^{\circ}_{s,t}x = \lambda^*_{s,t} + \gamma^*_{s,t}x.
        \end{split}
    \end{align}
Under SUTVA and the no anticipation assumptions, Equation \eqref{eq: lambda3} can be expressed as: 
    \begin{align}
    \label{eq: lambda2sutvafinal}
         \begin{split}
           & E[Y_{t^s-1}(0) \mid S_i = s, X_{t^s-1} = x] = \theta^{\circ}_{s,t^s-1} + \gamma^{\circ}_{s,t^s-1}x = \lambda^*_{s,t^s-1} + \gamma^*_{s,t^s-1}x.
        \end{split}
    \end{align}
Under SUTVA, Equations \eqref{eq: lambda2} and \eqref{eq: lambda4} can be expressed as:
    \begin{align}
    \label{eq: lambda3sutvafinal}
         \begin{split}
           & E[Y_{t} \mid S_i = s', X_{t} = x] =  \theta^{\circ}_{s',t} + \gamma^{\circ}_{s',t}x = \lambda^*_{s',t} + \gamma^*_{s',t}x.
        \end{split}
    \end{align}
    \begin{align}
    \label{eq: lambda4sutvafinal}
         \begin{split}
           & E[Y_{t^s-1} \mid S_i = s', X_{t^s-1} = x] =  \theta^{\circ}_{s',t^s-1} + \gamma^{\circ}_{s',t^s-1}x = \lambda^*_{s',t^s-1} + \gamma^*_{s',t^s-1}x.
        \end{split}
    \end{align}
    
Taking the difference between Equations \eqref{eq: lambda1sutvafinal} and \eqref{eq: lambda2sutvafinal} for the treated region, and subtracting the weighted average of the corresponding differences between Equations \eqref{eq: lambda3sutvafinal} and \eqref{eq: lambda4sutvafinal} across all relevant control regions $s' \in \mathbf{s}^c$, we get:
\begin{gather}
ATT_x(s,t) + \biggr( \theta^{\circ}_{s,t} - \theta^{\circ}_{s,t^s-1} \biggr) - \sum_{s' \in \mathbf{s}^c} \omega_{s',s,t} \biggr( \theta^{\circ}_{s',t} - \theta^{\circ}_{s',t^s-1}
\biggr) \nonumber \\ + \biggr[ \biggr( \gamma^{\circ}_{s,t} - \gamma^{\circ}_{s,t^s-1} \biggr) - \sum_{s' \in \mathbf{s}^c} \omega_{s',s,t} \biggr( \gamma^{\circ}_{s',t} - \gamma^{\circ}_{s',t^s-1} \biggr) \biggr]x. \nonumber
\\ = \biggr( \lambda^*_{s,t} - \lambda^*_{s,t^s-1} \biggr) - \sum_{s' \in \mathbf{s}^c} \omega_{s',s,t} \biggr( \lambda^*_{s',t} - \lambda^*_{s',t^s-1}
\biggr) \nonumber \\ + \biggr[ \biggr( \gamma^*_{s,t} - \gamma^*_{s,t^s-1} \biggr) - \sum_{s' \in \mathbf{s}^c} \omega_{s',s,t} \biggr( \gamma^*_{s',t} - \gamma^*_{s',t^s-1} \biggr) \biggr]x. \label{equation: DIDINTiptproof}
\end{gather}
Integrating both sides of Equation \eqref{equation: DIDINTiptproof} over the covariate distribution of the treated group yields:
\begin{gather}
\int ATT_x(s,t) dF_{X_t \mid S_i=s}(x) + \int \biggr\{\biggr( \theta^{\circ}_{s,t} - \theta^{\circ}_{s,t^s-1} \biggr) - \sum_{s' \in \mathbf{s}^c} \omega_{s',s,t} \biggr( \theta^{\circ}_{s',t} - \theta^{\circ}_{s',t^s-1}
\biggr) \nonumber \\ + \biggr[ \biggr( \gamma^{\circ}_{s,t} - \gamma^{\circ}_{s,t^s-1} \biggr) - \sum_{s' \in \mathbf{s}^c} \omega_{s',s,t} \biggr( \gamma^{\circ}_{s',t} - \gamma^{\circ}_{s',t^s-1} \biggr) \biggr]x \biggr\} dF_{X_t \mid S_i=s}(x) \nonumber
\\ = \int \biggr\{ \biggr( \lambda^*_{s,t} - \lambda^*_{s,t^s-1} \biggr) - \sum_{s' \in \mathbf{s}^c} \omega_{s',s,t} \biggr( \lambda^*_{s',t} - \lambda^*_{s',t^s-1}
\biggr) \nonumber \\ + \biggr[ \biggr( \gamma^*_{s,t} - \gamma^*_{s,t^s-1} \biggr) - \sum_{s' \in \mathbf{s}^c} \omega_{s',s,t} \biggr( \gamma^*_{s',t} - \gamma^*_{s',t^s-1} \biggr) \biggr]x \biggr\} dF_{X_t \mid S_i=s}(x). \label{equation: DIDINTiptproofbig}
\end{gather}

Under integrated parallel trends, or under conditional parallel trends since conditional parallel trends implies integrated parallel trends, the following integrated restriction equals zero.
\begin{gather}
    \int \biggr\{\biggr( \theta^{\circ}_{s,t} - \theta^{\circ}_{s,t^s-1} \biggr) - \sum_{s' \in \mathbf{s}^c} \omega_{s',s,t} \biggr( \theta^{\circ}_{s',t} - \theta^{\circ}_{s',t^s-1}
    \biggr) \nonumber \\ + \biggr[ \biggr( \gamma^{\circ}_{s,t} - \gamma^{\circ}_{s,t^s-1} \biggr) - \sum_{s' \in \mathbf{s}^c} \omega_{s',s,t} \biggr( \gamma^{\circ}_{s',t} - \gamma^{\circ}_{s',t^s-1} \biggr) \biggr]x \biggr\} dF_{X_t \mid S_i=s}(x) = 0.
\end{gather}
Since $ATT_x(s,t)$ is allowed to vary with $x$, taking its expectation over the covariate distribution of the treated group yields the unconditional $ATT(s,t)$.
\begin{equation}
\label{eq: attintforappendix}
ATT(s,t)=\int ATT_x(s,t) dF_{X_t\mid S_i=s}(x).
\end{equation}
Using the above expressions, Equation \eqref{equation: DIDINTiptproofbig} can be simplified to:
\begin{gather}
\int ATT_x(s,t) dF_{X_t \mid S_i=s}(x) = ATT(s,t) 
= \left(\lambda^*_{s,t}-\lambda^*_{s,t^s-1}\right)
-\sum_{s' \in \mathbf{s}^c}\omega_{s',s,t}
\left(\lambda^*_{s',t}-\lambda^*_{s',t^s-1}\right) \nonumber \\
+ \int \biggr[
\left(\gamma^*_{s,t}-\gamma^*_{s,t^s-1}\right)
-\sum_{s' \in \mathbf{s}^c}\omega_{s',s,t}
\left(\gamma^*_{s',t}-\gamma^*_{s',t^s-1}\right)
\biggr] x dF_{X_t \mid S_i=s}(x) = \tau^{DT*}_{s,t}. \label{equation: DIDINTcptfinal}
\end{gather}
Therefore, 
\begin{equation}
    \tau^{DT*}_{s,t} = ATT(s,t).
\end{equation}

\section{Identification under Flexible TWFE}
\label{appendix: twfeconsistencymod}

The population regression model for the flexible TWFE can be written as:
\begin{equation}
\label{eq: popmodtwfe_general}
    Y_{i,s,t} = \alpha_s^* + \delta_t^* + \beta_D^* D_{s,t} + f_{s,t}^*(X_{i,s,t}) + \epsilon^*_{i,s,t}.
\end{equation}
Here, $f_{s,t}^*(X_{i,s,t})$ is the functional form of the covariate under the relevant flexible TWFE specification, which depends on the type of CCC violation it addresses. For simplicity, we analyze the two-way flexible TWFE in this proof. The same argument applies to the region-varying and time-varying flexible TWFE when they are correctly specified. Let us first assume that integrated parallel trends holds, and the additive restriction $\theta^\circ_{s,t} = \alpha^*_s + \delta^*_t$ is valid. Therefore, the population regression model for the flexible TWFE can be written as:
\begin{equation}
\label{eq: popmodtwfe}
    Y_{i,s,t} = \alpha_s^* + \delta_t^* + \beta_D^* D_{s,t} + \gamma^*_{s,t} X_{i,s,t} + \epsilon^*_{i,s,t}.
\end{equation}

Taking the conditional expectation of the population regression model shown in Equation \eqref{eq: popmodtwfe} for the treated region $S_i = s$ for the post-intervention period $t$, and integrating over the treated region's covariate distribution gives:
\begin{equation}
\label{eq: modtwfediff1}
    \int \mathbb{E} \bigr[ Y_t \mid S_i = s, X_t = x \bigr] dF_{X_t \mid S_i = s}(x) = \alpha_s^* + \delta_t^* + \beta_D^* + \int \gamma^*_{s,t}x dF_{X_t \mid S_i = s}(x).
\end{equation}
$\beta_D^*$ remains in Equation \eqref{eq: modtwfediff1} since $D_{s,t} = 1$ for the treated region in the post-intervention period. For the treated region $S_i = s$ in the pre-intervention period $t^s-1$:
\begin{equation}
\label{eq: modtwfediff2}
    \int \mathbb{E} \bigr[ Y_{t^s-1} \mid S_i = s, X_{t^s-1} = x \bigr] dF_{X_t \mid S_i = s}(x) = \alpha_s^* + \delta_{t^s-1}^* + \int \gamma^*_{s,t^s-1}x dF_{X_t \mid S_i = s}(x).
\end{equation}
For the relevant control region $S_i = s'$ in the post-intervention period $t$:
\begin{equation}
\label{eq: modtwfediff3}
    \int \mathbb{E} \bigr[ Y_t \mid S_i = s', X_t = x \bigr] dF_{X_t \mid S_i = s}(x) = \alpha_{s'}^* + \delta_t^* + \int \gamma^*_{s',t}x dF_{X_t \mid S_i = s}(x).
\end{equation}
For a relevant control region $S_i = s' \in \mathbf{s}^c$ in the pre-intervention period $t^s-1$:
\begin{equation}
\label{eq: modtwfediff4}
    \int \mathbb{E} \bigr[ Y_{t^s-1} \mid S_i = s', X_{t^s-1} = x \bigr] dF_{X_t \mid S_i = s}(x) = \alpha_{s'}^* + \delta_{t^s-1}^* + \int \gamma^*_{s',t^s-1}x dF_{X_t \mid S_i = s}(x).
\end{equation}

Under linearity, the model for the untreated potential outcome is:
\begin{equation}
    Y_{i,s,t}(0)=\theta^\circ_{s,t}+\gamma^\circ_{s,t}X_{i,s,t}+\epsilon_{i,s,t}.
\end{equation}
The conditional expectation of the untreated potential outcome for a given value of $X_{i,s,t}=x$ for region $S_i=s$ is:
\begin{equation}
\label{eq: poexpectationmodtwfe}
    \mathbb{E}\bigr[Y_t(0)\mid S_i=s,X_t=x\bigr]=\theta^\circ_{s,t}+\gamma^\circ_{s,t}x.
\end{equation}

Under SUTVA and the definition of $ATT_x(s,t)$, Equation \eqref{eq: modtwfediff1} can be expressed as:
\begin{gather}
\int \mathbb{E}\bigr[Y_t\mid S_i=s,X_t=x\bigr]dF_{X_t\mid S_i=s}(x) \nonumber\\
=\int ATT_x(s,t) dF_{X_t\mid S_i=s}(x) +\theta^\circ_{s,t}+\int\gamma^\circ_{s,t}x\,dF_{X_t\mid S_i=s}(x) \nonumber\\
=\alpha_s^*+\delta_t^*+\beta_D^*+\int\gamma^*_{s,t}x\,dF_{X_t\mid S_i=s}(x).
\label{eq: modtwfetrue1}
\end{gather}
Under SUTVA and no anticipation, Equation \eqref{eq: modtwfediff2} can be expressed as:
\begin{gather}
\int \mathbb{E}\bigr[Y_{t^s-1}\mid S_i=s,X_{t^s-1}=x\bigr]dF_{X_t\mid S_i=s}(x) \nonumber\\
=\theta^\circ_{s,t^s-1}+\int\gamma^\circ_{s,t^s-1}x\,dF_{X_t\mid S_i=s}(x) \nonumber\\
=\alpha_s^*+\delta_{t^s-1}^*+\int\gamma^*_{s,t^s-1}x\,dF_{X_t\mid S_i=s}(x).
\label{eq: modtwfetrue2}
\end{gather}
Under SUTVA and no anticipation, Equation \eqref{eq: modtwfediff3} can be expressed as:
\begin{gather}
\int \mathbb{E}\bigr[Y_t\mid S_i=s',X_t=x\bigr]dF_{X_t\mid S_i=s}(x) \nonumber\\
=\theta^\circ_{s',t}+\int\gamma^\circ_{s',t}x\,dF_{X_t\mid S_i=s}(x) \nonumber\\
=\alpha_{s'}^*+\delta_t^*+\int\gamma^*_{s',t}x\,dF_{X_t\mid S_i=s}(x).
\label{eq: modtwfetrue3}
\end{gather}
Similarly, Equation \eqref{eq: modtwfediff4} can be expressed as:
\begin{gather}
\int \mathbb{E}\bigr[Y_{t^s-1}\mid S_i=s',X_{t^s-1}=x\bigr]dF_{X_t\mid S_i=s}(x) \nonumber\\
=\theta^\circ_{s',t^s-1}+\int\gamma^\circ_{s',t^s-1}x\,dF_{X_t\mid S_i=s}(x) \nonumber\\
=\alpha_{s'}^*+\delta_{t^s-1}^*+\int\gamma^*_{s',t^s-1}x\,dF_{X_t\mid S_i=s}(x).
\label{eq: modtwfetrue4}
\end{gather}

Taking the difference between Equations \eqref{eq: modtwfetrue1} and \eqref{eq: modtwfetrue2} for the treated region, and subtracting the weighted average of the corresponding differences between Equations \eqref{eq: modtwfetrue3} and \eqref{eq: modtwfetrue4} across all relevant control regions $s' \in \mathbf{s}^c$, we get:
\begin{gather}
\int ATT_x(s,t) dF_{X_t\mid S_i=s}(x) +\biggr( \theta^\circ_{s,t}-\theta^\circ_{s,t^s-1}\biggr)-\sum_{s'\in\mathbf{s}^c}\omega_{s',s,t}\biggr(\theta^\circ_{s',t}-\theta^\circ_{s',t^s-1}\biggr) \nonumber\\
+\int\biggr[\biggr(\gamma^\circ_{s,t}-\gamma^\circ_{s,t^s-1}\biggr)-\sum_{s'\in\mathbf{s}^c}\omega_{s',s,t}\biggr(\gamma^\circ_{s',t}-\gamma^\circ_{s',t^s-1}\biggr)\biggr]x\,dF_{X_t\mid S_i=s}(x) \nonumber\\
=\beta_D^*+\int\biggr[\biggr(\gamma^*_{s,t}-\gamma^*_{s,t^s-1}\biggr)-\sum_{s'\in\mathbf{s}^c}\omega_{s',s,t}\biggr(\gamma^*_{s',t}-\gamma^*_{s',t^s-1}\biggr)\biggr]x\,dF_{X_t\mid S_i=s}(x).
\label{eq: modbig}
\end{gather}

Under integrated parallel trends,
\begin{gather}
\biggr( \theta^\circ_{s,t}-\theta^\circ_{s,t^s-1}\biggr)-\sum_{s'\in\mathbf{s}^c}\omega_{s',s,t}\biggr(\theta^\circ_{s',t}-\theta^\circ_{s',t^s-1}\biggr) \nonumber\\
+\int\biggr[\biggr(\gamma^\circ_{s,t}-\gamma^\circ_{s,t^s-1}\biggr)-\sum_{s'\in\mathbf{s}^c}\omega_{s',s,t}\biggr(\gamma^\circ_{s',t}-\gamma^\circ_{s',t^s-1}\biggr)\biggr]x\,dF_{X_t\mid S_i=s}(x)=0.
\end{gather}
By the law of iterated expectations, 
\begin{gather}
    \int ATT_x(s,t) dF_{X_t\mid S_i=s}(x) = ATT(s,t).
\end{gather}

Using the above conditions, Equation \eqref{eq: modbig} can be written as:
\begin{gather}
ATT(s,t) =\beta_D^*+\int\biggr[\biggr(\gamma^*_{s,t}-\gamma^*_{s,t^s-1}\biggr)-\sum_{s'\in\mathbf{s}^c}\omega_{s',s,t}\biggr(\gamma^*_{s',t}-\gamma^*_{s',t^s-1}\biggr)\biggr]x\,dF_{X_t\mid S_i=s}(x).
\end{gather}

Now, we analyze the case where the treatment effect does not vary with the covariate. Under homogeneity of the treatment effect with respect to the covariate,
\begin{equation}
    ATT_x(s,t) = ATT(s,t) \quad \forall x.
\end{equation}

Under the additive TWFE restriction and correct specification,
\begin{equation}
    \theta^\circ_{s,t}=\theta^*_{s,t}=\alpha_s^*+\delta_t^*.
\end{equation}

Therefore, the intercept component of the integrated parallel trends condition is:
\begin{gather}
\biggr[(\theta^\circ_{s,t}-\theta^\circ_{s,t^s-1})-\sum_{s'\in\mathbf{s}^c}\omega_{s',s,t}(\theta^\circ_{s',t}-\theta^\circ_{s',t^s-1})\biggr] \nonumber\\
=\biggr[(\alpha_s^*+\delta_t^*)-(\alpha_s^*+\delta_{t^s-1}^*)\biggr]-\sum_{s'\in\mathbf{s}^c}\omega_{s',s,t}\biggr[(\alpha_{s'}^*+\delta_t^*)-(\alpha_{s'}^*+\delta_{t^s-1}^*)\biggr] \nonumber\\
=(\delta_t^*-\delta_{t^s-1}^*)-\sum_{s'\in\mathbf{s}^c}\omega_{s',s,t}(\delta_t^*-\delta_{t^s-1}^*)=0.
\end{gather}
where the final equality follows because $\sum_{s'\in\mathbf{s}^c}\omega_{s',s,t}=1$. For the correctly specified two-way flexible TWFE, and imposing $ATT_x(s,t)=ATT(s,t)$ for all $x$, the point-wise conditional mean for the treated region in the post-intervention period can be written as:
\begin{equation}
    ATT(s,t)+\theta^\circ_{s,t}+\gamma^\circ_{s,t}x
    =
    \alpha_s^*+\delta_t^*+\beta_D^*+\gamma_{s,t}^*x.
\end{equation}
Under the additive TWFE restriction, $\theta^\circ_{s,t}=\alpha_s^*+\delta_t^*$. Therefore,
\begin{equation}
    \bigr(\gamma_{s,t}^*-\gamma^\circ_{s,t}\bigr)x
    = ATT(s,t)-\beta_D^*.
\end{equation}
Since the right-hand side does not vary with $x$, while the equality must hold for all $x$, it follows that:
\begin{equation}
    \gamma_{s,t}^*=\gamma^\circ_{s,t}.
\end{equation}
The same argument applies to the treated pre-intervention period and to the relevant control regions in both periods, implying that
\[
\gamma^*_{s,t^s-1}=\gamma^\circ_{s,t^s-1}, \qquad
\gamma^*_{s',t}=\gamma^\circ_{s',t}, \qquad
\gamma^*_{s',t^s-1}=\gamma^\circ_{s',t^s-1}.
\]

Therefore, the integrated covariate slope component of the integrated parallel trends condition is:
\begin{gather}
\int\biggr[\gamma^\circ_{s,t}x-\gamma^\circ_{s,t^s-1}x-\sum_{s'\in\mathbf{s}^c}\omega_{s',s,t}\bigr(\gamma^\circ_{s',t}x-\gamma^\circ_{s',t^s-1}x\bigr)\biggr]dF_{X_t\mid S_i=s}(x) \nonumber\\
=\int\biggr[\gamma^*_{s,t}x-\gamma^*_{s,t^s-1}x-\sum_{s'\in\mathbf{s}^c}\omega_{s',s,t}\bigr(\gamma^*_{s',t}x-\gamma^*_{s',t^s-1}x\bigr)\biggr]dF_{X_t\mid S_i=s}(x).
\end{gather}

This is exactly the remaining component on the right-hand side of Equation \eqref{eq: modbig}. Since integrated parallel trends requires the sum of the intercept component and the integrated covariate slope component to equal zero, and the intercept component is zero under the additive TWFE restriction, it follows that:
\begin{equation}
\int\biggr[\gamma^*_{s,t}x-\gamma^*_{s,t^s-1}x-\sum_{s'\in\mathbf{s}^c}\omega_{s',s,t}\bigr(\gamma^*_{s',t}x-\gamma^*_{s',t^s-1}x\bigr)\biggr]dF_{X_t\mid S_i=s}(x)=0.
\end{equation}
Hence,
\begin{equation}
    \beta_D^*=ATT(s,t).
\end{equation}

We now relax the assumption that the treatment effect is homogeneous with respect to the covariate. Therefore, for at least some values of $x$,
\begin{equation}
    ATT_x(s,t) \neq ATT(s,t).
\end{equation}
The point-wise conditional mean for the treated region in the post-intervention period can then be written as:
\begin{equation}
    ATT_x(s,t)+\theta^\circ_{s,t}+\gamma^\circ_{s,t}x
    =
    \alpha_s^*+\delta_t^*+\beta_D^*+\gamma_{s,t}^*x.
\end{equation}
Under the additive TWFE restriction, $\theta^\circ_{s,t}=\alpha_s^*+\delta_t^*$. Therefore,
\begin{equation}
    \bigr(\gamma_{s,t}^*-\gamma^\circ_{s,t}\bigr)x
    =
    ATT_x(s,t)-\beta_D^*.
\end{equation}
Since $ATT_x(s,t)$ varies with $x$, while $\beta_D^*$ does not, the right-hand side of the above equation varies with $x$. Therefore, the left-hand side must also vary with $x$, which requires:
\begin{equation}
    \gamma_{s,t}^*-\gamma^\circ_{s,t}\neq 0.
\end{equation}
Hence,
\begin{equation}
    \gamma_{s,t}^*\neq\gamma^\circ_{s,t}.
\end{equation}
Therefore, in the non-trivial case where
\begin{equation}
    \int\biggr[\biggr(\gamma^*_{s,t}-\gamma^*_{s,t^s-1}\biggr)-\sum_{s'\in\mathbf{s}^c}\omega_{s',s,t}\biggr(\gamma^*_{s',t}-\gamma^*_{s',t^s-1}\biggr)\biggr]x\,dF_{X_t\mid S_i=s}(x) \neq 0.
\end{equation}
Equation \eqref{eq: modbig} implies that
\begin{equation}
    \beta_D^* \neq ATT(s,t).
\end{equation}

\section{Identification under Standard TWFE}
\label{appendix: twfeconsistency}

The population model corresponding to the standard two-way fixed effects specification can be written as:
\begin{equation}
\label{appendix: twfe}
    Y_{i,s,t}
    =
    \alpha_s^*
    +
    \delta_t^*
    +
    \beta_D^*D_{s,t}
    +
    \gamma^*X_{i,s,t}
    +
    \varepsilon^*_{i,s,t}.
\end{equation}
The key parameter of interest is $\beta_D$, which we derive in this section. For simplicity, we assume that there is a single time-varying covariate $X_{i,s,t}$. Let $\beta_D^*$ be the probability limit of $\beta_D$. The population OLS objective function is:
\begin{equation}
(\alpha_s^*,\delta_t^*,\beta_D^*,\gamma^*)
=
\arg\min_{\alpha_s,\delta_t,\beta_D,\gamma}
\mathbb{E}
\left[
\left(
Y_{i,s,t}
-
\alpha_s
-
\delta_t
-
\beta_DD_{s,t}
-
\gamma X_{i,s,t}
\right)^2
\right].
\end{equation}

The first-order condition with respect to $\beta_D$ is given by:
\begin{equation}
\mathbb{E}
\left[
D_{s,t}
\left(
Y_{i,s,t}
-
\alpha_s^*
-
\delta_t^*
-
\beta_D^*D_{s,t}
-
\gamma^*X_{i,s,t}
\right)
\right]
=
0.
\end{equation}
By distributing $D_{s,t}$ across the terms, using the linearity of expectations, and noting that $(D_{s,t})^2=D_{s,t}$ because it is binary, we obtain:
\begin{equation}
\mathbb{E}\left[D_{s,t}Y_{i,s,t}\right]
-
\mathbb{E}\left[D_{s,t}\alpha_s^*\right]
-
\mathbb{E}\left[D_{s,t}\delta_t^*\right]
-
\beta_D^*\mathbb{E}\left[D_{s,t}\right]
-
\mathbb{E}\left[\gamma^*D_{s,t}X_{i,s,t}\right]
=
0.
\end{equation}

Therefore, $\beta_D^*$ can be written as:
\begin{equation}
\label{appendix: betatwfe}
\beta_D^*
=
\frac{
\mathbb{E}\left[D_{s,t}Y_{i,s,t}\right]
-
\mathbb{E}\left[D_{s,t}\alpha_s^*\right]
-
\mathbb{E}\left[D_{s,t}\delta_t^*\right]
-
\mathbb{E}\left[\gamma^*D_{s,t}X_{i,s,t}\right]
}{
\mathbb{E}\left[D_{s,t}\right]
}.
\end{equation}

Because Equation \eqref{appendix: betatwfe} corresponds to a TWFE regression that includes covariates, the standard Goodman-Bacon decomposition cannot be applied directly to $\beta_D^*$. With covariates, the Frisch-Waugh-Lovell theorem requires the treatment indicator to be residualized with respect to the region fixed effects, period fixed effects, and covariates before characterizing the variation that identifies $\beta_D^*$. Since our objective in this section is to isolate the misspecification bias arising from imposing a common coefficient on $X_{i,s,t}$, we proceed directly from the population first-order condition in Equation \eqref{appendix: betatwfe}.

Under SUTVA, we know that the observed outcome is given by:
\begin{equation}
\label{appendix: sutva}
    Y_{i,s,t}
    =
    Y_{i,s,t}(0)
    +
    D_{s,t}
    \bigr(
    Y_{i,s,t}(1)
    -
    Y_{i,s,t}(0)
    \bigr).
\end{equation}

Plugging the above expression into Equation \eqref{appendix: betatwfe} and simplifying, we get:
\begin{equation}
\label{appendix: betatwfeplugged}
\begin{aligned}
\beta_D^*
={}&
\frac{
\mathbb{E}
\left[
D_{s,t}
\left(
Y_{i,s,t}(1)-Y_{i,s,t}(0)
\right)
\right]
}{
\mathbb{E}[D_{s,t}]
}
\\
&+
\frac{
\mathbb{E}\left[D_{s,t}Y_{i,s,t}(0)\right]
-
\mathbb{E}\left[D_{s,t}\alpha_s^*\right]
-
\mathbb{E}\left[D_{s,t}\delta_t^*\right]
-
\mathbb{E}\left[\gamma^*D_{s,t}X_{i,s,t}\right]
}{
\mathbb{E}[D_{s,t}]
}.
\end{aligned}
\end{equation}

Equation \eqref{appendix: betatwfeplugged} can be further decomposed into two components, as shown in Equation \eqref{appendix: betatwfesimplified}. The first term corresponds to the ATT, as shown in Equation \eqref{appendix: atttwfe}, which follows directly from its definition. The second term, as shown in Equation \eqref{appendix: biastwfe}, is the bias arising from model misspecification.
\begin{equation}
\label{appendix: betatwfesimplified}
    \beta_D^*
    =
    ATT
    +
    \text{Bias}_{D}.
\end{equation}
where
\begin{equation}
\label{appendix: atttwfe}
\begin{aligned}
    ATT
    &=
    \frac{
    \mathbb{E}
    \left[
    D_{s,t}
    \left(
    Y_{i,s,t}(1)-Y_{i,s,t}(0)
    \right)
    \right]
    }{
    \mathbb{E}[D_{s,t}]
    }
    \\
    &=
    \mathbb{E}
    \left[
    Y_{i,s,t}(1)-Y_{i,s,t}(0)
    \mid
    D_{s,t}=1
    \right].
\end{aligned}
\end{equation}
and
\begin{equation}
\label{appendix: biastwfe}
    \text{Bias}_{D}
    =
    \frac{
    \mathbb{E}\left[D_{s,t}Y_{i,s,t}(0)\right]
    -
    \mathbb{E}\left[D_{s,t}\alpha_s^*\right]
    -
    \mathbb{E}\left[D_{s,t}\delta_t^*\right]
    -
    \mathbb{E}\left[\gamma^*D_{s,t}X_{i,s,t}\right]
    }{
    \mathbb{E}[D_{s,t}]
    }.
\end{equation}

Now, we analyze the bias term shown in Equation \eqref{appendix: biastwfe} in more detail. Let the true untreated potential outcome $Y_{i,s,t}(0)$ be given by:
\begin{equation}
\label{appendix: truemodel}
    Y_{i,s,t}(0)
    =
    \alpha_s^{\circ}
    +
    \delta_t^{\circ}
    +
    \gamma^{\circ}_{s,t}X_{i,s,t}
    +
    \varepsilon^{\circ}_{i,s,t}.
\end{equation}
For simplicity, we assume that the intercept follows an additive region and time structure, such that
\begin{equation}
    \theta_{s,t}^{\circ}
    =
    \alpha_s^{\circ}
    +
    \delta_t^{\circ}.
\end{equation}

Substituting the true untreated potential outcome model in Equation \eqref{appendix: truemodel} into the bias expression in Equation \eqref{appendix: biastwfe} and simplifying yields:
\begin{equation}
\label{appendix: biastwfeaftersub}
\text{Bias}_{D}
=
\frac{
\mathbb{E}
\left[
D_{s,t}
\left\{
(\alpha_s^{\circ}-\alpha_s^*)
+
(\delta_t^{\circ}-\delta_t^*)
+
(\gamma_{s,t}^{\circ}-\gamma^*)X_{i,s,t}
\right\}
\right]
}{
\mathbb{E}[D_{s,t}]
}.
\end{equation}

Since $D_{s,t}$ is binary, Equation \eqref{appendix: biastwfeaftersub} can equivalently be written as:
\begin{equation}
\label{appendix: biastwfeconditional}
\text{Bias}_{D}
=
\mathbb{E}
\left[
(\alpha_s^{\circ}-\alpha_s^*)
+
(\delta_t^{\circ}-\delta_t^*)
+
(\gamma_{s,t}^{\circ}-\gamma^*)X_{i,s,t}
\mid
D_{s,t}=1
\right].
\end{equation}

To examine how the overall bias varies across treated region-period cells, we apply the Law of Iterated Expectations to Equation \eqref{appendix: biastwfeconditional}. For each treated region-period cell $(s,t)$, define:
\begin{equation}
\label{appendix: biasst}
\begin{aligned}
\text{Bias}_{s,t}
={}&
\mathbb{E}
\left[
(\alpha_s^{\circ}-\alpha_s^*)
+
(\delta_t^{\circ}-\delta_t^*)
+
(\gamma_{s,t}^{\circ}-\gamma^*)X_{i,s,t}
\mid
S_i=s,T=t,D_{s,t}=1
\right]
\\
={}&
(\alpha_s^{\circ}-\alpha_s^*)
+
(\delta_t^{\circ}-\delta_t^*)
+
(\gamma_{s,t}^{\circ}-\gamma^*)
\mathbb{E}
\left[
X_{i,s,t}
\mid
S_i=s,T=t,D_{s,t}=1
\right].
\end{aligned}
\end{equation}

The overall bias is therefore a weighted average of the region-period-specific biases among treated region-period cells:
\begin{equation}
\label{appendix: biastwfeLIE}
    \text{Bias}_{D}
    =
    \sum_{(s,t):D_{s,t}=1}
    \Pr
    \left(
    S_i=s,T=t
    \mid
    D_{s,t}=1
    \right)
    \text{Bias}_{s,t}.
\end{equation}

An important implication of using pre-trends as a plausibility check for conditional parallel trends is that researchers may inadvertently misspecify the model by imposing a common coefficient on $X_{i,s,t}$. Since the jointly necessary and sufficient conditions for conditional parallel trends are established as a novel result of this paper, the requirement that changes in the effects of covariates between the pre- and post-intervention periods be equal across treated and control regions has not previously been made explicit. As a result, researchers may interpret an apparently ``parallel" figure as justification for imposing a single common coefficient on $X_{i,s,t}$. Doing so can introduce a misspecification bias.

In the standard TWFE specification, the common population coefficient $\gamma^*$ is imposed across all region-period cells. Therefore, when $\gamma_{s,t}^{\circ}$ varies across regions or periods, $\gamma^*$ does not generally coincide with $\gamma_{s,t}^{\circ}$ for every $(s,t)$, and the corresponding region-period bias in Equation \eqref{appendix: biasst} need not equal zero.

In contrast, under a correctly specified flexible TWFE, the covariate coefficient is allowed to vary according to the relevant CCC violation. For the two-way flexible TWFE, for example, the corresponding population coefficient is $\gamma_{s,t}^*$. Under correct specification and the additive intercept restriction,
\begin{equation}
    (\alpha_s^{\circ}+\delta_t^{\circ})
    -
    (\alpha_s^*+\delta_t^*)
    =
    0
\end{equation}
and
\begin{equation}
    \gamma_{s,t}^{\circ}
    -
    \gamma_{s,t}^*
    =
    0.
\end{equation}
Therefore, the corresponding misspecification component is zero under the correctly specified flexible TWFE.

\section{Identification under Doubly Robust DID}
\label{Appendix: DRDID}

The Doubly-Robust DID estimand for repeated cross-sectional data is given by:
\begin{equation}
    \tau^{dr,rc}
    =
    \mathbb{E}
    \biggr[
    \biggr(
    \omega_1^{rc}(D_s,T_t)
    -
    \omega_0^{rc}(D_s,T_t,X_{i,s,t};\pi)
    \biggr)
    \biggr(
    OR_{i,s,t}
    \biggr)
    \biggr].
\end{equation}
where,
\begin{equation}
    \omega_1^{rc}(D_s,T_t)
    =
    \biggr(
    \frac{D_sT_t}{\mathbb{E}[D_sT_t]}
    -
    \frac{D_s(1-T_t)}{\mathbb{E}[D_s(1-T_t)]}
    \biggr)
\end{equation}
and
\begin{equation}
\begin{aligned}
    \omega_0^{rc}(D_s,T_t,X_{i,s,t};\pi)
    ={}&
    \frac{
    \frac{\pi(X_{i,s,t})(1-D_s)T_t}
    {1-\pi(X_{i,s,t})}
    }
    {
    \mathbb{E}
    \bigr[
    \frac{\pi(X_{i,s,t})(1-D_s)T_t}
    {1-\pi(X_{i,s,t})}
    \bigr]
    }
    \\
    &-
    \frac{
    \frac{\pi(X_{i,s,t})(1-D_s)(1-T_t)}
    {1-\pi(X_{i,s,t})}
    }
    {
    \mathbb{E}
    \bigr[
    \frac{\pi(X_{i,s,t})(1-D_s)(1-T_t)}
    {1-\pi(X_{i,s,t})}
    \bigr]
    }.
\end{aligned}
\end{equation}
The outcome-regression component is given by
\begin{equation}
\begin{aligned}
    OR_{i,s,t}
    ={}&
    T_t
    \left[
    Y_{i,s,t}
    -
    \mathbb{E}^*
    \left[
    Y_{i,s',t}
    \mid
    S_i=s',
    T_t=1,
    X_{i,s',t}=x
    \right]
    \right]
    \\
    &+
    (1-T_t)
    \left[
    Y_{i,s,t^s-1}
    -
    \mathbb{E}^*
    \left[
    Y_{i,s',t^s-1}
    \mid
    S_i=s',
    T_t=0,
    X_{i,s',t^s-1}=x
    \right]
    \right].
\end{aligned}
\end{equation}

The DR-DID estimand combines the outcome-regression and propensity-score components. However, as discussed in Section \ref{sec:CS-DID}, the interacted covariate structures considered in this paper can violate the overlap condition required for propensity-score weighting. We therefore do not analyze the propensity-score component further and instead focus on the outcome-regression component of the estimator. We assume that the outcome regression is correctly specified, such that
\begin{equation}
\begin{aligned}
&\mathbb{E}^*
\left[
Y_{i,s',t}
\mid
S_i=s',
T_t=1,
X_{i,s',t}=x
\right]
=
\mathbb{E}
\left[
Y_{i,s',t}
\mid
S_i=s',
T_t=1,
X_{i,s',t}=x
\right],
\\
&\mathbb{E}^*
\left[
Y_{i,s',t^s-1}
\mid
S_i=s',
T_t=0,
X_{i,s',t^s-1}=x
\right]
=
\mathbb{E}
\left[
Y_{i,s',t^s-1}
\mid
S_i=s',
T_t=0,
X_{i,s',t^s-1}=x
\right].
\end{aligned}
\end{equation}
The resulting outcome-regression estimand for repeated cross-sectional data is
\begin{equation}
\label{appendix: orestimand}
    \tau^{or,rc}
    =
    \mathbb{E}
    \left[
    \omega_1^{rc}(D_s,T_t)
    OR_{i,s,t}
    \right].
\end{equation}

When $T_t=1$,
\begin{equation}
\label{appendix: OR1}
    OR_{i,s,t}
    =
    Y_{i,s,t}
    -
    \mathbb{E}
    \left[
    Y_{i,s',t}
    \mid
    S_i=s',
    T_t=1,
    X_{i,s',t}=x
    \right].
\end{equation}
When $T_t=0$,
\begin{equation}
\label{appendix: OR2}
    OR_{i,s,t}
    =
    Y_{i,s,t^s-1}
    -
    \mathbb{E}
    \left[
    Y_{i,s',t^s-1}
    \mid
    S_i=s',
    T_t=0,
    X_{i,s',t^s-1}=x
    \right].
\end{equation}

Plugging Equations \eqref{appendix: OR1} and \eqref{appendix: OR2} into Equation \eqref{appendix: orestimand} and simplifying, we get:
\begin{align*}
\tau^{or,rc}
={}&
\mathbb{E}
\left[
\frac{D_sT_t}{\mathbb{E}[D_sT_t]}
\left\{
Y_{i,s,t}
-
\mathbb{E}
\left[
Y_{i,s',t}
\mid
S_i=s',
T_t=1,
X_{i,s',t}=x
\right]
\right\}
\right]
\\
&-
\mathbb{E}
\left[
\frac{D_s(1-T_t)}{\mathbb{E}[D_s(1-T_t)]}
\left\{
Y_{i,s,t^s-1}
-
\mathbb{E}
\left[
Y_{i,s',t^s-1}
\mid
S_i=s',
T_t=0,
X_{i,s',t^s-1}=x
\right]
\right\}
\right].
\end{align*}

The above expression can be written as:
\begin{equation}
\begin{aligned}
\tau^{or,rc}
={}&
\mathbb{E}
\left[
Y_{i,s,t}
-
\mathbb{E}
\left[
Y_{i,s',t}
\mid
S_i=s',
T_t=1,
X_{i,s',t}=x
\right]
\mid
D_s=1,T_t=1
\right]
\\
&-
\mathbb{E}
\left[
Y_{i,s,t^s-1}
-
\mathbb{E}
\left[
Y_{i,s',t^s-1}
\mid
S_i=s',
T_t=0,
X_{i,s',t^s-1}=x
\right]
\mid
D_s=1,T_t=0
\right].
\end{aligned}
\end{equation}

By SUTVA, we can write observed outcomes in terms of their potential outcomes. Therefore, $Y_{i,s,t}=Y_{i,s,t}(1)$, $Y_{i,s',t}=Y_{i,s',t}(0)$, $Y_{i,s,t^s-1}=Y_{i,s,t^s-1}(0)$, and $Y_{i,s',t^s-1}=Y_{i,s',t^s-1}(0)$. Hence,
\begin{align}
\tau^{or,rc}
={}&
\mathbb{E}
\left[
Y_{i,s,t}(1)
-
\mathbb{E}
\left[
Y_{i,s',t}(0)
\mid
S_i=s',
T_t=1,
X_{i,s',t}=x
\right]
\mid
D_s=1,T_t=1
\right]
\nonumber\\
&-
\mathbb{E}
\left[
Y_{i,s,t^s-1}(0)
-
\mathbb{E}
\left[
Y_{i,s',t^s-1}(0)
\mid
S_i=s',
T_t=0,
X_{i,s',t^s-1}=x
\right]
\mid
D_s=1,T_t=0
\right].
\end{align}

Adding and subtracting $Y_{i,s,t}(0)$ inside the first expectation and simplifying, we get:
\begin{align}
\label{appendix: orintermediate}
\tau^{or,rc}
={}&
\mathbb{E}
\left[
Y_{i,s,t}(1)-Y_{i,s,t}(0)
\mid
D_s=1,T_t=1
\right]
\nonumber\\
&+
\mathbb{E}
\left[
Y_{i,s,t}(0)
-
\mathbb{E}
\left[
Y_{i,s',t}(0)
\mid
S_i=s',
T_t=1,
X_{i,s',t}=x
\right]
\mid
D_s=1,T_t=1
\right]
\nonumber\\
&-
\mathbb{E}
\left[
Y_{i,s,t^s-1}(0)
-
\mathbb{E}
\left[
Y_{i,s',t^s-1}(0)
\mid
S_i=s',
T_t=0,
X_{i,s',t^s-1}=x
\right]
\mid
D_s=1,T_t=0
\right].
\end{align}

Let the true untreated potential outcome model be given by:
\begin{equation}
\label{appendix: orpo}
    Y_{i,s,t}(0)
    =
    \theta^{\circ}_{s,t}
    +
    \gamma^{\circ}_{s,t}x
    +
    \upsilon^{\circ}_{i,s,t}
\end{equation}
and
\begin{equation}
\label{appendix: Attor}
    ATT(s,t)
    =
    \mathbb{E}
    \left[
    Y_{i,s,t}(1)-Y_{i,s,t}(0)
    \mid
    D_s=1,T_t=1
    \right].
\end{equation}
When the treatment effect varies with the covariate,
\begin{equation}
    ATT(s,t)
    =
    \int
    ATT_x(s,t)
    dF_{X_{i,s,t}\mid D_s=1,T_t=1}(x).
\end{equation}

Plugging Equations \eqref{appendix: orpo} and \eqref{appendix: Attor} into Equation \eqref{appendix: orintermediate}, we get the following expression:
\begin{align}
\label{appendix: orcpt}
\tau^{or,rc}
={}&
ATT(s,t)
+
\biggr(
\bigr(\theta^{\circ}_{s,t}-\theta^{\circ}_{s,t^s-1}\bigr)
-
\bigr(\theta^{\circ}_{s',t}-\theta^{\circ}_{s',t^s-1}\bigr)
\biggr)
\nonumber\\
&+
\int
\bigr(\gamma^{\circ}_{s,t}-\gamma^{\circ}_{s',t}\bigr)
x\,
dF_{X_{i,s,t}\mid D_s=1,T_t=1}(x)
\nonumber\\
&-
\int
\bigr(\gamma^{\circ}_{s,t^s-1}-\gamma^{\circ}_{s',t^s-1}\bigr)
x\,
dF_{X_{i,s,t^s-1}\mid D_s=1,T_t=0}(x).
\end{align}

Under Assumption \ref{as: compositional}, the distribution of the covariates within the treated region remains unchanged across the pre- and post-intervention repeated cross-sections. Therefore,
\begin{equation}
    F_{X_{i,s,t}\mid D_s=1,T_t=1}(x)
    =
    F_{X_{i,s,t^s-1}\mid D_s=1,T_t=0}(x),
    \qquad
    \forall x.
\end{equation}
Assumption 1(b) in \cite[p.~103]{sant2020doubly} similarly requires the joint distribution of treatment assignment and the covariates to remain invariant over time, thereby ruling out compositional changes.

Therefore, Equation \eqref{appendix: orcpt} can be written as:
\begin{align}
\tau^{or,rc}
={}&
ATT(s,t)
+
\biggr(
\bigr(\theta^{\circ}_{s,t}-\theta^{\circ}_{s,t^s-1}\bigr)
-
\bigr(\theta^{\circ}_{s',t}-\theta^{\circ}_{s',t^s-1}\bigr)
\biggr)
\nonumber\\
&+
\int
\biggr[
\bigr(\gamma^{\circ}_{s,t}-\gamma^{\circ}_{s,t^s-1}\bigr)
-
\bigr(\gamma^{\circ}_{s',t}-\gamma^{\circ}_{s',t^s-1}\bigr)
\biggr]
x\,
dF_{X_{i,s,t}\mid D_s=1,T_t=1}(x).
\end{align}

Under Assumption \ref{as: compositional}, integrated parallel trends implies:
\begin{equation}
\begin{aligned}
&
\biggr(
\bigr(\theta^{\circ}_{s,t}-\theta^{\circ}_{s,t^s-1}\bigr)
-
\bigr(\theta^{\circ}_{s',t}-\theta^{\circ}_{s',t^s-1}\bigr)
\biggr)
\\
&+
\int
\biggr[
\bigr(\gamma^{\circ}_{s,t}-\gamma^{\circ}_{s,t^s-1}\bigr)
-
\bigr(\gamma^{\circ}_{s',t}-\gamma^{\circ}_{s',t^s-1}\bigr)
\biggr]
x\,
dF_{X_{i,s,t}\mid D_s=1,T_t=1}(x)
=
0.
\end{aligned}
\end{equation}
Hence,
\begin{equation}
    \tau^{or,rc}=ATT(s,t).
\end{equation}

Now let us relax Assumption \ref{as: compositional} while maintaining the integrated parallel trends assumption. Under integrated parallel trends,
\begin{equation}
\begin{aligned}
&
\biggr(
\bigr(\theta^{\circ}_{s,t}-\theta^{\circ}_{s,t^s-1}\bigr)
-
\bigr(\theta^{\circ}_{s',t}-\theta^{\circ}_{s',t^s-1}\bigr)
\biggr)
\\
&=
-
\int
\biggr[
\bigr(\gamma^{\circ}_{s,t}-\gamma^{\circ}_{s,t^s-1}\bigr)
-
\bigr(\gamma^{\circ}_{s',t}-\gamma^{\circ}_{s',t^s-1}\bigr)
\biggr]
x\,
dF_{X_{i,s,t}\mid D_s=1,T_t=1}(x).
\end{aligned}
\end{equation}

Substituting this expression into Equation \eqref{appendix: orcpt}, we get:
\begin{align}
\tau^{or,rc}
={}&
ATT(s,t)
+
\int
\bigr(
\gamma^{\circ}_{s,t^s-1}
-
\gamma^{\circ}_{s',t^s-1}
\bigr)
x\,
dF_{X_{i,s,t}\mid D_s=1,T_t=1}(x)
\nonumber\\
&-
\int
\bigr(
\gamma^{\circ}_{s,t^s-1}
-
\gamma^{\circ}_{s',t^s-1}
\bigr)
x\,
dF_{X_{i,s,t^s-1}\mid D_s=1,T_t=0}(x).
\end{align}

Therefore, when Assumption \ref{as: compositional} is violated, the two integrals need not cancel. In the non-trivial case where
\begin{align}
&
\int
\bigr(
\gamma^{\circ}_{s,t^s-1}
-
\gamma^{\circ}_{s',t^s-1}
\bigr)
x\,
dF_{X_{i,s,t}\mid D_s=1,T_t=1}(x)
\nonumber\\
&\qquad
-
\int
\bigr(
\gamma^{\circ}_{s,t^s-1}
-
\gamma^{\circ}_{s',t^s-1}
\bigr)
x\,
dF_{X_{i,s,t^s-1}\mid D_s=1,T_t=0}(x)
\neq 0,
\end{align}
integrated parallel trends alone is not sufficient for the outcome-regression estimator to identify the ATT. Hence,
\begin{equation}
    \tau^{or,rc}\neq ATT(s,t).
\end{equation}

	\section{New Jersey and New York - estimates}\label{app:mc-two-state}

\begin{figure}[!htbp]
		\centering
		\caption{New Jersey and New York: Expanded Estimator Comparison}
		\label{fig:mc-nj-density}
		\includegraphics[width=\textwidth]{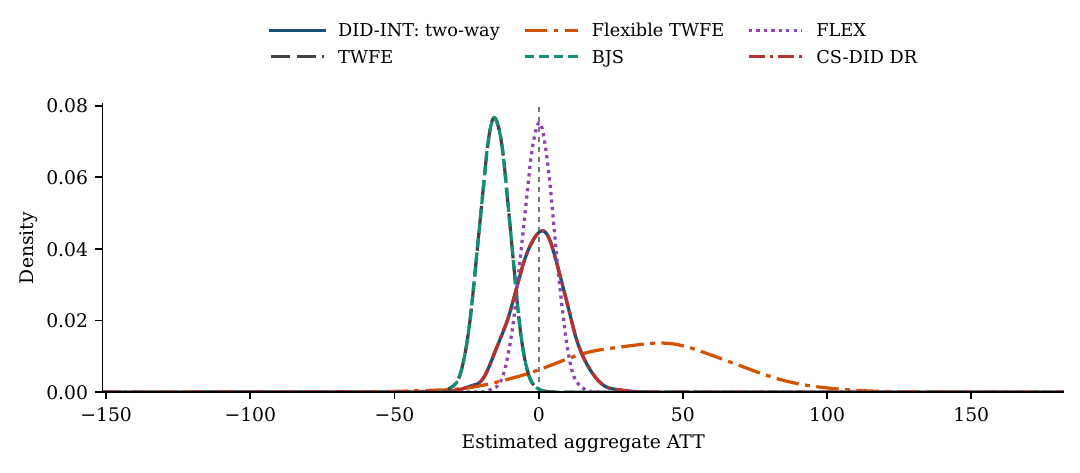}
		\par\smallskip
		{\footnotesize\raggedright\textit{Notes:}NJ--NY design: PT and CPT fail, while IPT holds. Densities use 2,500 replications; the reference line marks zero ATT.\par}
	\end{figure}

	Figure~\ref{fig:mc-nj-density} and Table~\ref{tab:mc-appendix} report estimates from the two state example where the returns to education increase over time in New York but not in New Jersey.  In this DGP CS-DID, FLEX, and DID-INT are unbiased, while BJS and TWFE are negatively biased, and  flexible TWFE is positively biased.
	
	\FloatBarrier\input{appendix_table.tex}\FloatBarrier
	
\end{document}

%% file: main_table_CPS.tex
\begin{table}[!htbp]
\centering
\caption{Monte Carlo Mean Error, Standard Deviation, and RMSE: CPS Covariate Signal (Panel I)}
\label{tab:mc-summary}
\begingroup\small\singlespacing
\setlength{\tabcolsep}{3pt}
\renewcommand{\arraystretch}{1.08}
\begin{tabular}{@{}lrrrrrr@{}}
\toprule
Estimator & {\footnotesize\shortstack[r]{Composition\\Shift}} & {\footnotesize\shortstack[r]{No\\Effect}} & {\footnotesize\shortstack[r]{Homogeneous}} & {\footnotesize\shortstack[r]{TE by\\Education}} & {\footnotesize\shortstack[r]{NY\\Premium}} & {\footnotesize\shortstack[r]{Two-Way\\Slopes}} \\
\midrule
\multicolumn{7}{@{}l}{\textit{Panel A: Monte Carlo Mean Error}} \\
DID-INT: homogeneous & -1.26 & -0.15 & -0.15 & 0.09 & -9.32 & 0.20 \\
DID-INT: state-varying & -0.14 & -0.15 & -0.14 & -0.44 & -9.32 & 1.28 \\
DID-INT: time-varying & -1.41 & -0.15 & -0.13 & 0.36 & -5.87 & -1.04 \\
DID-INT: two-way & -0.14 & -0.15 & -0.14 & -0.14 & 0.06 & 0.12 \\
TWFE & -3.80 & -0.02 & -0.02 & -0.22 & -11.27 & 15.68 \\
Flexible TWFE & 0.00 & -0.02 & -0.10 & -13.58 & 27.15 & 51.38 \\
BJS & -3.00 & -0.00 & -0.00 & -0.00 & -13.38 & -11.46 \\
FLEX & 1.14 & 1.23 & 0.78 & 0.78 & -0.75 & -9.19 \\
CS-DID DR & 2.19 & 1.00 & 1.28 & 1.01 & 0.05 & 2.17 \\
\midrule
\multicolumn{7}{@{}l}{\textit{Panel B: Standard Deviation}} \\
DID-INT: homogeneous & 5.58 & 5.57 & 5.58 & 5.58 & 5.55 & 5.44 \\
DID-INT: state-varying & 5.58 & 5.57 & 5.57 & 5.57 & 5.55 & 5.45 \\
DID-INT: time-varying & 5.59 & 5.57 & 5.59 & 5.59 & 5.60 & 5.45 \\
DID-INT: two-way & 5.65 & 5.57 & 5.60 & 5.60 & 5.73 & 5.49 \\
TWFE & 3.03 & 3.03 & 3.03 & 3.03 & 2.95 & 2.99 \\
Flexible TWFE & 16.36 & 3.03 & 16.21 & 16.21 & 15.96 & 15.61 \\
BJS & 3.26 & 3.26 & 3.26 & 3.26 & 3.20 & 3.17 \\
FLEX & 3.36 & 3.26 & 3.31 & 3.31 & 3.35 & 3.21 \\
CS-DID DR & 5.61 & 5.58 & 5.60 & 5.60 & 5.73 & 5.48 \\
\midrule
\multicolumn{7}{@{}l}{\textit{Panel C: RMSE}} \\
DID-INT: homogeneous & 5.72 & 5.58 & 5.58 & 5.58 & 10.85 & 5.44 \\
DID-INT: state-varying & 5.59 & 5.58 & 5.58 & 5.59 & 10.85 & 5.59 \\
DID-INT: time-varying & 5.76 & 5.58 & 5.59 & 5.60 & 8.12 & 5.54 \\
DID-INT: two-way & 5.65 & 5.58 & 5.60 & 5.60 & 5.73 & 5.49 \\
TWFE & 4.87 & 3.03 & 3.03 & 3.04 & 11.65 & 15.96 \\
Flexible TWFE & 16.36 & 3.03 & 16.21 & 21.14 & 31.49 & 53.70 \\
BJS & 4.43 & 3.26 & 3.26 & 3.26 & 13.76 & 11.89 \\
FLEX & 3.55 & 3.48 & 3.40 & 3.40 & 3.43 & 9.73 \\
CS-DID DR & 6.02 & 5.67 & 5.75 & 5.69 & 5.73 & 5.89 \\
\bottomrule
\end{tabular}\endgroup
\par\smallskip
\begin{minipage}{\textwidth}\footnotesize
\textit{Notes:} 
\textit{Notes:} All results use 2,500 replications. We report Monte Carlo mean error, SD, and RMSE. True ATT is zero except in TE by Education. Panel I uses the CPS-calibrated within-cell covariate $R^2$.
\end{minipage}
\end{table}

%% file: main_table_R2.tex
\begin{table}[!htbp]\ContinuedFloat
\centering
\caption{Monte Carlo Mean Error, Standard Deviation, and RMSE: $2\times$ CPS Covariate-Signal $R^2$ (Panel II)}
\label{tab:mc-sensitivity}
\begingroup\small\singlespacing
\setlength{\tabcolsep}{8pt}
\renewcommand{\arraystretch}{1.08}
\begin{tabular}{@{}lrrrr@{}}
\toprule
Estimator & {\footnotesize\shortstack[r]{Composition\\Shift}} & {\footnotesize\shortstack[r]{Homogeneous}} & {\footnotesize\shortstack[r]{NY\\Premium}} & {\footnotesize\shortstack[r]{Two-Way\\Slopes}} \\
\midrule
\multicolumn{5}{@{}l}{\textit{Panel A: Monte Carlo Mean Error}} \\
DID-INT: homogeneous & -1.91 & -0.15 & -14.32 & 0.26 \\
DID-INT: state-varying & -0.14 & -0.14 & -14.32 & 1.97 \\
DID-INT: time-varying & -2.15 & -0.13 & -9.04 & -1.72 \\
DID-INT: two-way & -0.14 & -0.14 & 0.06 & 0.12 \\
TWFE & -6.01 & -0.02 & -17.28 & 24.92 \\
Flexible TWFE & 0.00 & -0.10 & 41.31 & 81.47 \\
BJS & -4.75 & -0.00 & -20.54 & -18.24 \\
FLEX & 1.41 & 0.78 & -1.19 & -14.71 \\
CS-DID DR & 2.75 & 1.28 & 0.05 & 2.84 \\
\midrule
\multicolumn{5}{@{}l}{\textit{Panel B: Standard Deviation}} \\
DID-INT: homogeneous & 5.58 & 5.58 & 5.55 & 5.44 \\
DID-INT: state-varying & 5.58 & 5.57 & 5.55 & 5.45 \\
DID-INT: time-varying & 5.59 & 5.59 & 5.60 & 5.45 \\
DID-INT: two-way & 5.65 & 5.60 & 5.73 & 5.49 \\
TWFE & 3.03 & 3.03 & 2.95 & 2.99 \\
Flexible TWFE & 16.36 & 16.21 & 15.96 & 15.61 \\
BJS & 3.26 & 3.26 & 3.20 & 3.17 \\
FLEX & 3.36 & 3.31 & 3.35 & 3.21 \\
CS-DID DR & 5.61 & 5.60 & 5.73 & 5.48 \\
\midrule
\multicolumn{5}{@{}l}{\textit{Panel C: RMSE}} \\
DID-INT: homogeneous & 5.89 & 5.58 & 15.36 & 5.44 \\
DID-INT: state-varying & 5.59 & 5.58 & 15.36 & 5.79 \\
DID-INT: time-varying & 5.98 & 5.59 & 10.64 & 5.71 \\
DID-INT: two-way & 5.65 & 5.60 & 5.73 & 5.49 \\
TWFE & 6.73 & 3.03 & 17.53 & 25.10 \\
Flexible TWFE & 16.36 & 16.21 & 44.28 & 82.95 \\
BJS & 5.76 & 3.26 & 20.79 & 18.51 \\
FLEX & 3.65 & 3.40 & 3.55 & 15.06 \\
CS-DID DR & 6.25 & 5.75 & 5.73 & 6.17 \\
\bottomrule
\end{tabular}\endgroup
\par\smallskip
\begin{minipage}{\textwidth}\footnotesize
\textit{Notes:} \textit{Notes:} All results use 2,500 replications and zero true ATT. Panel II doubles the within-cell covariate $R^2$ while holding CPS noise fixed. 
\end{minipage}
\end{table}

%% file: final_didint_att_rejection.tex
\begin{tikzpicture}
\begin{groupplot}[group style={group size=3 by 2,horizontal sep=1.0cm,vertical sep=1.15cm},
width=0.30\textwidth,height=0.25\textheight,xmin=.8,xmax=19.2,ymin=0,ymax=0.185,
xlabel={$J=L$},ylabel={Rejection frequency},scaled y ticks=false,
yticklabel style={/pgf/number format/fixed,/pgf/number format/fixed zerofill,/pgf/number format/precision=2},
tick label style={font=\scriptsize},label style={font=\small},title style={font=\small},
legend style={font=\scriptsize,draw=none,at={(.98,.98)},anchor=north east},grid=major,grid style={gray!20}]
\nextgroupplot[title={(a) $S=8$},xtick={1,3,7,11,15,19},xlabel={}]
\addplot[color=blue!70!black,mark=*,line width=1pt,mark size=1.4pt] coordinates {(1,0.08166000) (2,0.04432000) (3,0.05644000)};
\addlegendentry{Jackknife}
\addplot[color=orange!85!black,mark=square*,line width=1pt,mark size=1.4pt] coordinates {(1,0.04812000) (2,0.04766000) (3,0.04680000)};
\addlegendentry{RI-beta}
\addplot[black!60,dashed,no marks,domain=.8:19.2] {.05};
\nextgroupplot[title={(b) $S=16$},xtick={1,3,7,11,15,19},xlabel={},ylabel={}]
\addplot[color=blue!70!black,mark=*,line width=1pt,mark size=1.4pt] coordinates {(1,0.12226000) (2,0.07036000) (3,0.05372000) (4,0.04768000) (5,0.04466000) (6,0.05084000) (7,0.08902000)};
\addplot[color=orange!85!black,mark=square*,line width=1pt,mark size=1.4pt] coordinates {(1,0.04338000) (2,0.04862000) (3,0.04872000) (4,0.04978000) (5,0.04938000) (6,0.04928000) (7,0.04788000)};
\addplot[black!60,dashed,no marks,domain=.8:19.2] {.05};
\nextgroupplot[title={(c) $S=24$},xtick={1,3,7,11,15,19},xlabel={},ylabel={}]
\addplot[color=blue!70!black,mark=*,line width=1pt,mark size=1.4pt] coordinates {(1,0.14386000) (2,0.08288000) (3,0.06622000) (4,0.05786000) (5,0.05132000) (6,0.04666000) (7,0.04702000) (8,0.04794000) (9,0.05322000) (10,0.06298000) (11,0.10790000)};
\addplot[color=orange!85!black,mark=square*,line width=1pt,mark size=1.4pt] coordinates {(1,0.04956000) (2,0.04956000) (3,0.04932000) (4,0.04826000) (5,0.04912000) (6,0.04836000) (7,0.05000000) (8,0.04944000) (9,0.05102000) (10,0.04876000) (11,0.04656000)};
\addplot[black!60,dashed,no marks,domain=.8:19.2] {.05};
\nextgroupplot[title={(d) $S=32$},xtick={1,3,7,11,15,19}]
\addplot[color=blue!70!black,mark=*,line width=1pt,mark size=1.4pt] coordinates {(1,0.15258000) (2,0.09358000) (3,0.07508000) (4,0.06474000) (5,0.06072000) (6,0.05566000) (7,0.05046000) (8,0.04972000) (9,0.04808000) (10,0.04754000) (11,0.04966000) (12,0.05342000) (13,0.05926000) (14,0.06936000) (15,0.11750000)};
\addplot[color=orange!85!black,mark=square*,line width=1pt,mark size=1.4pt] coordinates {(1,0.05046000) (2,0.04896000) (3,0.04866000) (4,0.04998000) (5,0.05198000) (6,0.04928000) (7,0.04776000) (8,0.04952000) (9,0.04898000) (10,0.04908000) (11,0.04794000) (12,0.04924000) (13,0.05054000) (14,0.04870000) (15,0.04842000)};
\addplot[black!60,dashed,no marks,domain=.8:19.2] {.05};
\nextgroupplot[title={(e) $S=36$},xtick={1,3,7,11,15,19},ylabel={}]
\addplot[color=blue!70!black,mark=*,line width=1pt,mark size=1.4pt] coordinates {(1,0.15392000) (2,0.09584000) (3,0.07740000) (4,0.06868000) (5,0.05766000) (6,0.05462000) (7,0.05336000) (8,0.05162000) (9,0.05056000) (10,0.04818000) (11,0.04918000) (12,0.04840000) (13,0.04940000) (14,0.05468000) (15,0.05890000) (16,0.07178000) (17,0.12124000)};
\addplot[color=orange!85!black,mark=square*,line width=1pt,mark size=1.4pt] coordinates {(1,0.04852000) (2,0.04916000) (3,0.04816000) (4,0.04892000) (5,0.04728000) (6,0.04908000) (7,0.05094000) (8,0.04862000) (9,0.04878000) (10,0.04822000) (11,0.04936000) (12,0.04760000) (13,0.04782000) (14,0.05018000) (15,0.04896000) (16,0.04954000) (17,0.04852000)};
\addplot[black!60,dashed,no marks,domain=.8:19.2] {.05};
\nextgroupplot[title={(f) $S=40$},xtick={1,3,7,11,15,19},ylabel={}]
\addplot[color=blue!70!black,mark=*,line width=1pt,mark size=1.4pt] coordinates {(1,0.16090000) (2,0.10008000) (3,0.08114000) (4,0.06956000) (5,0.06258000) (6,0.06036000) (7,0.05622000) (8,0.05184000) (9,0.05184000) (10,0.05040000) (11,0.05018000) (12,0.04758000) (13,0.04928000) (14,0.04936000) (15,0.05250000) (16,0.05580000) (17,0.06118000) (18,0.07422000) (19,0.12920000)};
\addplot[color=orange!85!black,mark=square*,line width=1pt,mark size=1.4pt] coordinates {(1,0.04850000) (2,0.04876000) (3,0.04984000) (4,0.04862000) (5,0.04782000) (6,0.05028000) (7,0.05056000) (8,0.04900000) (9,0.04994000) (10,0.04908000) (11,0.05028000) (12,0.04904000) (13,0.04946000) (14,0.04848000) (15,0.04948000) (16,0.04908000) (17,0.04854000) (18,0.04830000) (19,0.04736000)};
\addplot[black!60,dashed,no marks,domain=.8:19.2] {.05};
\end{groupplot}
\end{tikzpicture}

%% file: appendix_table.tex
\begin{table}[!htbp]\centering
\caption{New Jersey--New York: Numerical Comparison}
\label{tab:mc-appendix}
\begingroup\small\singlespacing
\begin{tabular}{lrrr}\toprule
Estimator & Bias & Standard deviation & RMSE \\\midrule
DID-INT (two-way) & 0.13 & 8.83 & 8.83 \\
TWFE & -15.31 & 4.96 & 16.10 \\
Flexible TWFE & 36.81 & 27.71 & 46.07 \\
BJS & -15.31 & 4.96 & 16.10 \\
FLEX & -0.05 & 5.10 & 5.10 \\
CS-DID DR & 0.14 & 8.83 & 8.83 \\
\bottomrule\end{tabular}\endgroup
\par\smallskip
{\footnotesize\textit{Notes:} The same 2,500 replications as Figure~\ref{fig:mc-nj-density}; true ATT is zero and there are no failed estimates. CPS-noise SD is 466.6383. Flexible TWFE translates education by the frozen New Jersey share, $p_{NJ}=0.45885665$, and leaves other covariates uncentered.\par}
\end{table}